\documentclass[%
 reprint,
 amsmath,amssymb,
pra,
]{revtex4-2}

\usepackage{graphicx}
\usepackage{color}
\usepackage{dcolumn}
\usepackage{bm}
\usepackage{hyperref}

\begin{document}

\preprint{APS/PRA}

\title{Forbidden trajectories for path integrals}
\thanks{Published in Phys. Rev. A 107, 032207, 2023, \url{https://doi.org/10.1103/PhysRevA.107.032207}}

\author{Janusz E. Jacak}
 \email{janusz.jacak@pwr.edu.pl}
\affiliation{Deparment of Quantum Technologies, Wroc{\l}aw University of Science and Technology, Wybrze\.ze Wyspia\'nskiego 27, 50-370 Wroc{\l}aw, Poland}%



\begin{abstract}
The problem of the availability of trajectories for the Feynman path integral is considered. Forbidden trajectories for single particle integrals are featured in the case of quantum tunneling across barriers. In the case of multiparticle systems of indistinguishable identical particles, some  limits for the  availability of cyclotron braid trajectories are demonstrated, which leads to the explanation of statistics and correlation in quantum Hall systems of interacting 2D electrons. The homotopy-type restrictions for trajectories close to general-relativity singularities are discussed with indication of quantum properties of black holes manifesting themselves at quasar luminosity or at neutron star merger collapses. The related supplementation to conventional models of accretion disk luminosity in the close vicinity of the event horizon of a super massive black hole is proposed and compared with observations. 

\end{abstract}

\maketitle


\section{Introduction}
The Feynman path integral, originally defined for a single particle \cite{feynman1964},
gives the evolution operator matrix element in the position representation $\langle z_1|e^{i\int_{t_1}^{t_2}dt\hat{H}(t)/\hbar}|z_2\rangle$, expressed 
by the functional integral \cite{chaichian1}, $I(z_1,t_1;z_2,t_2)=\int d\lambda e^{i S[\lambda(z_1,t_1;z_2,t_2)]/\hbar}$, where $S[\lambda]=\int_{t_1}^{t_2} {\cal{L}}(\lambda(z_1,t_1;z_2,t_2))dt$ is the classical action for the trajectory $\lambda(z_1,t_1;z_2,t_2)$ with the initial point in the configuration space $z_1$ at time instant $t_1$ and the final point $z_2$ at $t_2$. The action is the time integral of the Lagrangian ${\cal{L}}$ for a selected trajectory and is the functional of trajectories with fixed initial and final conditions. 
		This functional is minimal (extremal) for the  real classical trajectory defined by the Euler-Lagrange equation for the extreme of the action functional, leading  to the classical Hamilton equations with the Hamilton function ${\cal{H}}$ corresponding to the Hamiltonian $\hat{H}$. The functional integral $I(z_1,t_1;z_2,t_2)$ is taken over all {\it possible}  classical trajectories linking points $z_1$ and $z_2$ including equally the extremal and all nonextremal ones, and the squared modulus of $I$ gives the probability of quantum transition of the particle between points $(z_1,t_1)$ and $(z_2,t_2)$. The functional integral $I$ is conventionally called a propagator. The equivalence of quantization by Feynman path integration and of canonical quantization in Hilbert space has been evidenced \cite{feynman1964,chaichian1} and the  functional integration quantization occurs very useful and is 
widely applied in the field theory and in condensed matter \cite{chaichian2}. 

The path integral quantization is explicitly nonlocal  in distinction to the Schr\"odinger equation. The latter is a differential equation for the wave function at a certain potential, but some global topological constraints must be additionally imposed beyond the Hamiltonian form, which does not display nonolocal topological conditions.  For example, for bosonic or fermionic particles the Hamiltonian is the same and the constraint imposed on the symmetry or antisymmetry of the wave function in position representation is an external condition. The same holds for choosing commutation or anticommutation algebra of field operators in second quantization representation. Path integral quantization is, however, more general and  allows for explicit inclusion of nonlocal factors, which have roots in topological properties of classical trajectories of particles. The algebraic topology \cite{spanier1966,mermin1979} can describe properties of classical trajectories in a system and some related global properties cause quantum effects readable via the path integral, supporting in this way related external conditions needed to be imposed on the equivalent Schr\"odinger equation. 

In the present paper we will demonstrate several examples of  topological effects in quantum systems, which are evident in path integral quantization but are not  explicit in the Schr\"odinger equation without additional restrictions. First, we will address an elementary example related to single particle quantum tunneling.

Next, we will present more complex topological effects related to multiparticle systems of indistinguishable particles, their quantum statistics and quantum correlations conditioned by particle interaction and governed  by the global homotopy properties of corresponding  configuration spaces
and restrictions imposed onto classical trajectories. The experimental illustration of topological behavior possible to be explicitly studied via path integral quantization will be referred to quantum Hall physics in 2D spaces and to specific behavior of multiparticle systems near gravitational singularity of a black hole, which may have significant high-energy consequences. 

In the following section the examples of constraints imposed on trajectories in path integrals are presented. The supporting detailed calculations and additional comments are shifted to appendixes.

\section{Which trajectories enter a path integral ?}

In the path integral for a single particle \cite{feynman1964},
\begin{equation}
	\label{gwn500}
	I(z_1,t_1;z_2,t_2)=\int d\lambda e^{i \int_{t_1}^{t_2}{\cal{L}}[\lambda(z_1,t_1;z_2,t_2)]/\hbar},
\end{equation}
the summation over  trajectories (integration with the measure $d\lambda$ in the space of trajectories)
concerns all accessible classical trajectories linking the initial point $z_1$ in the configuration space of this particle at time $t_1$ and the final point $z_2$ at time instant $t_2$. Lagrangian ${\cal{L}}=T-V$, where $T$ is the kinetic energy and $V$ is the potential, integrated over time gives  the action $S[\lambda]$ -- a functional over the domain of  trajectories.  
This functional is minimal (extremal) for classical trajectory defined by the Euler-Lagrange equation, $\frac{d}{dt}\left(\frac{\partial \cal{L}}{\partial \dot z}\right)-\frac{\partial {\cal{L}} }{\partial z }=0$,  according to the least action principle. 
The family of trajectories contributing to the Feynman path integral includes also arbitrary not extremal paths, but those which are classically accessible for the potential   $V(z)$ and for the topology of the configuration space. The contribution to the propagator  $I$ of all trajectories  with the weight $e^{iS[\lambda]/\hbar}$ reproduces the quantum behavior of the particle, the same as that in the Schr\"odinger formulation with the Hamiltonian $\hat{H}=-\frac{\hbar^2 \nabla^2}{2m}+V(z)$ (depending on configuration space, $z$ can be appropriate dimensionality vector defining position of the particle) \cite{feynman1964}. The quantum Feynman path integral is analogous to the Wiener classical path integral \cite{wienera}
 applied to the Brownian motion \cite{chaichian1}. In distinction to the well-defined Wiener measure for summation of probabilities \cite{wienera,chaichian1}, the summation of complex probability amplitudes in a quantum Feynman integral \cite{feynman1964} precludes a proper measure definition \cite{chaichian1}, and the approach is more heuristic. Commonly accepted is the discretization method for explicit summation over trajectories \cite{feynman1964,chaichian1}. This method clearly shows which trajectories contribute to the sum -- arbitrary continuous piecewise  paths of segments being minimal (extremal) for boundary conditions on discrete consecutive internal points (approximated even by straight line sectors, without any loss of generality at an infinitesimal time step of discretization), provided, however, that the real extreme trajectory exists for each segment. This is an important restriction for the discretization procedure, because next the integration over all internal points  is performed in limits of piecewise trajectory existence. For quadratic Langrangians these integrals are of Gaussian type, which attain the analytic form only for infinite limits. However, if the piecewise trajectory cannot be constructed in some coordination space region [due to features of the potential $V(z)$ or topological restrictions imposed] the limits cannot always be  infinite. This problem has been noticed by Pauli \cite{paulib,chaichian1} and also earlier for Wiener path integrals for Brownian motion with inaccessible regions \cite{chaichian1}.

Classical trajectories cannot enter  potential barriers of arbitrary height but can be virtually placed in energy region beyond the potential barrier as nonextremal paths for selected classical initial and final conditions, which can be also discretized without any restrictions (note that the discretization produces continuous but nondifferentiable paths, similarly as for the Wiener measure).  These trajectories lead to  quantum tunneling across the classically inaccessible region expressed by the Feynman path integral, the same tunneling effect as that via the solution of the Schr\"odinger equation. Nevertheless, in the case of an infinitely high barrier none classical trajectories beyond the barrier exist and the propagator across the barrier is zero \cite{gwn102}, which is equivalent to the fact  that quantum particles do not tunnel through infinitely high potential barriers and cannot penetrate such infinite vertical potential walls. This simple example shows that not all trajectories contribute to the Feynman path integrals, but only those which  are classically accessible (even piecewise) for a given potential and topological restrictions. In another example, the 1D oscillator potential $x^2$ is the parabola also of infinite height, which, however, does not restrict the possibility of arbitrary piecewise trajectory construction (as parabola is infinitely wide), contrary to a rectangular infinite well or potentials ranged by vertical asymptotes. Instructive would be also the infinite potential of a Dirac delta type, which despite its infinity allows a nonzero transition across it. This is clear due to the fact that the Dirac delta is not a function but a distribution and at best can be considered as the limit of the series of finite height and width ordinary functions with integer 1, which do not impose restrictions on trajectory existence in a path integral. The limiting Dirac delta is nonzero only in a single point (because of limiting value 1 of the integral it is not a function, however) and this point can be selected as fixed one of the discretization points, which does not conflict with the existence of trajectories on the left and on the right. Using the property for multiplication for path integral \cite{feynman1964} chaps. 2-5 (with integration over the intermediate time), one gets nonzero probability transmission from the left to the right.  Despite that the infinite barrier of a finite width also can be considered as limit of finite height functions, the nonzero width persists in the limit, and twice applying the multiplication property on walls of the barrier we get, however, a multiplicative zero factor from the inside of the barrier (without any accessible trajectories there),  which causes the whole  propagator to vanish.

The constraint to only accessible trajectories in path integrals has profound consequences, generally of topological character, which cannot be easily noticeable in  Schr\"odinger equation formulation. In the elementary example of vanishing of the wave function at  infinite potential barriers, this property is evident in path integral quantization but in the Schr\"odinger equation approach it needs a  study of limiting behavior of finite height barriers.  
The more complicated situations can occur in collective multiparticle systems of identical particles, when the restriction to only accessible classical trajectories in path integrals has experimentally verified consequences -- such  topological effects will be discussed 
in the following section.

\section{Path integrals for multiparticle systems of identical indistinguishable particles}

If one considers a $N$ particle system, then the trajectories contributing to path integrals are $N$-strand  bunches in the multiparticle configuration space. Particles can mutually interact and their dynamics can be restricted additionally by some external potential or other factors like a magnetic field, which can influence availability of trajectories. Each particle contributes  with its individual trajectory to the whole  multistrand classical paths.
If particles are distinguishable, then the configuration space of $N$-particle system is, 
\begin{equation}
	P_N=M^N-\Delta,
\end{equation}
where $M$ is the space (mathematically -- manifold) on which all $N$ particles are located, $M^N=M\times M\times \dots \times M$ is the $N$-fold product of the space $M$ to account individual trajectories of all particles, and the space $M$ is accessible for each particle equally. The set $\Delta$ is the diagonal subset of $M^N$ with coordinates of at least two particles coinciding. $\Delta$ is removed from the coordination space $P_N$ to ensure particle number conservation. 
In this case classical multiparticle  trajectories are the collections of  single particle individual paths restricted by the single-particle potential or a magnetic field and inter-particle interaction, avoiding  crossing of individual single-particle paths in $M$ in the same moment, as the diagonal subset $\Delta$ is removed from the space $P_N$. This does not cause topological quantum effects beyond those for the single particle case.  

The situation changes, however, if particles are identical and indistinguishable. Quantum indistinguishability of identical particles is not explicitly imprinted in multiparticle Hamiltonian $\hat{H}= \sum_i^N\left(\frac{-\hbar^2 \nabla_i^2}{2m}+V(z_i)\right)+\frac{1}{2}\sum_{i,j,i\neq j}U(z_i,z_j)$, and some external constraint must be imposed onto the Schr\"odinger equation (regarding the symmetry or antisymmetry of a multiparticle wave function as for bosons or fermions, correspondingly). This constraint is of an external character and relates to the notion of indistinguishability of particles against their mutual position swapping expressed by changes of their numeration, which should not cause any physically observable effect. Thus, the square of the wave function module must be assumed not to change at particle exchanges, which admits either a plus or minus sign of the wave function at exchanges of particle coordinate pairs, provided that the double exchange of the particle pair restores the initial situation. 
This conventional reasoning and the related intuitive definition of particle indistinguishability is, however, misleading for 2D manifolds $M$ and can be applied  exclusively to 3D physical space (or for higher dimension manifolds $M$). For a 2D manifold a double exchange of identical indistinguishable particles is topologically distinct from the neutral operation. This needs more precise consideration of particle indistinguishability. 

The rigorous inclusion of the indistinguishability of identical particles must be done in classical topological terms in the definition of the multiparticle configuration space, which for $N$  indistinguishable particles attains the form,
\begin{equation}
	\label{gwn90}
	F_N=(M^N-\Delta)/S_N,
\end{equation}
with the quotient structure by $S_N$ -- the permutation group of $N$-elements. This displays the immunity of the configuration space against arbitrary swapping of indistinguishable particles expressed by a permutation of their numbering. In other words,  points in  $F_N$ space are unified if they differ by numbering of particles only  (in $P_N$ space such points are different). The space $F_N$ does not have an intuitive geometrical visualisation in contrast to $P_N$. It is counter-intuitive to any imagination of the multidimensional configuration space with various distributions  of differently numbered particles represented, however, as the single point in $F_N$. A multiparticle trajectory between point distributions which differ in particle numbering only is thus a closed trajectory loop in $F_N$ space (in $P_N$ is open). 

Closed trajectory loops in the space $F_N$ form disjoint classes of closed multistrand continuous trajectories starting from a certain point in $F_N$ and finishing in the same point.  Closed multistrand loops in $F_N$  joining the same positions of $N$ particles distinctly numbered are called as braids.  These  braids are open multistrand trajectories in $P_N$, but are closed loops in $F_N$. Closed loops in any topological space (arc-connected) ${\cal{A}}$ can be characterized in a topological sense by the first homotopy group $\pi_1({\cal{A}})$ of this space \cite{spanier1966}. The $\pi_1({\cal{A}})$ group collects disjoined classes of inequivalent loops, which can be deformed one onto another without cutting within the class but cannot between different classes.     The space ${\cal{A}}$ is multiply-connected if the $\pi_1({\cal{A}})$ group is nontrivial. Otherwise, when this group is trivial. i.e.,    $\pi_1({\cal{A}})=\{\varepsilon\}$ (where $\varepsilon$ is neutral element in the group),  the space ${\cal{A}}$ is simply-connected.  If the space ${\cal{A}}$ is the multiparticle configuration space of indistinguishable particles, $F_N$, then  $\pi_1(F_N)$ is called as the braid group (more precisely, full braid group in distinction to pure braid group $\pi_1(P_N)$). The full braid group is usually nontrivial for $N\geq 2$ (except for some special cases, which will be discussed later) and its form depends strongly on the dimension of the manifold $M$. 

\begin{figure}
	\centering
	\includegraphics[width=0.65\columnwidth]{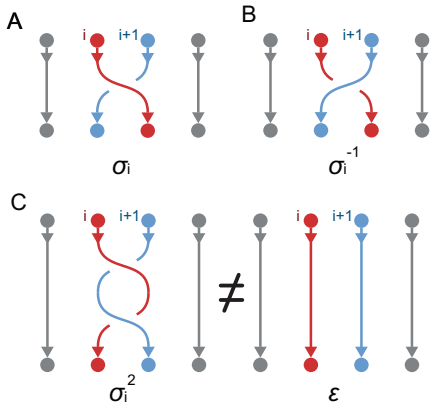}
	\caption{\label{braidpodst} Braid generators $\sigma_i$, $i=1,\dots,N-1$ define exchanges of $i$-th particle with $(i+1)$-th one, while other particles are at rest (at arbitrary but fixed particle numbering). Graphical presentation of $\sigma_i$ (in A) and $\sigma_i^{-1}$ (in B) are depicted. Braids must be closed loops in the space $F_N$, thus in the illustration the initial and final particle orderings are considered as unified. The braid $\sigma_i^2$ (in C), though does not change the particle ordering, is not a neutral element in the group $\varepsilon$, unless the manifold $M$ has the dimension $>2$.    	}
\end{figure}

\begin{figure}
	\centering
	\includegraphics[width=0.80\columnwidth]{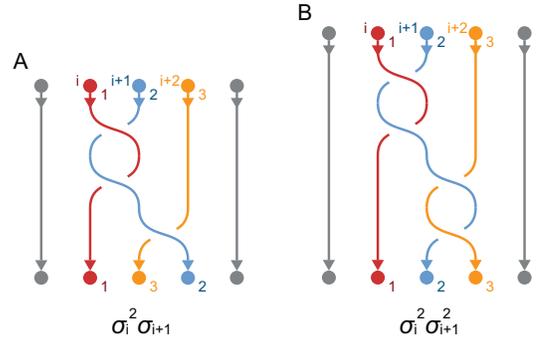}
	\caption{\label{czysta} The full braid group is the group $\pi_1(F_N)$, whereas the group $\pi_1(P_N)$ is called the pure braid group. An ordering of braided particles can change for $\pi_1(F_N)$ (left example), but for $\pi_1(P_N)$ must be conserved (right example). The pure braid group is a subgroup of the full braid group.  	}
\end{figure}

For a 3D manifold $M$ (or of higher dimension) the braid group $\pi_1(F_N)=S_N$ \cite{birman,mermin1979}, i.e., the full braid group is here always the ordinary finite permutation group with $N!$ elements. However, for 2D manifolds, the related braid groups are infinite and more complicated. For $M=R^2$ (2D plane) the full braid group is the so-called Artin group \cite{artin1947,birman}, which is an infinite and multicyclic group, i.e., is generated by a finite set of generators. These  generators can be chosen as $\sigma_i$, for $i=1, \dots ,N-1$ satisfying the following conditions \cite{birman};
\begin{equation}
	\label{warunki}
	\begin{array}{l}
		\sigma_i \sigma_{i+1} \sigma_i =\sigma_{i+1} \sigma_i \sigma_{i+1}, \;\;\; \textrm{for} \;1\leq i\leq N-2,\\
		\sigma_i \sigma_j =\sigma_j \sigma_i, \;\;\; \textrm{for} \; 1\leq i,j \leq N-1, \;|i-j|\geq 2.\\
		\end{array}
\end{equation} 
The generators $\sigma_i$ can be selected as elementary exchanges of positions of the $i$-th particle with the $(i+1)$-th one, at arbitrary but fixed numbering of all particles \cite{birman,mermin1979}, cf. Fig. \ref{braidpodst}. In the case of the 2D manifold $M$, $\sigma_i^2\neq \varepsilon$ contrary to 3D $M$, for which $\sigma_i^2=\varepsilon$. The latter condition makes the full braid group simple in 3D and equal to the permutation group $S_N$, despite that both the permutation and Artin groups are multicyclic generated by $\sigma_i$, elementary exchanges of neighbors (at arbitrary though fixed numbering of particles), but with different conditions imposed onto  these generators in both cases.
Full braid groups allow for a precise  definition of exchanges of identical indistinguishable particles on arbitrary manifolds. A particle swapping can be visualised in the space $P_N$ as shown in an example in Fig. \ref{czysta} and assuming that start and final points (before and after particle swapping) are the same point in $F_N$, then such a picture illustrates graphically a braid from $F_N$. Note, that the true braids in $P_N$ must also be  closed loops in $P_N$, as in any $\pi_1$ group, and thus must link the same distributions of particles with the same particle numbering. The braids in $F_N$ are not braids in $P_N$ except for braids in $F_N$ linking the initial and final points also with identical numbering, thus $\pi_1(P_N)$ is a subgroup of $\pi_1(F_N)$ \cite{birman,mermin1979}) (Fig. \ref{czysta}).

\begin{figure}
	\centering
	\includegraphics[width=1.0\columnwidth]{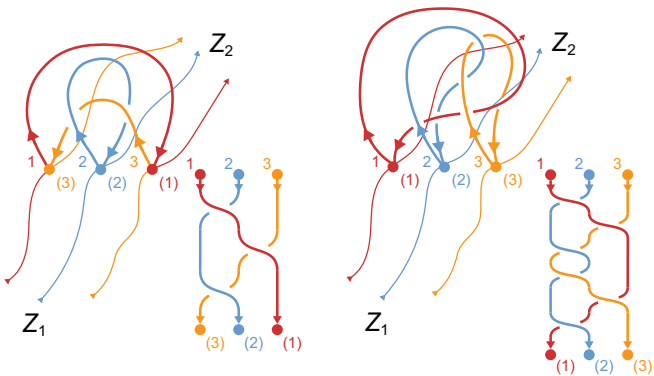}
	\caption{\label{gow11111} 
		In $F_N$ space a trajectory linking two different points $Z_1$ and $Z_2$ is the multistrand bunch of single-particle paths. The numeration of particles can be varied on the way, i.e., in any intermediate point of the open trajectory an arbitrary braid from $\pi_1(F_N)$ can be attached. In principle, an arbitrary finite number of braids can be attached to multistrand trajectory in various points on the way between initial and final points in $F_N$ -- however, according to the group structure of $\pi_1(F_N)$, the attachment of all these braids is equivalent to the attachment of the group multiplication of all of them, which is also a braid.  Braids are nonhomotopic, thus the space of trajectories decomposes into disjoined sectors numbered by various braids. In the illustration  different braids are attached to the same 3-strand trajectory in $F_3$ space.  Resulting trajectories cannot be continuously deformed one into another without cutting.  
	}
\end{figure}

Braids---elements from the full braid group---have a fundamental significance for the definition of multiparticle trajectories contributing to path integrals for identical indistinguishable particle systems. Open trajectories linking various points in $F_N$ space  can be attached at arbitrary point on their way by some closed loop -- the braid, which means that particles  can change their numbering on the way, cf. Fig. \ref{gow11111}. The attachment of an arbitrary finite number of braids to various points of the trajectory in $F_N$ is equivalent to the attachment of a single braid -- the group product  of all added braids. As braids are nonhomotopic, i.e., they cannot be continuously deformed one onto another without cutting, thus the whole space of multiparticle trajectories decomposes into disjoint nonhomotopic sectors. The discontinuity between sectors precludes the definition of the  measure $d\lambda$ for the path integral on the whole domain of trajectories. Instead, separate measures can be defined only on disjoined sectors of the domain and the contributions of all sectors must be finally added  up with arbitrary unitary weight factors (the unitarity of these weight factors is required to maintain the structure of the path integral displaying the causality in quantum mechanics \cite{lwitt-1}). Sectors in the domain of multiparticle trajectories of indistinguishable particles are  numbered by braids, thus the weight factors form the unitary scalar representation (1DUR, 1D unitary representation) of the full braid group \cite{mermin1979,lwitt-1,sud}. The Feynman path integral for the system of $N$ identical indistinguishable particles  attains thus the form,
\begin{equation}
	\label{gwn501}
	I(Z_1,t_1;Z_2,t_2)=\sum_l e^{i\alpha_l}\int d\lambda_l e^{i S[\lambda_l(Z_1,t_1;Z_2,t_2)]/\hbar},
\end{equation}
where 
points $Z_1=(z_1^1,\dots, z_N^1) $ and $Z_2=(z_1^2,\dots, z_N^2)$ are two {\it different} points in multidimensional configuration space $F_N$ of $N$ indistinguishable particles, which define the start and final points for the propagator $I(Z_1,t_1;Z_2,t_2)$ at time instants $t_1$ and $t_2$, respectively. 
 The discrete index $l$ enumerates braids in the full braid group and $e^{i \alpha_l}$, $\alpha_l\in[0,2\pi)$ is the scalar unitary representation of $l$-th braid (the element of 1DUR of the full braid group). Braid groups are generated by the finite number of generators, thus are countable or finite and  $l$ is a discrete index.
 The trajectory in $F_N$, $\lambda_l(Z_1,t_1;Z_2,t_2)$, is the trajectory between $Z_1$ and $Z_2$ but with attached the $l$-th braid from the full braid group at some intermediate point of this trajectory. Braids are nonhomotopic, thus the measure in the path space can be only defined separately on each disjoined sector of the path space numbered by $l$ -- such family of measures for path integration  is denoted by $d\lambda_l$.  

1DURs of braid groups are precisely defined, though the same braid group can have a variety of distinct 1DURs. Each one defines a different quantization of the same classical system. There are therefore as many different quantum particles corresponding to the same classical ones as there are different 1DURs of the full braid group for these classical particles.

For 3D manifolds $M$ the full braid group is always $S_N$ \cite{birman,mermin1979}. There exist only two 1DURs of $S_N$ \cite{birman,mermin1979},  defined on the group generators $\sigma_i\rightarrow \left\{ \begin{array}{l} e^{i0}\\ e^{i\pi}\end{array}\right.,  $ leading to bosons and fermions, respectively.

The Artin group \cite{artin1947} -- the full braid group for $M=R^2$ has, however, an infinite number of  1DURs \cite{birman,wu,imbo,sud}, $\sigma_i\rightarrow e^{i\alpha}$ with $\alpha\in[0,2\pi)$ corresponding to so-called anyons \cite{wilczek} (including bosons for $\alpha=0$ and fermions for $\alpha=\pi$). Note, that 1DURs of Artin group are uniform on its generators, i.e., 1DURs do not depend on the index $i$ of the generator, which follows from the first equation (\ref{warunki}). If one takes 1DUR of both sides of this equation, then bearing in mind  that scalar representations commute (are Abelian), one obtains  1DUR$(\sigma_i)$=1DUR$(\sigma_{i+1})$, which means independence of 1DUR($\sigma_i$) from $i$.   

Exchanges of indistinguishable particles defined by braid groups and their 1DURs display quantum statistics of particles and in the case of anyons are referred to fractional statistics \cite{wilczek,wu,sud}. 

The equivalence between path integral quantization and wave function formulation leads to 
restrictions imposed onto multiparticle wave functions by 1DURs of the related braid group \cite{sud,imbo}. If coordinates of the wave function $\Psi(z_1,\dots,z_N)$ (these coordinates represent classical positions of particles on $M$) are exchanged in the fashion defined by some selected braid from $\pi_1(F_N)$, then the multiparticle wave  function must gain a phase factor equal to 1DUR of this particular braid \cite{imbo}. This clarifies symmetry or antisymmetry of wave functions for bosons and fermions, and the phase shift $e^{i\alpha},\;\alpha\in[0,2\pi)$ at the exchange of a pair of coordinates for anyons. 

Despite that the exchange of particles looks like the exchange of two coordinate indices (simple permutation) in the multiparticle wave function $\Psi(z_1,\dots, z_N)$, we must, however, know the braid which actually  realizes this exchange in the related configuration space $F_N$, which essentially depends on the manifold $M$. Only for 3D manifolds $M$ (and for higher dimensionality) these braids are  simple permutations, but for 2D manifolds, not.  In the latter case, a simple permutation of coordinate indices in multiparticle wave function must cause a phase shift of this wave function $e^{i\alpha_l}$, which is the 1DUR of the braid ($l$-th one) actually realising the exchange of particles in planar geometry  prescribed by the simple permutation of coordinate indices. The actual form of this braid (being not the permutation in 2D) and its 1DUR (also different than that of  permutation) force the analytic form of the multiparticle wave function to  comply with these transformation requirements. Thus, we see that the form of braids, their availability in various physical situations and their 1DURs are essential for external constraints needed to be imposed on the solution
of the multiparticle Schr\"odinger equation (not fermionic or bosonic, in general), which is equivalent to the explicit role of 1DURs of the braid groups in  path integrals given by Eq. (\ref{gwn501}).

Braid groups depend also on topological factors imposed on multiparticle systems besides the dimensionality of $M$. A simple example of such factors is an infinitely high potential wall which can occur in some middle place of the manifold $M$. The system of $N$ particles in this way can be divided into subsystems $N^{'}+N^{''}=N$ on both sides of the wall. As any trajectories linking particles form the group $N^{'}$ and $N^{''}$ are not available, thus the full braid group $\pi_1(F_N)$ cannot be defined as the space $F_N$ starts to be not arc-connected and none quantum statistics can assigned to all $N$ particles simultaneously. Quantum statistics in this case can be assigned separately to subsets of $N^{'}$ and $N^{''}$ particles. If, however, the wall is of a finite height, then the trajectories linking $N^{'}$ and $N^{''}$ particles are admissible and the common statistics of all $N$ particles is restored. 

Other examples of forbidden trajectories induced by topological factors can be associated with a magnetic field in 2D systems of interacting particles and for 3D particles in folded spacetime close to gravitational singularity. Both these examples will be presented in the following section.

\section{Inaccessible trajectories for charged interacting particles in 2D at perpendicular magnetic field presence}

\begin{figure}
 	\centering
 	\includegraphics[width=1\columnwidth]{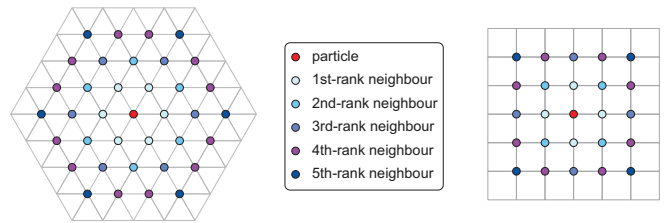}
 	\caption{\label{wignerlattice} Two possible classical 2D Wigner lattices - the hexagonal and regular ones. In both lattices the consecutive next-nearest neighbors are marked in different colors. The Bravais elementary cells in both lattices have the same surface size, $\frac{S}{N}$ (because per the elementary cell falls one particle in both lattices), but the hexagonal lattice is more convenient energetically (minimises the Coulomb interaction).}
 \end{figure}

\begin{figure}
 	\centering
 	\includegraphics[width=0.85\columnwidth]{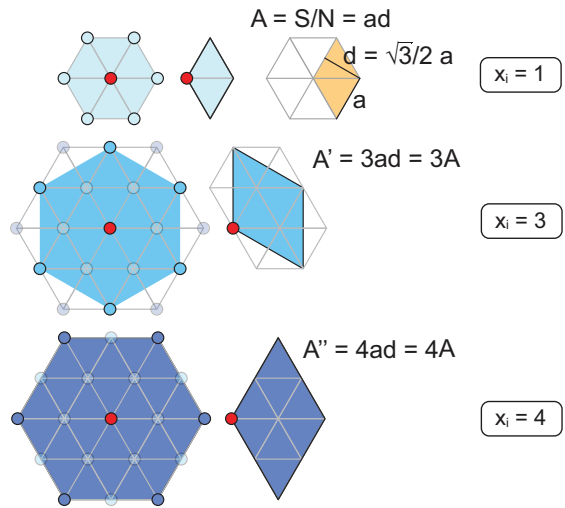}
 	\caption{\label{hex} The Bravais elementary  cells for the hexagonal Wigner 2D lattice for nearest, next-nearest and next-next-nearest neighbors. The sizes of elementary cells for consecutive rank neighbors determine the possible values of $x_i$, which can enter the commensurability condition (\ref{gwn11}). Note that first-rank next-nearest neighbor  sublattice contains only $\frac{1}{3}$ of all $N$ electrons, while in the regular lattice (cf. Fig. \ref{reg}) the first rank next-nearest neighbor sublattice contains $\frac{N}{2}$ electrons.   	}
 \end{figure}

\begin{figure}
 	\centering
 	\includegraphics[width=0.85\columnwidth]{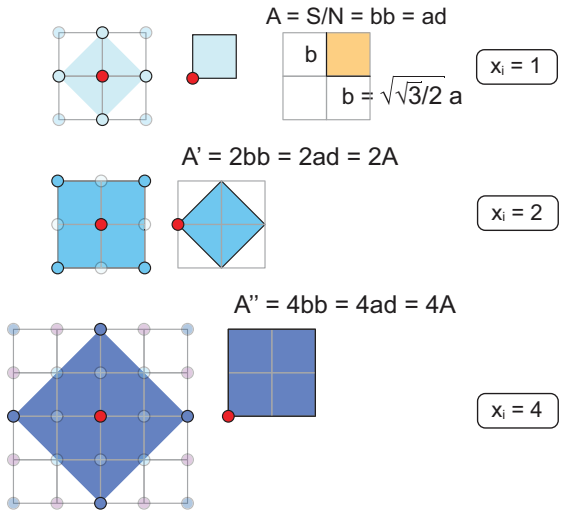}
 	\caption{\label{reg} The Bravais elementary cells for the regular Wigner 2D lattice for nearest, next-nearest and next-next-nearest neighbors ($a$ is the distance between electrons in hexagonal lattice and $d$ is the height in the triangle as in Fig. (\ref{hex})). The sizes of elementary cells for consecutive rank neighbors determine the possible values of $x_i$, which can enter the commensurability condition (\ref{gwn11}).  	}
 \end{figure}

 The magnetic field is not a potential one but can restrict trajectories of particles. Charged particles (let us say electrons) in a magnetic field spiral along the field direction and no other trajectories are attainable at the magnetic field presence for free particles. Braids are classes of homotopic trajectories. Trajectories which belong to the same class can be continuously deformed one into another without cutting  but if they belong to different classes, they are nonhomotopic (cannot be deformed one into another). The  braids inside  homotopy classes can be, however, restricted by topological factors.  In the case of 2D particles, their cyclotron orbits are flat, and therefore of finite size (in 3D not, as the drift along the magnetic field direction is not ranged spatially). They can be deformed but the availability of deformation can be limited in some special cases. As braids describe exchanges of particle numbering only, thus braids of size of finite cyclotron orbits must fit particle positions. In the 2D gas  mutual positions of particles are arbitrary, thus finite size of planar cyclotron braids does not impose any restriction. However, charged particles (e.g., electrons) repulse themselves and at $T=0$ K form a uniform classical distribution on 2D positive uniform jellium -- the classical 2D Wigner crystal with hexagonal Bravais lattice or a regular one \cite{ashcroftmermin} -- cf. Fig. \ref{wignerlattice}. The Bravais lattices minimise the interaction Coulomb energy of charged particles (electrons) distributed on positive uniform jellium, which is an electrostatic problem at $T=0$ K, when the classical kinetic energy is zero. The hexagonal lattice attains the minimal energy.   The  regular one has slightly greater Coulomb energy (classical kinetic energy in both cases is zero at $T=0$ K), but allows a convenient regular distribution of next-nearest neighbors (of first rank) including  half of all particles, whereas the hexagonal lattice gives only one third of particles as next-nearest neighbors (of first rank), as illustrated in Figs \ref{hex} and \ref{reg}.
 
Cyclotron braids, even if deformed,  must fit to Wigner crystal distribution of interacting particles in 2D, which can be expressed as the commensurability of cyclotron orbit size (the surface size of orbits) with elementary cells of Bravais lattice including  also Bravais sublattices of next-nearest neighbors of consecutive ranks (here the advantage of regular lattice manifests itself for next-nearest neighbors of first rank). Otherwise, when cyclotron  orbits do not commensurate with particle distribution -- neither for nearest neighbors nor for next-nearest neighbors of arbitrary rank, then braid trajectories are inaccessible and the braid group cannot be defined. 

The size of classical cyclotron 2D trajectories is defined by the Bohr-Sommerfeld rule. This rule precisely determines the surface field of 1D classical phase space encircled by a classical trajectory. The Bohr-Sommerfeld rule can be applied to components of 2D kinematic momentum of a single electron at constant perpendicular magnetic field, $P_x =-i\hbar \frac{\partial }{\partial x}$ and $P_y=-i\hbar\frac{\partial}{\partial y}-eBx$ ($e$ is the electron charge), which define the effective 1D phase space,  $Y=\frac{P_x}{eB}$ and $P_y$, because $[Y,P_y]_-=i\hbar$ (independently of magnetic field gauge, though here we used the Landau gauge for  concreteness, $\mathbf{A}=(0,Bx,0)$, $\mathbf{B}=rot\mathbf{A}=(0,0,B)$). The smallest surface in this effective phase space (according to the Bohr-Sommerfeld rule) determines the smallest classical cyclotron orbit size, $\frac{h}{eB}$ -- it is the surface of the quantum of magnetic flux $\frac{h}{e}$ conserved despite any deformation, $h=2\pi \hbar$ (details are shifted to Appendix \ref{A}).  This invariant surface size of the 2D cyclotron orbit plays the role of a topological factor limiting braid groups and their 1DURs. 

For magnetic field sufficiently large that the gap between Landau levels (LLs),  proportional to the magnetic field, is much larger than Coulomb interaction between electrons, one can consider separately electrons filling consecutive LLs in a multiparticle system with 
the size of cyclotron orbits $(2n+1)\frac{h}{eB}$, where $n$ is the Landau index (cf. Appendix \ref{A}). Moreover, one can distinguish between two opposite spin orientations of electrons as the Zeeman splitting is of the order of the gap between LLs and opposite spin electron subsystems can be considered separately. Spin orientation does not influence the cyclotron orbit size and does not change homotopy of trajectories. Due to indistinguishability of electrons, cyclotron braids must be consistent with LLs structure and magnetic field flux quantization. This consistency expresses itself via the commensurability condition of cyclotron orbit size and surface of classical Wigner crystal Bravais cells including sublattices of next-nearest neighbors of consecutive ranks, which ensures  possibility of the exchanging of interacting 2D electrons. Such a commensurability condition is the homotopy invariant for a particular LL. Note that LL structure and related size of cyclotron orbits are single-particle properties (Landau quantization of single 2D electron in perpendicular magnetic field).   Such a prequantization has been justified for path integration quantization of a single particle in a magnetic field via the equivalence to geometrical quantization \cite{gwn100,gwn101}.
As we aim to consider quantum statistics and correlations in a multi-electron system in a magnetic field, we adopt the similar concept of prequantization of the single particle. This allows for the identification of homotopy invariants separately for particular LLs including spin. 

The commensurability  condition of a Wigner crystal elementary cell (existing only for interacting particles) with  cyclotron orbits (or braids) is a multiparticle property. For fixed number of electrons $N$ per surface $S$ of the sample and the rigidly defined the surface size of cyclotron orbits in particular LLs (via Bohr-Sommerfeld constraint imposed on  surface of the classical orbit, cf. Appendix \ref{A}) one can get all admissible LL filling rates $\nu=\frac{N}{N_0}$ ($N_0=\frac{BSe}{h}$ is the degeneracy of LLs --  a single-particle property) via the commensurability condition for braids in the classical Wigner lattice of these electrons. This reproduces the hierarchy of quantum Hall effect observed experimentally, including integer (IQHE, integer quantum Hall effect \cite{klitzing1980}) and fractional (FQHE, fractional quantum Hall effect \cite{tsui1982}).
 
The simplest commensurability  condition for electrons in the lowest LL (LLL) with spin orientation along the magnetic field has the form,
\begin{equation}
	\label{gwn5}
	\frac{S}{N}=\frac{h}{eB},
\end{equation}
which means that the size of the elementary cell in Bravais lattice of classical Wigner crystal, $\frac{S}{N}$, perfectly fits to the surface size of single-loop cyclotron orbit of electrons in the LLL -- i.e., the quantum of magnetic field flux $\frac{h}{e}$ divided by the magnetic field $B$ (in the above equation $S$ is the sample surface and $N$ is the number of indistinguishable electrons). In this case the braids, which must be half of cyclotron loops \cite{annals2021}, can exchange the closest neighboring electrons. The commensurability condition (\ref{gwn5}) is a homotopy invariant, it  is robust against braid deformations, i.e., it confines topologically the homotopy of trajectories. If one takes into account that the degeneracy of Landau levels is $N_0=\frac{BSe}{h}$, the equation (\ref{gwn5}) can be rewritten as $\frac{N}{N_0}=1$. Thus we see that the cyclotron commensurability condition (\ref{gwn5}) occurs at the completely filled LLL, i.e., at filling factor $\nu=\frac{N}{N_0}=1$. However, this is not only complete filling of this level (which would happen even in the noninteracting gas), but (\ref{gwn5}) expresses the strong correlations in the interacting planar system of electrons, for which a Wigner crystal can be defined. The wave function corresponding to this correlated state must transform at coordinate exchanges as 1DUR of the full braid group (Artin group in this case). Assuming original electrons to be fermions, this 1DUR is $\sigma_i\rightarrow e^{i\pi}$. As the multiparticle wave function in the lowest Landau level must be analytical function \cite{prange}, thus the only possible form of this function is,
\begin{equation}
	\label{gwn6}
	\Psi_{\nu=1}(z_1,\dots,z_N)={\cal{A}} \prod_{i>j}^N(z_i-z_j)e^{-\sum_i^N|z_i|^2/4l_B^2},
\end{equation}
	where ${\cal{A}}$ is the normalization constant, $l_B=\sqrt{\frac{\hbar}{eB}}$ is the magnetic length and the envelope function $e^{-\sum_i^N|z_i|^2/4l_B^2}$ is the interaction independent factor for any multiparticle wave function in the LLL (because the single-particle state in the LLL at cylindrical gauge  of magnetic field is $f(z)e^{-|z|^2/4l_B^2}$, where $f(z)$ is arbitrary analytic function \cite{prange}, which can be chosen as $\sim z^n$, $n=0,\dots,N_0-1$ displaying Landau level degeneracy). The polynomial part of the wave function (\ref{gwn6}) is uniquely derived from 1DUR of the full braid group in the interacting system. The 1DUR, $\sigma_i\rightarrow e^{i\pi}$ forces the monomial $(z_i-z_{i+1})$ in the wave function, because  the exchange of $i$-th particle with $(i+1)$-th one is the rotation on the complex plane of the complex number $z_i-z_{i+1}$ by the angle $\pi$, which must produce in the wave function the phase shift $e^{i\pi}$ -- just as the monomial $z_i-z_{i+1}$.  This reasoning can be  generalized onto $(z_i-z_j)$, as the exchange of $i$-th particle with arbitrary $j$-th one is realized by the braid $\sigma_i\sigma_{i+1} \dots  \sigma_{j-1}\dots\sigma_{i+1}^{-1}\sigma_i^{-1}$, which has also 1DUR $e^{i\pi}$.
  
  One can notice that the polynomial in Eq. (\ref{gwn6})  is just the Vandermonde determinant of the Slater function of $N=N_0$ non interacting  fermions in the LLL (the envelope part is the same as in (\ref{gwn6})), 
\begin{equation}
	\label{vandermonde}
	\left|\begin{array}{l}1,z_1,\dots,z_1^{N_0-1}\\
		1,z_2,\dots,z_2^{N_0-1}\\
		\dots\\
		1,z_N,\dots,z_N^{N_0-1}\\
	\end{array}\right|=\prod_{i>j}^N(z_i-z_j).
\end{equation}
Due to (\ref{vandermonde}) we see that the same multiparticle wave function (\ref{gwn6})
is the ground state of interacting particles (here electrons), i.e., for strongly correlated state at $\nu=1$ corresponding to IQHE  and, on the other hand, is the ground state of the gaseous system of noninteracting particles at $\nu=1$, which is not IQHE state. The energies of this same state in the Hilbert space are different in both these cases, as they correspond to two different Hamiltonians,
\begin{equation}
	\label{gwn8}
	\begin{array}{l}
	\hat{H}_1=\sum_{i=1}^N \frac{(-i\hbar\nabla_i-e\mathbf{A}_i)^2}{2m}+\sum_{i,j,i>j}^N\frac{e^2}{4\pi\epsilon\epsilon_0 
	|\mathbf{r}_i-\mathbf{r}_j|}\\
+\frac{\rho_0^2}{2}\int_Sd^2r\int_Sd^2r'\frac{e^2}{4\pi\epsilon\epsilon_0 |\mathbf{r}-\mathbf{r'}|}\\
-\rho_0\sum_{i=1}^N \int_Sd^2r\frac{e^2}{4\pi\epsilon\epsilon_0 |\mathbf{r}-\mathbf{r}_i|},\\
\end{array}
\end{equation}
for interacting  electron system, and 
\begin{equation}
	\label{gwn9}
	\hat{H}_2=\sum_{i=1}^N \frac{(-i\hbar\nabla_i-e\mathbf{A}_i)^2}{2m},
\end{equation} 
for gaseous noncorrelated particles. In (\ref{gwn8}) the first term is the kinetic energy (the same as in (\ref{gwn9})), the next three  terms describe electron-electron, jellium-jellium and electron-jellium interactions, respectively. In above formulas, $\mathbf{A}_i$ is the vector potential of magnetic field $B$ for $i$-th electron; it is convenient to chose it in symmetric gauge, $\mathbf{A}_i=\frac{1}{2}(-By_i,Bx_i,0)$ for circular geometry of the sample with the surface $S$; $\rho_0=\frac{N}{S}$ is the uniform charge density of the jellium and $\epsilon_0$ and $\epsilon$ are dielectric constant and material permittivity, respectively.  For correlated state of the system described by the Hamiltonian (\ref{gwn8}) the energy is negative counted from the bottom of the LLL and depends on $N$ (can be calculated, cf. e.g., Ref. \cite{montecarlo1,montecarlo2}), whereas for the gaseous system is zero (in the same scale) and particle independent. The lowering of the energy due to correlations (\ref{gwn5}) of interacting electrons is small and stabilizes this state only if this energy gain is not washed out by the temperature chaos $k_BT$ ($k_B$ is the Boltzmann constant). Thus IQHE is observable at extremely low temperatures of the scale of the energy gain due to correlations, much lower than the gap between LLs. The "accidental" coincidence of the multiparticle wave function for the correlated state (\ref{gwn6}) with the gaseous Stater determinant for completely filled lowest Landau level in gaseous system follows from the same full braid group for the gas and for the correlations due to interaction (\ref{gwn5}) occurring at complete filling of the LLL. 

However, if the LLL is fractionally filled ($N<N_0$), then such a coincidence disappears and the multiparticle wave functions for interacting electrons are different than those for noninteracting gaseous particles. For interacting electrons   the pattern of braid correlation changes and the corresponding wave function is governed by  the 1DUR of a different braid group associated with different correlations. In the gaseous system none correlations exist (as the Wigner crystal does not exist without interaction) except for fermionic (for electrons) representation of full braid group of gaseous system, which results in the Slater function for arbitrary filling rate $\nu<1$ (with the degeneracy $N_0 \choose N$, the number of combinations of  $N$ elements from $N_0$). The wave function of interacting $N$ electrons is, however, different. In general, it is a linear combination of all $N_0 \choose N$ degenerated states of a gas, which span the appropriate subspace in the Hilbert space. For some specific filling fractions the multiparticle states will be correlated states with lower energy (the energy gain due to correlations for Hamiltonian (\ref{gwn8})), provided that for this filling rate some commensurability of cyclotron orbit with Wigner crystal is admitted.  These specific filling rates create the so-called hierarchy of FQHE \cite{annals2021} and are observed experimentally \cite{tsui1982,pan2003}. 

An example of such a situation is the state at fractional filling $\nu=\frac{1}{q}$ ($q$ odd integer) of the LLL described by the multiparticle Laughlin wave function \cite{laughlin2}. To identify theoretically other filling rates for correlated states (FQHE states) many various approaches have been proposed, like the Haldane-Halperin model \cite{hh1,hh2} of daughter states of consecutive anyon generations (anyons are associated with quasiparticles or quasiholes for Laughlin state \cite{laughlin2}), the composite fermion (CF) model of hypothetical quasiparticles created from electrons and pinned to them auxiliary fictitious magnetic field quanta \cite{jain} or the Halperin model of multicomponent Laughlin states \cite{halperin1983}. The anyonic model \cite{hh1,hh2} has been abandoned early, as the size of daughter wave functions quickly exceeds the sample size, CF fermion model partly elucidates the FQHE hierarchy in the LLL, but not of so-called enigmatic states in the LLL and fails in higher LLs. Moreover, the trial wave functions in CF model \cite{jain2007} do not keep the required symmetry and can be treated as only variational functions.  The Halperin multicomponent model perfectly agrees with energy gain (from experiment and from numerical simulation of toy systems) for some chosen  correlated states, but is unable to predict in advance which states it concerns. 

\begin{figure}
	\centering
	\includegraphics[width=0.65\columnwidth]{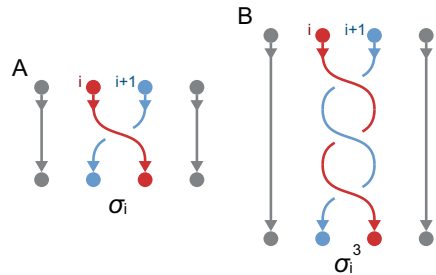}
	\caption{\label{sigma3} The braid $\sigma_i^{2k+1}$ with additional $k$ loops is also the exchange of  $i$-th particle with $(i+1)$-th one -- in the illustration for $k=1$ (right). In the case when $\sigma_i$ cannot be implemented as being too short, the braids $\sigma_i^{2k+1}$, $i=1,\dots,N-1$ take the role of  braid group generators  (in the simplest case $\sigma_i^3$).  They generate a cyclotron braid subgroup.	}
\end{figure}

In order to account for all LL filling rates at which the FQHE correlations occur one must consider available cyclotron braid trajectories for interacting electrons in 2D. 
To be more specific, let us consider larger magnetic field than that $B$ at which IQHE occurred, let say $3B$ at which the size of the single-loop cyclotron orbit in the LLL is $\frac{h}{e3B}$ and is far too small to match even nearest neighbors in the Wigner crystal lattice, $\frac{h}{e3B}<\frac{S}{N}$ (if compared to (\ref{gwn5}) for the same $S$ and $N$). Such short cyclotron braids (with size of single-loop cyclotron orbits) are thus inaccessible, i.e., the simplest exchange braids $\sigma_i$ cannot be  defined for electrons in the LLL at the field $3B$. Nevertheless, some other braids (also exchanges of nearest neighbors, cf. Fig. \ref{sigma3}), $\sigma_i^3$, $i=1, \dots ,N-1$ can be of larger size (cf. Appendix \ref{A}) sufficient to match nearest electrons in the Wigner lattice. The braids $\sigma_i^3$, $i=1,\dots,N-1$ (more generally, $\sigma_i^{2k+1}$)  can generate the new braid group -- the subgroup of the initial full braid group and called a cyclotron braid subgroup. This is related to the fact that the multi-loop cyclotron trajectory in the LLL has a larger size than the single-loop one in the case when single-loop cyclotron orbits and  braids built of their pieces  are too short and cannot match even nearest particles  (cf. Appendix \ref{A}). The multi-loop cyclotron orbit with $2k+1$ loops ($k$ integer)  has the size $ \frac{(2k+1)h}{eB}$ in the LLL  in the case when too small single-loop orbits cannot define  braids $\sigma_i$. The related new elementary braids --  half of the multi-loop cyclotron orbits, $\sigma_i^{2k+1}$, are the exchanges of closest electrons with additional $k$ loops (in the example, for  field 3B, $k=1$). The proof of this fact follows from the Bohr-Sommerfeld constraint imposed on orbits in multiply connected space \cite{pra}. This proof is briefly presented in Appendix \ref{A} and it displays the essential difference between $\sigma_i^{2k+1}$ in two situations -- in the case when $\sigma_i$ can be implemented, then the braid $\sigma_i^{2k+1}$ is simply $(2k+1)$ times repeating of $\sigma_i$ (and  $\sigma_i^{2k+1}$ is of the same size as $\sigma_i$ in this case); however, when $\sigma_i$ cannot be implemented as too short (is the forbidden trajectory), the braid $\sigma_i^{2k+1}$ is not multiple exchange, as $\sigma_i$ does not exist. In the latter case $\sigma_i^{2k+1}$ ($i=1,\dots, N-1$) take the role of the elementary exchanges of closest neighboring electrons, $i$-th with $(i+1)$-th one, and have a larger size. 

The braid subgroup generated by $\sigma_i^{2k+1}$ (called as the cyclotron braid subgroup \cite{annals2021}) has different 1DURs than the full braid group, and these new 1DURs decide on braid symmetry and shape of the corresponding multiparticle wave function. For exemplary 3B magnetic field, the multiparticle wave function attains the form of Laughlin function,
\begin{equation}
	\label{gwn16}
\Psi_{\nu=1/3}(z_1,\dots,z_N)={\cal{B}} \prod_{i>j}^N(z_i-z_j)^3e^{-\sum_i^N|z_i|^2/4l_B^2},
\end{equation}	
	and the filling fraction is $1/3$ (because degeneracy of LLs grows linearly with magnetic field and gives $\nu=\frac{N}{N_0}=\frac{1}{3}$ for $3B$). 
	
	The detailed mathematically rigorous derivation of the Laughlin function is presented in Ref. \cite{sr2022}.
In short -- when the full braid group had the 1DUR $\sigma_i\rightarrow e^{i\alpha}$, then the projective  1DUR for the cyclotron subgroup must be $\sigma^{2k+1}_i\rightarrow e^{i(2k+1)\alpha}$ (for original fermionic electrons one must choose $\alpha=\pi$). This projective 1DUR defines the symmetry of the polynomial part of the multiparticle wave function of interacting 2D electrons at field $(2k+1)B$ (in the simplest case, 3B). This polynomial must be of the form of Jastrow polynomial $\prod_{i>j}^N(z_i-z_j)^{2k+1}$ as in Laughlin function (\ref{gwn16}), because the monomial $(z_i-z_j)^{2k+1}$ is induced by the 1DUR of the braid $\sigma_i^{2k+1}$ (i.e., $\sigma_i^{2k+1}\rightarrow e^{i(2k+1)\pi}$) at the rotation of the complex number $(z_i-z_j)$ by the angle $\pi$ (exchange of $x_i$ with $x_j$ coordinates of multiparticle wave-function). Thus this monomial must be of the form $(z_i-z_j)^3$ (for $k=1$) as 1DUR$(\sigma_i^3)=e^{3\pi}$.  

The cyclotron subgroup (when inaccessible braids $\sigma_i$ are removed from the braid group) describes simultaneously the new commensurability condition (new correlation pattern, new homotopy invariant),
\begin{equation}
	\label{gwn10}
	\frac{(2k+1)h}{eB}=\frac{S}{N},
\end{equation}
For example,  for $k=1$  the condition (\ref{gwn10}) is satisfied at $3B$ (cf. Eq. (\ref{gwn5})), for which the degeneracy of the LLL equals to $N_0=\frac{3BS e}{h}$ and $\nu=\frac{N}{N_O}=\frac{1}{3}$ -- this is the most pronounced  manifestation of FQHE. The generalisation to other $k$ is evident	and leads to the derivation of Laughlin functions for $\nu=\frac{1}{q}, \; q=2k+1$ \cite{sr2022}. 

In the case when the $q$-loop cyclotron orbit does not fit to the nearest neighbors in the Wigner lattice, one can consider the commensurability of each loop separately taking into account also next-nearest neighbors in the  Wigner lattice. 
In this way  one can define the general commensurability condition \cite{annals2021},
\begin{equation}
	\label{gwn11}
	\frac{SB}{N}=\frac{h}{x_1e}\pm \frac{h}{x_2e}\pm \dots \pm\frac{h}{x_qe},
\end{equation}
which  gives the most general filling fraction for various phases of FQHE in the LLL, i.e., the complete FQHE hierarchy in the LLL,
\begin{equation}
	\label{gwn12}
\nu=\frac{N}{N_0}=\left(\frac{1}{x_1}\pm\frac{1}{x_2}\pm\dots \pm \frac{1}{x_q}\right)^{-1},		
\end{equation} 
where $N_0=\frac{BSe}{h}$ is the degeneracy of LLs (the single-particle property). In above formulas the particular  $x_i=1,2,3, \dots$, ($i=1,\dots, q$, $q=2k+1$) is the ratio of all electrons to next-next nearest neighbors  in the Wigner lattice; for consecutive ranks of next-nearest neighbors each of $x_i$ can attain independently of $i$ the values 1 (for nearest neighbors in hexagonal lattice), 2 (for first rank
  next-nearest neighbors in regular lattice), 3 (for first rank  next-nearest neighbors in hexagonal lattice),  4 (for second rank next-nearest neighbors in hexagonal lattice)  and so on (cf. Figs \ref{hex} and \ref{reg}). The $\pm$ between components in (\ref{gwn11}) and (\ref{gwn12}) display two possible circulations of consecutive loops in $q=2k+1$-loop cyclotron orbit, congruent or inverse (in the shape of an eighth, in the latter case).
The FQHE hierarchy (\ref{gwn12}) perfectly fits experimental data in GaAs and graphene \cite{pan2003,sr2022,sr1ccc,dean2011,amet,bil}. For $x_1=\dots=x_q=1$ (and  $+$ instead of all $\pm$) one gets $\nu=\frac{1}{q}$ as for Laughlin functions \cite{laughlin2}, ${x_i}$ larger than 1 define subsets of differently correlated electrons in a similar manner as in Halperin multicomponent state \cite{halperin1983}. The specific choice, $x_1=\dots=x_{q-1}=1$ and $x_q=y\geq 1$ (and $\pm$ before only last term) gives the CF hierarchy \cite{jain}, $\nu=\frac{1}{(q-1)y\pm1}$. More than single $x_i$ larger than 1 describe the so-called enigmatic FQHE states in the LLL inaccessible for CF model \cite{annals2021}.  

Various form of cyclotron braid subgroup conditioned by the availability of specific cyclotron braids corresponding to the general commensurability pattern (\ref{gwn11}) explains the origin of FQHE hierarchy and elucidates former partial heuristic models, like Halperin multicomponent generalisation of Laughlin function  \cite{halperin1983} or CF model  \cite{jain}. In Halperin theory \cite{halperin1983} the trial wave function in the form of multicomponent Laughlin function for the electron system  variationally divided into subsystems, corresponds  actually to 
subsets  of next-nearest  neighbors  of various rank in the electron Wigner lattice commensurate with selected loops of multi-loop orbit  as given in (\ref{gwn11}). This condition defines precisely the related generators of the cyclotron subgroups and also uniquely shapes (via 1DURs of particular cyclotron braid subgroups) the multiparticle wave functions  \cite{annals2021} -- as shortly summarised in Appendix \ref{B}. The subsets of electrons creating sublattices of particular rank of next-nearest neighbors in Wigner crystal are just those subsets of electrons featured in the  Halperin multicomponent wave functions \cite{halperin1983}.
In CF model the Jain's hierarchy $\nu=\frac{1}{(q-1) y \pm 1}$ \cite{jain} is the sub-case of (\ref{gwn12}) for a specific choice  $x_1=x_2=\dots =x_{q-1}=1$ and $x_q=y$ (and  last $\pm$ maintained). Only in this case the multiloop cyclotron orbit structure can be imitated by fictitious auxiliary field flux quanta ($q-1$ of them) attached to electrons in CF model. Jain suggested further \cite{jain} that his parameter  $y$ is the Landau index in the model artificial system of noninteracting spinless fermions in resultant magnetic field reduced by average field of fluxes pinned to electrons. This is, however, confusing, as actually $y=x_q$ and is linked to the rate of next nearest neighbors in Wigner lattice, $x_q$. This integer, similarly as other $x_i$  in Eq. (\ref{gwn11}), equals to $\frac{N}{N'}$ where $N'$ is the number of next-nearest neighbors of some rank  in the Wigner lattice of $N$ electrons. This mixing of the sense of the integer $y$ causes in CF model the next error -- the CF wave function is artificially assumed in the form of gaseous function in $y$-th completely filled LL in Jain's model system of spinless gaseous fermions and projected onto the LLL (to remove poles existing in gaseous functions in higher LLs but precluded in the LLL also for interacting electrons). Such a trial CF wave function has a correct rank of the polynomial part (for $y=1,2,3,4$) because $y$-th spinless LL is filled with $\frac{N}{y}$ electrons (similar as the number of next-nearest neighbors for $x_q=y$),  but it does not keep the braid symmetry exhibited by exact multielectron wave functions (as given in Appendix \ref{B}). Thus the CF trial wave functions  can be treated as the  variational wave functions only (the variational degree of freedom corresponds here to the arbitrariness in the definition of the projection onto the LLL from higher LLs in CF construction). The proper-symmetry wave functions for hierarchy (\ref{gwn12}) are presented in \cite{annals2021} (cf. Appendix \ref{B}).

We see thus, that FQHE is the result of  changing the accessibility of  cyclotron braid trajectories  when the magnetic field varies. The same occurs for FQHE in higher LLs, which agrees with FQHE observation in higher LLs \cite{jetpllaaa} in GaAs and in graphene, including bilayer graphene \cite{sr1ccc,bil1}. Discussion of cyclotron orbits and related braids availability explains all up to date observed features related to FQHE in GaAs 2DES in graphene monolayer and bilayer and also in fractional
Chern topological insulators without any magnetic field (substituted, however, by the Berry field). This is visualised both in the shape of multiparticle  wave functions and, equivalently, in quantum statistics (and correlations) described in the framework of path integral quantization.  The universal aspect of similar FQHE manifestations in different materials, like GaAs 2DES, graphene monolayer or bilayer and Chern topological insulators,  is linked with the forbidden braid trajectories in varying  homotopy classes of trajectories in response to a similar topological factor (magnetic field or Berry field).

The results presented in this paragraph and related to FQHE are invoked here in order to demonstrate the experimentally verifiable effects of inaccessibility of some trajectories in path integrals. They were formerly presented in even more issues for a variety of materials used in Hall experiments (GaAs and graphene, for both single layers and double layers). Interesting is to mention on experiment which allows the observation of the homotopy change of trajectories by a macroscopic factor on demand – a vertical electric field applied to double layer Hall system (bilayer graphene in this experiment \cite{maher}), which on demand can block hopping of carriers between layers in one way along the electrical field. This, however, completely precludes hopping of closed multi-loop trajectory between layers (all loops must be placed thus in a single layer, but without the electrical field, loops can be shared between both layers) and the effect is visible in the experiment as the change of FQHE hierarchy upon switching on and off this electrical field (experiment is simple as the voltage required is of order of 1 V  \cite{maher}.

\section{Homotopy of trajectories close to general-relativistic gravitational singularity}

Let us now consider constraints imposed onto classical trajectories close to the gravitational singularity associated with the notion of a black hole. The related folded spacetime can be described in terms of general-relativistic metric satisfying the Einstein equations for gravitation.
The  Schwarzschild metric \cite{schwarzschild} is the solution of general-relativistic Einstein equations for  a spherical   nonrotating and uncharged body with the mass $M$, 
\begin{equation}
	\label{metryka1}
	\begin{array}{lll}
	-c^2d\tau^2&=&-\left(1-\frac{r_s}{r}\right)	c^2dt^2+\left(1-\frac{r_s}{r}\right)^{-1}dr^2\\
	&+&r^2(d\theta^2+sin^2\theta d\phi^2),\\
	\end{array}
\end{equation}
where $\tau$ is the proper time, $t$ is the time measured infinitely far from the massive body, $r,\phi,\theta$ are rigid spherical coordinates, the same as for a remote observer, $r_s=\frac{2GM}{c^2}$ is the Schwarzschild radius defining the event horizon  ($G$ is the gravitation constant, $c$ is the light velocity in the vacuum). Eq. (\ref{metryka1})  defines the curved spacetime for $r>R$, where $R$ is the radius of a central body and for $R=0$ is addressed to a black hole without charge and angular momentum.  The Schwarzschild metric has a singularity at $r=0$, which is an intrinsic curvature singularity surrounded by the event horizon sphere with the radius $r_s$ -- the region from which neither matter nor light can escape. The Schwarzschild metric has also a singularity on the event horizon $r=r_{s}$, because of the second term in Eq. (\ref{metryka1}). This singularity is, however, apparent (called as coordinate singularity) and disappears when changing to other coordinates.  The metric (\ref{metryka1}) is, however, defined only on the exterior region $r>r_{s}$ or on the interior region $r<r_{s}$ -- in the latter region, the  time-like intervals became space-like and conversely. This is also an artificial property related to specific choice of coordinates -- in this case, the choice of ordinary spacetime coordinates the same as of a remote observer. If, however, one changes to another coordinate system corresponding to a different slicing of the same folded spacetime onto its spatial and time 	parts, the ostensible singularity at the event horizon disappears, but the horizon is still and universally defined as 	
the boundary surface of an area close to the central singularity  from which any matter or light cannot escape. In the metric (\ref{metryka1})  a remote observer can notice the matter falling onto event horizon of the black hole infinitely long, i.e., the falling matter  achieves the event horizon at $t\rightarrow \infty$ and it never crosses the horizon for this observer.  If one changes to other coordinates, e.g., using proper time instead of $t$,  the matter can smoothly pass the event horizon within a finite period of the proper time and then also within a finite period of the proper time any motion terminates in the central singularity.   This has been demonstrated in various  coordinates in metrics proposed by  Lemaitre, Eddington–Finkelstein, Kruskal–Szekeres, Novikov or Gullstrand–Painlevé  \cite{novikov,kruskal,szekeres}. 
Each of these specific metrics displays a different slicing of the same curved spacetime   into space and time its  components, with emphasizing of the Kruskal-Szekeres metric, being the maximally extended solution of the Einstein equations (analytic in the whole accessible domain) \cite{kruskal,szekeres}. However, there does not exist a rigid time-independent space coordinate system describing simultaneously the outer and inner regions with respect to $r_s$. Hence, the stationary metric (\ref{metryka1}), the Schwarzschild metric,  describes the falling of matter on the event horizon in infinite time in ordinary coordinates of a remote observer. In this metric, the spatial inner volume  below the event horizon is zero, though in nonstationary coordinates it is nonzero \cite{novikov,kruskal,szekeres} and particles pass the event horizon and terminate their motion in central singularity within a finite period of the proper time.  

The outer vicinity of the  event horizon, $r>r_s$, is well described by (\ref{metryka1}) in the stationary  system of coordinates $(t,r,\theta,\phi)$, the same as for a remote observer. In this region one can thus use the same path integral (\ref{gwn500}) or (\ref{gwn501}), without having to modify the coordinates of position and time.
 This convenient opportunity allows for the easy study of trajectories expressed in ordinary rigid and stationary coordinates and to observe the change of the homotopy of these trajectories as they approach the event horizon.
In this case a topological factor which precludes availability of trajectories except for spirals with single cross-points is the
curvature of gravitationally folded spacetime close to the singularity, as will be demonstrated below.

The qualitative change of the trajectory homotopy takes place at the so-called photon sphere rim at the distance from the central singularity $1.5 r_s$, i.e., $0.5 r_s$ beyond the horizon. The photon sphere rim in the metric (\ref{metryka1}) is defined by the innermost unstable circular orbit for any particle, even massless. Any particle, massive or massless, unavoidable spirals onto the event horizon if it passes the photon sphere rim in an inward direction. Beneath the photon sphere rim none circular orbits  are possible. This is quite different in comparison to Newton gravitation center, for which circular orbits are available at arbitrary small distance from the point-like gravitation center. For  Newton gravitation center various conic section trajectories are accessible for particles arbitrarily close to the center.  Even though for large distance from the gravitation center the Schwarzschild trajectories can be approximated by  conic sections (slightly modified with some additional precession of elliptical orbits, like that observed for Mercury in Sun gravitation), in close vicinity of the event horizon the difference in accessible trajectories is essential. Conic sections allow formation of  closed arbitrarily small local loops  from pieces of various trajectories because conic sections  can cross in two points (like circle with ellipse, hyperbole or parabola). This is in contrast to spirals below the innermost unstable circular orbit, where accessible spirals can intersect in a single point only.      Outside  the photon sphere the situation changes and closed trajectories are possible because deformed conic sections are admissible here. The innermost unstable circular orbit (the photon sphere rim) separates two different space regions with different classes of homotopy of trajectories. The proof of the above is presented in Appendix \ref{D}. 

An exchange of two particle positions in some manifold requires the existence of two oppositely directed different trajectory sectors joining these points, which together form a local closed loop in this manifold. 
The absence of such closed trajectory loops for particles beneath the photon sphere rim  precludes the possibility of mutual interchanges of particle positions in systems of $N$ identical indistinguishable particles located on some manifold $M^* \subset \tilde{R^3}$ ($\tilde{R^3}$ is the  3D position space $R^3$ folded according to the metric (\ref{metryka1})), and $M^*$ is the subset of the rotationally symmetrical region with $r\in(r_s,1.5r_s)$. Particles located in $M^*$ have admissible trajectories exclusively in the form of short spirals directed onto the event horizon. No other trajectories are possible here. It is unreachable to form a closed loop from pieces of these spirals, as different spirals defined by Eqs (\ref{promien}) and (\ref{phase}) can intersect only in a single point (in contrast to conic-section-like trajectories) -- cf. Eq. (\ref{phase}) in Appendix \ref{D} with the integral  in limits $r\in(r_s,1.5 r_s)$.  The homotopy type of trajectories cannot be changed by local interaction and trajectories for particle exchanges are not available here also for interacting particles.  The topological constraint induced by the space curvature in folded spacetime close to the gravitational singularity essentially limits the availability for trajectories.  If on a manifold $M^*$ $N$ identical indistinguishable particles are located, then the multiparticle configuration space of these particles, $F_N=(M^{*N}-\Delta)/S_N$ becomes simply-connected, whereas for any $M$ beyond the sphere $r=1.5r_s$, $F_N=(M^N-\Delta)/S_N$ is multiply-connected.

For a simply-connected space  the first homotopy group of this space is trivial, while for multiply-connected it is nontrivial \cite{spanier1966}.   For any manifold located in a spherical region $r\in (r_s,1.5r_s) $ the related braid group is trivial but for $r>1.5r_s$ is  equal to $S_N$, as usual in 3D.
 For $r>1.5r_s$ the fermionic or bosonic quantum statistics can be assigned \cite{mermin1979,birman} in contrary to the region $r\in(r_s,1.5r_s)$.

\begin{figure}
	\centering
	\includegraphics[width=0.8\columnwidth]{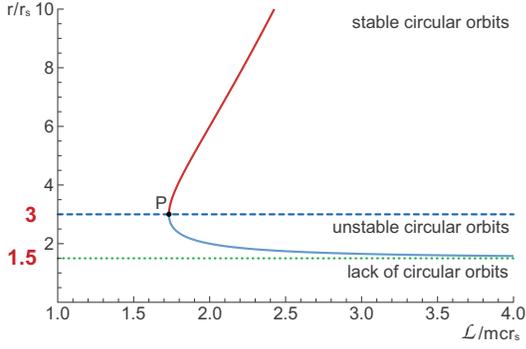}
	\caption{\label{photosphere} Radii of stable (red) and unstable (blue) circular orbits in Schwarzschild geometry. The innermost circular stable orbit occurs at $r=3r_s$ (point $P$ on the level of dashed line in blue colour), whereas the innermost unstable circular orbit occurs at $r=1.5 r_s$ (asymptotic dotted line in green colour). Below $r=1.5r_s$ none circular orbits exist.  	}
\end{figure}

The more detailed calculations from Appendix \ref{D} are illustrated in Fig. \ref{photosphere} -- the upper curve (red one in this figure) gives positions of stable circular orbits of a particle with small mass $m$, $m\ll M$ in the metric (\ref{metryka1}) (with respect to angular momentum of the particle ${\cal{L}}$, one of the motion integers) and the lower curve (blue one) gives positions of unstable circular orbits (also with respect to ${\cal{L}}$). 
The upper curve (the red one)  terminates in  the point $P$ at $r=3r_s$. This point defines   the innermost stable circular orbit. It is  at $r=3r_s$, ${\cal{L}}=\sqrt{3}mcr_s$ and energy (the second integer of a particle motion in gravitational field) ${\cal{E}}_0=\sqrt{\frac{8}{9}}mc^2$ (point $P$ in Fig. \ref{photosphere}) -- cf. the derivation in Appendix \ref{D}. 
The position of the innermost unstable circular orbit is at $r=1.5 r_s$ for ${\cal{L}} \rightarrow \infty $ and ${\cal{E}}_0\rightarrow \infty$ -- it is an asymptotic value defined by the horizontal dotted  line in Fig. \ref{photosphere}, marked  in  green colour. The orbit $r=1.5 r_s$ is also the unstable circular orbit for photons (by taking the limit $m=0$), thus it defines the  photon sphere rim in Schwarzschild metric.

\begin{figure}
	\centering
	\includegraphics[width=1\columnwidth]{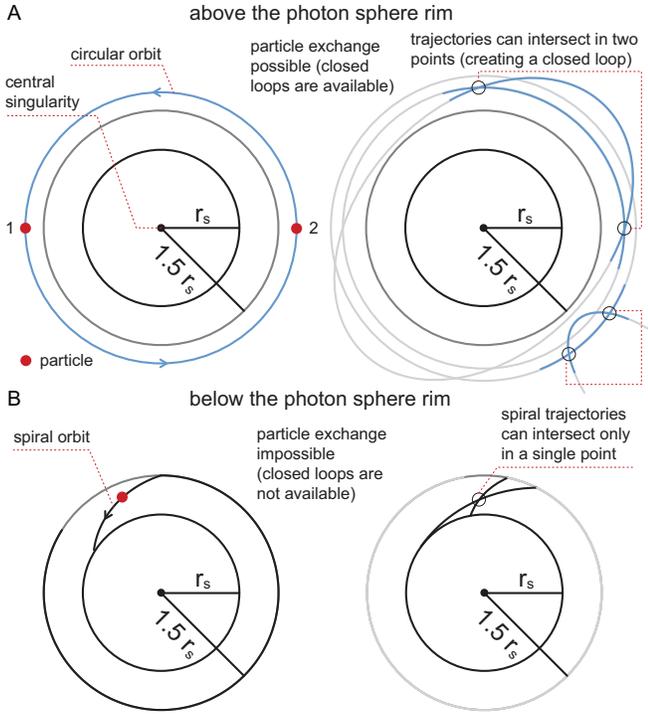}
	\caption{\label{gc} Simplified pictorial illustration of the change of trajectory homotopy at passing the innermost unstable circular orbit of a black hole. If conic section-type trajectories are available then the particle position interchanges are possible. Conic section-type trajectories beyond the photon sphere can intersect in two points and local closed loops can be constructed from pieces of such trajectories (upper picture). When only short spiral trajectories are admitted beneath the photon sphere rim and particles unavoidably fall towards the event horizon (lower picture),  then particles cannot mutually interchange  positions (these spirals can intersect in only one point and do not create loops needed for particle exchange).	Beyond the photon sphere  circular orbits are possible, which topologically allows particle exchanges, beneath the innermost unstable circular orbit at $r=1.5r_s$ (the rim of the photon sphere) none of circular orbits are accessible.}
\end{figure}

The homotopy class of trajectories is immune  to a change to distinct curvilinear coordinates at another choice of metric for the same gravitational singularity.
The homotopy of trajectories is the same in arbitrary equivalent metric, precluding particle interchanges below the innermost unstable circular orbit.  
Hence, for the inner of the event horizon, where the  dynamics of particles is also completely controlled by the central singularity and particles must  unavoidably travel  to the singularity point along short  spirals towards the origin despite any value of energy and angular momentum of particular particles and any strength of interparticle interaction, the exchanges of particles are inaccessible.   This is visible in Kruskal-Szekeres \cite{kruskal,szekeres} or  Novikov \cite{novikov} coordinates. Interparticle interaction does not change the  trajectory homotopy though locally can deform trajectories from its free particle shape, it cannot, however, produce closed cycles from one-way directed spirals  governed in an overwhelming manner by central singularity both beneath the event horizon and beyond it up to the sphere with the radius of the  innermost  unstable circular orbit.  This topological property  --
the qualitative change of the trajectory homotopy at the innermost unstable circular orbit is schematically illustrated  in Fig. \ref{gc}.

Because of the division by the permutation group $S_N$  in the definition of  $F_N$ space  (\ref{gwn90}), this space is not intuitive and differently numbered particle configurations are unified to the same point in $F_N$, which is counter-intuitive. Despite of  such a limitation in the intuitive geometrization of braids, there must be available  trajectories in $M$ that are able to connect particle pairs at some fixed but arbitrary their numbering. In particular, they are needed to be possible for the exchange of two particles along two distinct oppositely directed trajectories in $M$ joining these two particles   when other particles are at rest. This pair of trajectories creates a local loop in $M$. Due to indistinguishability of identical particles, if such a possibility exists for two selected particles, then it holds also for any other pair of particles.
The existence of circular orbits (for $r\geq 1.5r_s$) ensures trajectory topology sufficient to particle interchange. Two particles located at ends of an arbitrary  diameter of a circle can exchange their positions along such semicircular trajectories (as illustrated in Fig. \ref{gc}). The existence of circular trajectories (even if deformed by interparticle interaction) ensures in topological sense the possibility of the nontrivial braid group implementation.

The closure  of single particle orbits in $M$ does not mean the closure of loops in $F_N$, but some pieces of closed single-particle  orbits in $M$ (e.g., semi-circles of circular geodesics)  allow for the organization of elementary braids (of generators of the braid group $\sigma_i$, i.e., of  exchanges of particles $i$-th with $(i+1)$-th one at some fixed numbering of particles, conserving simultaneously positions of the rest of particles \cite{birman,mermin1979}). In other words, the generators of the braid group, $\sigma_i$, $i=1,\dots,N-1$, are $N$-strand trajectory bunches exchanging only $i$-th particle with $(i+1)$-th one when the other particles remain at rest, at  arbitrary but fixed particle numbering. This exchange of $i$-th and $(i+1)$-th particles  must be, however, available in the manifold $M$. It means that for a trajectory linking particle $i$-th with $(i+1)$-th one in $M$, it must exist another trajectory linking inversely particle $(i+1)$-th with $i$-th one, i.e., individual particle trajectories must be able to intersect at two points.  

Below the innermost unstable circular orbit (at $r=1.5 r_s$ for ${\cal{L}}\rightarrow \infty$) none closed orbit is possible. The dominating term in the effective potential (cf. Appendix \ref{D}, Eqs (\ref{differential}) and (\ref{potencjal})) is here $\sim -\frac{{\cal{L}}^2}{r^3}$ which causes an unavoidable spiral  movement towards the event horizon  of any particle despite its energy and angular momentum and local mutual interaction between particles. The phase shift for such spirals is all the more limited the larger ${\cal{E}}_0$ for given  ${\cal{L}}$.  From pieces of these short spirals it is impossible to form any closed local loop required to local particle interchanges (spirals given by Eq. (\ref{phase}) cannot intersect in two points simultaneously, in contrast to conic sections).  

Outside the photon sphere closed loops are possible because trajectories are quasi-conic-sections which admit local closed loops formed from their pieces which allows implementation of  braids  realising  particle interchanges. The photon sphere rim separates thus two distinct regions in the neighborhood of the black hole -- the inner of the photon sphere simply-connected and outer multiply-connected. The change of the homotopy class of the multiparticle configuration space causes local removal of the quantum statistics which cannot be assigned for a trivial braid group (if $\pi_1(F_N)=\{\varepsilon\}$ then  1DUR must be $e^{i\alpha}=1$, because $\varepsilon {\cdot} \varepsilon =\varepsilon$
 for the neutral group element $\varepsilon$, which gives $2 \alpha=\alpha$ and $\alpha=0$; this is not bosonic statistics as ${\varepsilon}$ does not describe any particle interchange). In particular, no fermions or bosons can be assigned if particles cannot interchange.

\section{Quantum collective transition beneath the innermost unstable circular orbit of a black hole}
1DURs of the braid group determine quantum statistics of particles for which classical multiparticle  trajectory loops in the configuration space $F_N$ (exchanges of particles) are defined by this braid group \cite{wilczek,mermin1979,lwitt-1}. For $\pi_1(F_N)=S_N$, as for arbitrary 3D $M$ beyond the photon sphere, there exist two distinct scalar unitary representations, 
$\sigma_i\rightarrow e^{i0}=1$ and $\sigma_i\rightarrow e^{i\pi}=-1$ defined on the generators of $S_N$,
$\sigma_i$ ($i=1,\dots,N-1$), which are exchanges of $i$-th and $(i+1)$-th particles. These two representations define bosons and fermions, respectively. The coincidence of these representations with the unitary irreducible  representations of the rotation group O(3) covered by SU(2) in 3D space, leads to the Pauli theorem on spin and statistics \cite{pauli}. The linkage between representations of both groups is detailed in Appendix \ref{C}.
For $\pi_1(F_N)=\{\varepsilon\}$ as for $M^*$ beneath the photon sphere rim of the black hole, only existing representation is $\varepsilon\rightarrow 1$ (by virtue of $\varepsilon \cdot \varepsilon=\varepsilon$) and it does not define any quantum statistics, neither bosons nor fermions, because trivial braid $\varepsilon$ is not the exchange of particles.

Thus, for particles beneath the photon sphere rim (the sphere with the radius of the innermost unstable circular orbit in metric (\ref{metryka1})) any quantum  statistics of these particles cannot be assigned. This does not violate the Pauli theorem on spin and statistics, as the latter is not defined for a trivial braid group without any linkage to rotations in 3D governed over particle spin. In particular,    fermions  cannot be assigned and particles with half spin lose their statistics.

The other theorem by  Pauli -- the so-called exclusion principle  asserts that quantum particles of fermionic type  cannot share any common single-particle quantum state. Fermions cannot approach a space region already occupied by another fermion, and thus they mutually repulse themselves. This is called the quantum degeneracy repulsion and the related  pressure is the origin of stopping the collapse of white dwarfs or neutron stars. In the former case the degeneracy pressure of electrons plays the role \cite{chandrasekhar}, whereas in the latter case of neutrons \cite{tolman,volkoff,olandau}. 

The exclusion principle for fermions leads also to the formation of the so-called Fermi sphere in the case of large number of identical fermions in some volume, when the chemical potential $\mu$ (the  energy increase by adding  a single particle to the considered multiparticle system) is much greater than the temperature of the system in energy scale, $k_BT$, where $k_B$ is the Boltzmann constant. In such a case, referred to as a quantum degenerate Fermi system, the particles are forced to occupy some consecutive in energy single-particle states one by one, resulting in the great accumulation of the energy. For example, free electrons in a normal metal in room temperature with the typical concentration of order of $10^{23}$ (of order of Avogadro number) per cm$^3$ constitute the large Fermi sphere with Fermi radius in momentum space $p_F \simeq 1.5 \times 10^{-24} $ kg m/s and with accumulated total energy $\sim 3\times10^{10}$ J/m$^3$. This energy cannot be released in normal conditions because all fermions are blocked in their single-particle stationary states by Pauli exclusion principle, i.e., all lower single-particle states are occupied and thus there is no room for fermions to jump  from  higher energy states to  lower ones.

 The situation changes, however, when quantum statistics cannot be defined below the photon sphere rim of a black hole. When quantum statistics cannot be assigned in some space region then  in the system of indistinguishable fermions the Fermi sphere collapses at entering this space region. The energy accumulated in the Fermi sphere can be released here.

The Fermi momentum $p_F$, the radius of the Fermi sphere in the degenerate homogeneous  quantum liquid of fermions depends only on particle concentration \cite{ashcroftmermin,abrikosov1975},
\begin{equation}
	\label{fm}
	p_F=\hbar(3\pi^2 \rho)^{1/3},
\end{equation}
where $\hbar=1.05 \times 10^{-34} $ J s is the reduced Planck constant and $\rho=\frac{N}{V}$ is the concentration of fermions, i.e., $N$ is the number of particles in the spatial volume $V$.
The Fermi momentum is independent of interaction of fermions according to Luttinger theorem \cite{luttinger,abrikosov1975}.   This follows from the fact that the phase space volume  $V \frac{4}{3} \pi p_F^3$ with the position space volume $V$ and the volume of momentum-space  of spherical shape $\frac{4}{3}\pi p_F^3$, corresponds to $n  =\frac{V8 \pi p_F^3}{3h^3}$ of single-particle quantum states according to   Bohr-Sommerfeld rule \cite{landau1972} (here additional factor $2$ accounts for doubling of states  for $\frac{1}{2}$ spin of fermions, and $h=2\pi \hbar$ is the Planck constant). If all these states are filled, i.e., when $n=N$, then one gets (\ref{fm}). The formula for Fermi momentum, as quasiclassically derived by  Bohr-Sommerfeld rule (immune to interaction), is independent of interaction of fermions in Fermi liquid even with arbitrary strong particle interaction. With Fermi momentum the Fermi energy is linked -- the kinetical energy of a particle with Fermi momentum (the chemical potential of fermions at $T=0$ K is equal to the Fermi energy and weakly changes with the temperature growth \cite{abrikosov1975}). 

The whole Fermi sphere collects the energy per spatial  volume $V$ (neglecting interaction of fermions),
\begin{equation}
	\label{en}
	\begin{array}{l}
		E=\sum_{\mathbf{p}} \varepsilon(\mathbf{p}) f(\varepsilon(\mathbf{p}))\\
		= \frac{V}{(2\pi \hbar)^3}\int d^3\mathbf{p}\varepsilon(\mathbf{p})f(\varepsilon(\mathbf{p}))\\
		=\int_0^{p_F}dp \int_0^{\pi} d \theta\int_0^{2\pi} d\phi  p^2 sin\theta \varepsilon(\mathbf{p}) \frac{V}{(2\pi \hbar)^3}\\
		=\frac{V}{2 \pi^2 \hbar^3}\int_0^{p_F} dp p^2 \varepsilon(p),\\
	\end{array}	
\end{equation}
where the sum runs  over occupied states only, what is guaranteed by Fermi-Dirac distribution function $f(\varepsilon(\mathbf{p}))=\frac{1}{e^{(\varepsilon(\mathbf{p})-\mu)/k_BT}+1}\rightarrow_{T\rightarrow 0}1-\Theta(\varepsilon(\mathbf{p})-\varepsilon_F)$ (here $\Theta(x)$ is the Heaviside step function and $\varepsilon_F=\varepsilon(p_F)=\mu(T=0)$, $\mu$ is the chemical potential),  $p,\theta, \phi$ are spherical variables in momentum space and $\varepsilon(\mathbf{p})$ is the kinetical energy of a fermion equal to $\frac{p^2}{2m}$ (in nonrealtivistic case), $\sqrt{p^2c^2+m^2c^4}-mc^2$ (in relativistic case) or $cp$ (in ultrarelativistic case). The factor $\frac{V}{(2\pi \hbar)^3}$ is the density of quantum states, i.e., the number of single-particle quantum states in the element of the phase space $Vd^3\mathbf{p}$. The energy estimation (\ref{en}) holds also for nonzero temperatures, if $k_BT\ll\mu\simeq\varepsilon_F$ (i.e., when Fermi liquid is quantumly degenerated). For example, in metals $\varepsilon_F\simeq 90000$ K (in units assuming $k_B=1$) and the Fermi sphere practically does not change even at melting temperature.

In another example -- a neutron star at Tolman-Oppenheimer-Volkoff limit \cite{tolman,volkoff} with the density of order of $10^{18}$ kg/m$^3$ (i.e., of the order of 2.3 Sun mass compressed to the compact neutron star with radius of ca. $10$ km), the neutron Fermi sphere energy attains the range of $10^{47}$ J, just as energy of frequently observed cosmic short giant gamma-ray bursts (assuming the isotropy of their sources) --  cf. Table \ref{tab-kw1}. The Fermi energy in this case 
$\varepsilon_F\simeq 3.7 \times 10^{12}$ K (at $k_B=1$ units), which is much greater than the supposed temperature of the neutron star,  of order of $10^6$ K.
\begin{table*}
	\caption{Fermi momentum $p_F$ (acc. to Eq. (\ref{fm})),  Fermi energy $\varepsilon_F=\varepsilon(p_F)$ (for relativistic case of kinetic energy $\varepsilon(\mathbf{p})=\sqrt{c^2p^2+m_n^2c^4}-m_nc^2$, $m_n=1.675\times 10^{-27}$ kg -- the mass of neutron) and total energy of the Fermi sphere $E$ (acc. to Eq. (\ref{en})) released at the decay of  statistics for  exemplary density $\xi$ and radius $r$ of a neutron star merger ($n$ total number of neutrons in the star).}             
	\label{tab-kw1}      
	\centering                          
	\begin{tabular}{c c c c c c}        
		\hline\hline                 
		$\xi$ [kg/m$^3$]  & $r$ [km] & $n$& $p_F$ [kg m/s]& $\varepsilon_F$ [GeV] & $E$ [J] \\    
		\hline                        
		$5\times10^{18}$& $10$ & $1.13\times 10^{58}$&$4.52 \times 10^{-19}$ &$0.32$ & $1.84 \times 10^{47}$ \\
		\hline 
		$2 \times 10^{18}$&10&$4.7 \times 10^{57}$&$3.37 \times 10^{-19}$&$0.2$&$4.8 \times 10^{46} $\\
		\hline
		$1.0 \times 10^{19}$&$8$&$1.08 \times 10^{58}$&$5.57\times 10^{-19}$&$0.58$&$3 \times 10^{47}$\\
		\hline
		$2.5 \times 10^{18}$&$8$&$2.97 \times 10^{57}$&$3.62 \times 10^{-19}$&$0.24$&$3.5 \times 10^{46}$\\
		\hline                                  
	\end{tabular}
\end{table*}

The same holds for any quantumly degenerated fermion system even at high temperature, i.e., when the chemical potential -- the energy increase caused by addition a single particle to multiparticle statistical system (at $T=0$ the chemical potential equals to $\varepsilon_F=\varepsilon(p_F)$ and weakly depends on the temperature growth) -- is much greater than the actual thermal energy $k_B T$, $k_B=1.38 \times 10^{-23}$ J/K is the Boltzmann constant, $T$ is the absolute temperature in the system.

The coincidence of the energy stored in the Fermi sphere of neutrons in a neutron star at Tolman-Oppenheimer-Volkoff limit with the energy of short giant gamma-ray bursts may support the concept that the source (yet unknown) of some of these bursts is a collapse of Fermi sphere of neutrons when the total star is compressed to the volume inside its own photon sphere. At the decay of the Fermi sphere of neutrons, the latters liberated from the Pauli exclusion principle constraint fall apart on electrons and protons, charge particles interacting with the electromagnetic field. Rapidly allowed jumping of these particles onto their ground state at the decay of the neutron Fermi sphere, will release giant flux of isotropic electromagnetic radiation along the Fermi golden rule for quantum transitions \cite{landau1972},  with dominant component of gamma radiation because of large value of Fermi energy in this case (cf. Table \ref{tab-kw1}).  

The next  example of the decay of Fermi sphere of fermions at passing the photon sphere rim of a black hole would be a quantum contribution to quasar luminosity. Quasars are sources of giant electromagnetic radiation, so they are visible from cosmological distances with luminosity exceeding hundreds times luminosity of a whole large galaxy. There are already known  over million quasars, the majority at a distance between 3 and 13 billion light years from us. Quasars are considered as supermassive active black holes (with mass of order of $10^9$ Sun mass) consuming a surrounding matter, which fall down onto the gravitational singularity in the form of a vast accretion disk. Various channels of gravitational energy transfer to radiation from the accretion disk are taken into account, mostly due to temperature radiation of hot gas of electrons and ions heated to millions K in due of matter compression. nonthermal components of radiation are related with bremsstrahlung or inverse Compton effect in the dense plasma close to the event horizon of the black hole, conventionally assumed  at the distance not closer than $6r_s$. 
    
If one considers an accretion disk of a quasar, then the density of electron and proton plasma (assuming accretion of neutral hydrogen) grows with falling of the matter towards the Schwarzschild horizon. The originally diluted neutral gas (of a hydrogen cloud)  ionises itself due to friction in the accretion disk and eventually becomes a degenerate Fermi liquid despite the high local temperature. This Fermi liquid attains an ultra-high concentration in an increasingly flattened and compressed region in vicinity of the event horizon. For two component plasma both Fermi spheres  of electrons and protons contribute to the energy storage. At the same concentration of electrons and protons (due to the neutrality condition of plasma in the disc) electrons accumulate larger kinetic energy because of lighter mass (the relativistic kinetic energy of the electron and  proton is $\varepsilon(\mathbf{p})=\sqrt{c^2p^2 +m^2c^4}-mc^2$ with $m=m_e=9.1 \times 10^{-31} $ kg and $m=m_p=1.67 \times 10^{-27}$ kg, respectively). 
The energy accumulated in the Fermi spheres of electrons and protons can be released in the form of the electromagnetic radiation if the Pauli exclusion principle is locally waived  at the rim of the photon sphere, due to local decay of quantum statistics. Charged particles couple to electromagnetic field and the collapse of their Fermi spheres is accompanied with the emission of photons in agreement with Fermi golden rule \cite{landau1972} for quantum transitions between initial states of particles in the Fermi sphere and their ground state.    

In the case of quasars surrounded by the abundance of the matter as in the cosmological epoch, when accretion of the gas is limited only by the uppermost density of matter compression (similar to the density of neutron star at Tolman-Oppenheimer-Volkoff limit or to the density of atom nuclei), the continuous decay of Fermi spheres of electrons and protons in accretion plasma crossing the photon sphere of supermassive central black hole releases  photons with total  energy in amount of 30 \% of infalling mass (quasars consume typically  ca. 0.1 Earth mass per second and 30 \% of this mass is converted into radiation). This radiation contributes to the luminosity of quasars with ca. $10^{40}$ W from the close vicinity of the event horizon (at the rim of the photon sphere at $r=1.5 r_s$ from the singularity) in better agreement with observations than only radiation from more distant regions of the accretion disk. The detailed quantitative estimation, including general gravitation correction to the density at the innermost unstable circular orbit of the black hole is presented in Appendix \ref{E}.

The topological effect of the change of homotopy class of particle trajectories at passing the innermost unstable circular orbit of a black hole seems to modify also  premises for the discussion of the information paradox for black holes and related problems. Since the illuminating papers of Hawking \cite{hawking} and Bekenstein \cite{bekenstein}, who pointed on the entropy behavior of black holes in view of second law of thermodynamics, the problem of the fate of information encoded in matter falling into a black hole is still open. The proposed temperature of black holes and the Hawking-Unruh radiation \cite{hawking,unruh} are considered as breakthrough quantum properties of black holes. However, it has been proved \cite{hawking1} that this radiation is fully random, which causes the information paradox. Trials to solve it via the quantum entanglement of particle-antiparticle pairs at the event horizon and the escape of a particle associated by the falling to the inside of the black hole of its antiparticle partner, encountered, however, problems with entanglement properties. Hawking radiation \cite{hawking} has been  resolved  to the entanglement of two particles -- escaping and falling into a black hole. Simultaneously, new Hawking radiation must be entangled with the old Hawking radiation, which leads to a conflict with the principle called "monogamy of entanglement". To avoid this problem, the entanglement must somehow get immediately broken between the infalling particle and the outgoing particle. Breaking this entanglement would release large amounts of energy, thus creating a searing black hole firewall at the black hole event horizon \cite{polczinski}. This resolution causes, however, a violation of Einstein's equivalence principle, which states that free-falling is indistinguishable from floating in empty space. No entanglement between the emitted particle and previous Hawking radiation would require black hole information loss, a controversial violation of unitarity.
	
	 Polchinski with co-authors \cite{polczinski} stated  that the paradox  may eventually force to give up one of three time-tested principles: Einstein's equivalence principle, unitarity, or existing quantum field theory. The firewall concept is one of a possible information paradox solution. The Hawking-Unruh radiation has not been observed as of yet and the firewall idea is still controversial as it breaks Einstein's equivalence principle \cite{polczinski}. Moreover, the holographic formulation for quantum gravitation by Maldacena \cite{maldac} strengthened the belief that the information should be conserved during the falling of matter into  black holes in accordance with unitary quantum evolution. Thus, the fundamental premises of quantum mechanics and of gravitation remain still in conflict in the case of black hole quantum behavior, which poses a significant problem for future quantum gravitation theory \cite{polczinski,maldac}.
	
	In the scenario proposed in the present paper, the matter deprived  of quantum statistics at passing the photon sphere rim, does not form a conventional system of particles but a specific quantum state of particles which cannot mutually interchange their positions, resembling an idealised perfect crystal in the third law of thermodynamics. For a remote observer the entropy of these particles would be even zero, if speculating they form a pure quantum state. Only due to Unruh effect \cite{unruh} the mixed state would be noticed by an accelerating observer (in particular, in a proper system attached to falling particles). The related Hawking-Unruh radiation is random \cite{hawking1}, completely independent of a matter state but is related to the kinematic  property in general relativity and the equivalence principle still holds at passing  the event horizon. Instead at the rim of the photon sphere the radiation is released due to the decay of quantum statistics and waving out of Pauli exclusion principle. This radiation is emitted according to the Fermi golden rule in an unitary manner. Thus, the entropy of the falling matter could be taken away with this radiation before crossing  the event horizon. This radiation is somewhat similar to a firewall, but in distinction to that proposed by Polchinski and co-authors \cite{polczinski}, is visible for remote observer (the Polchinski's firewall must be invisible for a remote observer as it would associate the entanglement breaking at passing the event horizon, which is unobservable for a distant observer).
	
We do not go here to a conclusion as the problem of the entropy change at local disappearance of quantum statistics needs a separate consideration, but we can point only on some related aspects, 1) the information in condensed matter is encoded mainly in matter organised according to Pauli exclusion principle, 2) the local waving out of quantum statistics restrictions allows the reducing of entangled states described by nonseparable symmetric and antisymmetric multiparticle wave functions of indistinguishable particles to separable states, 3) at passing the photon sphere rim by a single particle in a pure state nothing happens, which does not violate the classical Einstein's equivalence rule for such a particle.  

\section{Supplement to the conventional models of accretion disk and comparison with observations}

The collapse of Fermi spheres of  degenerate Fermi systems  at passing  the innermost unstable circular orbit in Schwarzschild metric  (photon sphere rim of a nonrotating uncharged black hole) does not conflict with the conventional models of the accretion disks of quasars \cite{sunaev,novikov}. The latters base on Shakura-Sunyaev classical hydrodynamic approach to plasma in accretion disk \cite{sunaev}, where the inverse transfer of the orbital momentum is modelled by friction and turbulence factors allowing  to increase plasma internal energy on the cost of the black hole gravitational energy \cite{salpet,zeld,ksiazka}.  Then hot plasma irradiates energy by thermal radiation \cite{sunaev,merloni}. This gives, however, only soft photons, and to elucidate giant luminosity of quasars (or micro-quasars) in X-ray range the mechanism of inverse Compton scattering of soft photons on hot electrons or ions is invoked \cite{shapiro}. The model of hot accretion disk, originally proposed by Shapiro et al. for micro-quasar Cignus X-1 with 15 Sun mass black hole \cite{shapiro} assumes extremely high temperature of the inner part of the accretion disk ($10^9$ K for electrons and $10^{11}$ K for ions) to assure sufficient energy of charge carriers needed to Comptonization of soft photons (the identification of a sufficiently abundant soft photon source is not clear, however). Though the Comptomization mechanism in this hot accretion disk model is adjusted to X-ray radiation luminosity of Cignus X-1, the generalisation of the model to extremely luminous giant quasars with supermassive black holes ($\sim 10^9$ Sun mass or larger) \cite{supereddington} is problematic, since the temperature $10^9 - 10^{11}$ K  of hot plasma in vast disc seems to be unrealistic. Nevertheless, some numerical simulations of developments of  Shakura-Sunyaev and Thorne-Novikov  model \cite{sunaev,novikov}  done recently by Fragile et al. \cite{6p} allow to match to observable luminosities of some not distant ($z<0.3$) active binary black hole objects \cite{farahzest}, but rather not to super-luminous remote quasars \cite{supereddington}. Higher radiation efficiency has been modelled within conventional classical magneto-hydrodynamic approach by inclusion of hypothetical giant magnetic component to accretion plasma in the case of a spinning black hole \cite{24p}, assuming, however, unrealistic  accretion mass rate to gain sufficiently large luminosity. 

All hydrodynamic or magneto-hydrodynamic models of matter accretion onto a black hole are applicable, however, only relatively far from the event horizon of the black hole. In all such models the inner edge of the accretion disk is assumed to be located well above the photon sphere of the black hole (the latter coincides with radius of the innermost unstable circular orbit in Schwawrzschild metric, at $r=1.5 r_s$, $r_s= \frac{2GM}{c^2}$), i.e., even more distant than the innermost stable circular orbit with the radius $3r_s$ (conventionally assumed the inner edge  of the accretion disk is at $\sim 6 r_s$ \cite{shapiro}). 

The collapse of the Fermi spheres described in the present paper takes place at the rim of the photon sphere at $r=1.5 r_s$, in the region completely neglected in conventional hydrodynamic                                                   models of matter accretion \cite{sunaev,novikov,6p,24p}, thus this mechanism does not interfere with thermal (including bremsstrahlung) and Comptonization mechanisms for radiation emission from more distant parts of the accretion disk (typically for $r>6 r_s$). Inclusion of the quantum effect of Fermi sphere collapse at the photon sphere rim ($r=1.5 r_s$) can, however, considerably  supplement developments of Shakura-Sunyaev-Thorne-Novikov  classical hydrodynamic approach \cite{sunaev,novikov} not applicable in close vicinity of the event horizon. The release of high energy photons (depending on the accretion mass rate, the black hole mass and governed by the Fermi momentum in compressed plasma) due to Fermi sphere collapse can add to the total luminosity of quasars and micro-quasars, allowing for the avoidance of some parameter-fitting  problems and shortages of conventional classical models \cite{6p,24p}. 

The rejection of the quantum statistics in a system of indistinguishable identical particles at passing the rim of the photon sphere is a general property of any black hole, regardless of its mass.
The energy per unit volume stored  in the Fermi sphere of a degenerate Fermi system is a monotonic function of the density of matter. Its maximum is attained for the uppermost possible density of compressed electron-hadron plasma, as in extremely bright quasars or in the case of neutron star mergers exceeding the Tolman-Oppenheimer-Volkoff  stability limit (the uppermost density of compressed Fermi liquid is of order of the density of hadrons in atom nuclei).
In the case of quasars the collapse of electron and proton Fermi spheres in the stream of ionised matter in accretion disk  gives the steady luminosity $\sim 10^{40}$ W for  $\sim 10^9$ Sun mass supermassive black holes consuming ca. 10 Sun mass per year ($0.1$ Earth mass per second) for a long time (as assessed in Appendix \ref{E}). In the  case of a neutron star merger which exceeds Tolman-Oppenheimer-Volkoff limit of ca. $2.3$ Sun mass compressed to uppermost density of hadrons, the merger  rapidly collapses  due to the relief of internal pressure caused by local recall of Pauli exclusion principle, when the entire star is smaller than its own photon sphere. The released energy due to collapse of neutron Fermi sphere in the neutron star merger reaches  $10^{47}$ J -- cf. Table \ref{tab-kw1} (this energy  partly escapes from the photon sphere of a rising black hole in the form of a short giant burst of gamma-rays). In both these extremal cases of the matter compression (in super-luminous quasars and at collapse of neutron star mergers)  the efficiency of mass to radiation conversion is ca. 30 \%  not achievable in any other physical mechanism except for matter-antimatter annihilation (the efficiency of nuclear fusion in stars is only of order of $0.7$ \% for mass to energy conversion rate). 

In the case of not extremal matter compression in the accretion disk of a black holes (as in many active galactic nuclei or in micro-quasars with lower rate of the matter influx) the radiation emitted due to Fermi sphere collapse is less intensive (the mass to energy conversion rate is lower than 30 \%) and softer,  but still contributes to the total luminosity and can help to explain radiation properties of observed binary black hole systems \cite{farahzest,supereddington}. In particular, the Fermi sphere collapse can help to elucidate the relatively short-lasting  brightening of active galactic nuclei, like the recently observed for AGN 1ES 1927+654 \cite{ga1flare}. The 100-fold increase of its luminosity within a few months period would be associated with accidental increase of the matter consumption rate during the corresponding time. If this accidental matter influx is not extremal, then photons emitted due to the Fermi sphere collapse  may not reach over-MeV energy and are not able to produce electron-positron pairs in the ergosphere of this spinning black hole. However, the massive isotropic flux of lower energy photons caused by the Fermi sphere collapse  of electrons and protons may push electron-positron pairs created in the ergosphere according to Blandford-Znajek electromagnetic mechanism \cite{znajek} towards the event horizon, lowering in this way their evaporation to jets across ergosphere nodes. The model by Blandford-Znajek \cite{znajek} gives the theory of jet formation for  spinning Kerr-like black holes, where due to dragging  of reference frame in Kerr metric the magnetic field carried with the accretion matter  rotates. The rotation of the ergosphere causes the magnetosphere inside it to rotate, the outgoing flux of angular momentum results in extraction of energy from the black hole. The magnetic field beams in the form of jets and electrons and positrons diffuse across nodes of the ergosphere and next are  highly accelerated in magnetic field beams in jets producing intensive X-ray radiation. The hypothetical source of electron-positron pairs is a strong electrical field created by the rotating magnetic field frozen in the ergosphere.  The sufficient intensity of the electric field to generate particle-antiparticle pairs in the ergosphere is, however, speculative. A collapse of Fermi spheres in plasma approaching the Kerr black hole would be helpful here, as it can supply the abundance of over-MeV photons in the case of extreme matter influx, when the energy of these released photons can reach GeV level and can produce large number of electron-positron pairs, which are able to power up jets besides the Blandford-Znajek mechanism. Nevertheless, in the case when the released photons are sub-MeV at smaller rate of matter consumption by a black hole, then they cannot produce  additional electron-positron pairs inside the ergosphere but can push towards the event horizon those created according to Blandford-Znajek mechanism, reducing their supply to jets.        This could explain   the temporal change in the radiation spectrum during brightening episode consisting  in optical 100 fold increase of the optical luminosity and simultaneous lowering of X-ray radiation  (the latter probably due to reduction of the amount of electrons and positrons in the jet, the source of X-ray radiation in jets), without need to speculate on remagnetization  of the AGN and quenching of its jets by oppositely magnetised gas cloud  during this episode \cite{ga1flare}. When the Fermi energy in accreting plasma does not exceed MeV, then the collapse of Fermi sphere can produce an increase of the luminosity with lower frequency (just as has been observed for AGN 1ES 1927+654) and, simultaneously, can temporarily reduce the intensity of X-ray radiation from jets, also in agreement with observations.

\section*{Conclusions} 

The important role of the availability of classical trajectories for Feynman path integrals is demonstrated. This method of quantization is especially convenient to identify quantum effects rooted in topological homotopy type constraints imposed onto  trajectories which  enter functional integrals. Despite some popular belief, not all trajectories can contribute to path integrals, but only those  which are classically accessible. This is visible in path integral approach to single particle tunneling  across barriers  and in path integral generalisation to multiparticle systems. 
 For  indistinguishable identical particle systems, path integrals  are especially sensitive to  constraints imposed on trajectories by their topological homotopy properties. The paths in this case are linked to so-called braids displaying indistinguishability of particles leading to quantum statistics and correlation phenomena. In topologically rich systems of interacting 2D electrons exposed to a strong magnetic field, the  availability of braids is changing with varying  magnetic field value. Braid trajectories are admitted only at  the commensurability of multi-loop cyclotron orbits with distribution of interacting electrons in 2D, which results in the definition of complete FQHE hierarchy. The consistency of such topological approach with  experimental observations of  FQHE in conventional semiconductor 2D systems, in graphene monolayer and bilayer and even in fractional Chern topological insulators (in which the magnetic field is substituted by the Berry field) evidences a topological character of quantum Hall physics.   

Another consequence of the  homotopy restrictions imposed on classical trajectories occurs in spacetime folded by general-relativity gravitational singularity. The homotopy transition between multiply- and simply-connected configuration space of identical indistinguishable particle systems at passing the photon sphere rim of a black hole results in high energy phenomena, which can help clarify the astrophysical observations. This homotopy transition results in a quantum  mechanism of transfer of gravitation energy into radiation in the accretion disk of a black hole with the efficiency up to 30 \% mass to energy conversion rate, which is actually encountered in  super-luminous quasars and is beyond the explanation ability of conventional models of quasar luminosity. The same mechanism can be responsible for the creation of some kind of  giant gamma-ray bursts (the source of them is unknown as of yet) at neutron star merger collapses, when also ca. 30 \% of the merger mass is converted into radiation. The agreement with these two independent astrophysical high energy phenomena  supports the model of Fermi sphere decay in the vicinity of a black hole due to the change of the availability of  braid trajectories  for indistinguishable particles approaching the event horizon.

The spin and statistics are closely related as pointed out in Pauli theorem. This theorem holds not only in 3D where exclusively fermion and boson type statistics are available, but also in 2D where fractional anyon statistics are admitted, in agreement, on the other hand, with nonquantised spin in 2D (related with U(1) rotational symmetry). In 2D case of charged indistinguishable particles exposed to a magnetic field (or Berry field in topological Chern insulators) the Pauli theorem must be again generalised to account for specific quantum statistics responsible for fractional quantum Hall effect (including composite fermions and their generalisations) and fractional Chern insulators. Spin of particles is governed by a rotational group independently of constraints imposed on braid properties of particles as is illustrated in the case of general relativistic gravitation singularity. Similarly for a single particle the spin can be well-defined but not the quantum statistics. From these examples it follows that quantum statistics of the same classical particles can change in response to the topological constraints imposed on the multiparticle system. This reflects the fundamental role of the availability or restrictions regarding the identical particle exchanges and expressed in homotopy braid group terms, which for a multiparticle system of indistinguishable particles can vary in response to topological constraints and can modify quantum statistics of the same classical counterparts.



\begin{thebibliography}{77}%
\makeatletter
\providecommand \@ifxundefined [1]{%
 \@ifx{#1\undefined}
}%
\providecommand \@ifnum [1]{%
 \ifnum #1\expandafter \@firstoftwo
 \else \expandafter \@secondoftwo
 \fi
}%
\providecommand \@ifx [1]{%
 \ifx #1\expandafter \@firstoftwo
 \else \expandafter \@secondoftwo
 \fi
}%
\providecommand \natexlab [1]{#1}%
\providecommand \enquote  [1]{``#1''}%
\providecommand \bibnamefont  [1]{#1}%
\providecommand \bibfnamefont [1]{#1}%
\providecommand \citenamefont [1]{#1}%
\providecommand \href [0]{\@secondoftwo}%
\providecommand \href [0]{\begingroup \@sanitize@url \@href}%
\providecommand \@href[1]{\@@startlink{#1}\@@href}%
\providecommand \@@href[1]{\endgroup#1\@@endlink}%
\providecommand \@sanitize@url [0]{\catcode `\\12\catcode `\$12\catcode
  `\&12\catcode `\#12\catcode `\^12\catcode `\_12\catcode `\%12\relax}%
\providecommand \@@startlink[1]{}%
\providecommand \@@endlink[0]{}%
\providecommand \url  [0]{\begingroup\@sanitize@url \@url }%
\providecommand \@url [1]{\endgroup\@href {#1}{\urlprefix }}%
\providecommand \urlprefix  [0]{URL }%
\providecommand \Eprint [0]{\href }%
\providecommand \doibase [0]{}%
\providecommand \selectlanguage [0]{\@gobble}%
\providecommand \bibinfo  [0]{\@secondoftwo}%
\providecommand \bibfield  [0]{\@secondoftwo}%
\providecommand \translation [1]{[#1]}%
\providecommand \BibitemOpen [0]{}%
\providecommand \bibitemStop [0]{}%
\providecommand \bibitemNoStop [0]{.\EOS\space}%
\providecommand \EOS [0]{\spacefactor3000\relax}%
\providecommand \BibitemShut  [1]{\csname bibitem#1\endcsname}%
\let\auto@bib@innerbib\@empty
\bibitem [{\citenamefont {Feynman}\ and\ \citenamefont
  {Hibbs}(1964)}]{feynman1964}%
  \BibitemOpen
  \bibfield  {author} {\bibinfo {author} {\bibfnamefont {R.~P.}\ \bibnamefont
  {Feynman}}\ and\ \bibinfo {author} {\bibfnamefont {A.~R.}\ \bibnamefont
  {Hibbs}},\ }\href {} {\emph {\bibinfo {title} {Quantum Mechanics and
  Path Integrals}}}\ (\bibinfo  {publisher} {McGraw-Hill},\ \bibinfo {address}
  {New York},\ \bibinfo {year} {1964})\BibitemShut {NoStop}%
\bibitem [{\citenamefont {Chaichian}\ and\ \citenamefont
  {Demichev}(2001{\natexlab{a}})}]{chaichian1}%
  \BibitemOpen
  \bibfield  {author} {\bibinfo {author} {\bibfnamefont {M.}~\bibnamefont
  {Chaichian}}\ and\ \bibinfo {author} {\bibfnamefont {A.}~\bibnamefont
  {Demichev}},\ }\href {} {\emph {\bibinfo {title} {Path Integrals in
  Physics Volume I Stochastic Processes and Quantum Mechanics}}}\ (\bibinfo
  {publisher} {IOP Publishing Ltd},\ \bibinfo {address} {Bristol;
  Philadelphia},\ \bibinfo {year} {2001})\BibitemShut {NoStop}%
\bibitem [{\citenamefont {Chaichian}\ and\ \citenamefont
  {Demichev}(2001{\natexlab{b}})}]{chaichian2}%
  \BibitemOpen
  \bibfield  {author} {\bibinfo {author} {\bibfnamefont {M.}~\bibnamefont
  {Chaichian}}\ and\ \bibinfo {author} {\bibfnamefont {A.}~\bibnamefont
  {Demichev}},\ }\href {} {\emph {\bibinfo {title} {Path Integrals in
  Physics Volume II Quantum Field Theory, Statistical Physics and other Modern
  Applications}}}\ (\bibinfo  {publisher} {IOP Publishing Ltd},\ \bibinfo
  {address} {Bristol; Philadelphia},\ \bibinfo {year} {2001})\BibitemShut
  {NoStop}%
\bibitem [{\citenamefont {Spanier}(1966)}]{spanier1966}%
  \BibitemOpen
  \bibfield  {author} {\bibinfo {author} {\bibfnamefont {E.}~\bibnamefont
  {Spanier}},\ }\href {} {\emph {\bibinfo {title} {Algebraic topology}}}\
  (\bibinfo  {publisher} {Springer-Verlag},\ \bibinfo {address} {Berlin},\
  \bibinfo {year} {1966})\BibitemShut {NoStop}%
\bibitem [{\citenamefont {Mermin}(1979)}]{mermin1979}%
  \BibitemOpen
  \bibfield  {author} {\bibinfo {author} {\bibfnamefont {N.~D.}\ \bibnamefont
  {Mermin}},\ }\bibfield  {title} {\bibinfo {title} {The topological theory of
  defects in ordered media},\ }\href {https://doi.org/10.1103/RevModPhys.51.591} {\bibfield  {journal} {\bibinfo
  {journal} {Rev. Mod. Phys.}\ }\textbf {\bibinfo {volume} {51}},\ \bibinfo
  {pages} {591} (\bibinfo {year} {1979})}\BibitemShut {NoStop}%
\bibitem [{\citenamefont {Wiener}(1921)}]{wienera}%
  \BibitemOpen
  \bibfield  {author} {\bibinfo {author} {\bibfnamefont {N.}~\bibnamefont
  {Wiener}},\ }\bibfield  {title} {\bibinfo {title} {The average of an analytic
  functional},\ }\href {https://doi.org/10.1073/pnas.7.9.253} {\bibfield
  {journal} {\bibinfo  {journal} {PNAS}\ }\textbf {\bibinfo {volume} {7}},\
  \bibinfo {pages} {253} (\bibinfo {year} {1921})}\BibitemShut {NoStop}%
\bibitem [{\citenamefont {Pauli}(1973)}]{paulib}%
  \BibitemOpen
  \bibfield  {author} {\bibinfo {author} {\bibfnamefont {W.}~\bibnamefont
  {Pauli}},\ }\href {} {\emph {\bibinfo {title} {Pauli Lectures on Physics
  vol 6, ch. 7}}}\ (\bibinfo  {publisher} {MIT Press},\ \bibinfo {address}
  {Cambridge, MA},\ \bibinfo {year} {1973})\BibitemShut {NoStop}%
\bibitem [{\citenamefont {Nevels}\ \emph {et~al.}(1993)\citenamefont {Nevels},
  \citenamefont {Wu},\ and\ \citenamefont {Huang}}]{gwn102}%
  \BibitemOpen
  \bibfield  {author} {\bibinfo {author} {\bibfnamefont {R.~D.}\ \bibnamefont
  {Nevels}}, \bibinfo {author} {\bibfnamefont {Z.}~\bibnamefont {Wu}},\ and\
  \bibinfo {author} {\bibfnamefont {C.}~\bibnamefont {Huang}},\ }\bibfield
  {title} {\bibinfo {title} {Feynman path integral for an infinite potential
  barrier},\ }\href {https://doi.org/10.1103/PhysRevA.48.3445} {\bibfield  {journal} {\bibinfo  {journal} {Phys.
  Rev. A}\ }\textbf {\bibinfo {volume} {48}},\ \bibinfo {pages} {3445}
  (\bibinfo {year} {1993})}\BibitemShut {NoStop}%
\bibitem [{\citenamefont {Birman}(1974)}]{birman}%
  \BibitemOpen
  \bibfield  {author} {\bibinfo {author} {\bibfnamefont {J.~S.}\ \bibnamefont
  {Birman}},\ }\href {} {\emph {\bibinfo {title} {Braids, Links and
  Mapping Class Groups}}}\ (\bibinfo  {publisher} {Princeton UP},\ \bibinfo
  {address} {Princeton},\ \bibinfo {year} {1974})\BibitemShut {NoStop}%
\bibitem [{\citenamefont {Artin}(1947)}]{artin1947}%
  \BibitemOpen
  \bibfield  {author} {\bibinfo {author} {\bibfnamefont {E.}~\bibnamefont
  {Artin}},\ }\bibfield  {title} {\bibinfo {title} {Theory of braids},\
  }\href {https://doi.org/10.2307/1969218} {\bibfield  {journal} {\bibinfo  {journal} {Annals of Math.}\
  }\textbf {\bibinfo {volume} {48}},\ \bibinfo {pages} {101} (\bibinfo {year}
  {1947})}\BibitemShut {NoStop}%
\bibitem [{\citenamefont {Laidlaw}\ and\ \citenamefont
  {DeWitt}(1971)}]{lwitt-1}%
  \BibitemOpen
  \bibfield  {author} {\bibinfo {author} {\bibfnamefont {M.~G.}\ \bibnamefont
  {Laidlaw}}\ and\ \bibinfo {author} {\bibfnamefont {C.~M.}\ \bibnamefont
  {DeWitt}},\ }\bibfield  {title} {\bibinfo {title} {Feynman functional
  integrals for systems of indistinguishable particles},\ }\href {https://doi.org/10.1103/PhysRevD.3.1375}
  {\bibfield  {journal} {\bibinfo  {journal} {Physical Review D}\ }\textbf
  {\bibinfo {volume} {3}},\ \bibinfo {pages} {1375} (\bibinfo {year}
  {1971})}\BibitemShut {NoStop}%
\bibitem [{\citenamefont {Sudarshan}\ \emph {et~al.}(1988)\citenamefont
  {Sudarshan}, \citenamefont {Imbo},\ and\ \citenamefont {Govindarajan}}]{sud}%
  \BibitemOpen
  \bibfield  {author} {\bibinfo {author} {\bibfnamefont {E.~C.~G.}\
  \bibnamefont {Sudarshan}}, \bibinfo {author} {\bibfnamefont {T.~D.}\
  \bibnamefont {Imbo}},\ and\ \bibinfo {author} {\bibfnamefont {T.~R.}\
  \bibnamefont {Govindarajan}},\ }\bibfield  {title} {\bibinfo {title}
  {Configuration space topology and quantum internal symmetries},\ }\href
  {https://doi.org/10.1016/0370-2693(88)91294-4} {\bibfield  {journal} {\bibinfo  {journal} {Phys. Lett. B}\ }\textbf
  {\bibinfo {volume} {213}},\ \bibinfo {pages} {471} (\bibinfo {year}
  {1988})}\BibitemShut {NoStop}%
\bibitem [{\citenamefont {Wu}(1984)}]{wu}%
  \BibitemOpen
  \bibfield  {author} {\bibinfo {author} {\bibfnamefont {Y.~S.}\ \bibnamefont
  {Wu}},\ }\bibfield  {title} {\bibinfo {title} {General theory for quantum
  statistics in two dimensions},\ }\href {https://doi.org/10.1103/PhysRevLett.52.2103} {\bibfield  {journal}
  {\bibinfo  {journal} {Phys. Rev. Lett.}\ }\textbf {\bibinfo {volume} {52}},\
  \bibinfo {pages} {2103} (\bibinfo {year} {1984})}\BibitemShut {NoStop}%
\bibitem [{\citenamefont {Imbo}\ \emph {et~al.}(1990)\citenamefont {Imbo},
  \citenamefont {Imbo},\ and\ \citenamefont {Sudarshan}}]{imbo}%
  \BibitemOpen
  \bibfield  {author} {\bibinfo {author} {\bibfnamefont {T.~D.}\ \bibnamefont
  {Imbo}}, \bibinfo {author} {\bibfnamefont {C.~S.}\ \bibnamefont {Imbo}},\
  and\ \bibinfo {author} {\bibfnamefont {C.~S.}\ \bibnamefont {Sudarshan}},\
  }\bibfield  {title} {\bibinfo {title} {Identical particles, exotic statistics
  and braid groups},\ }\href {https://doi.org/10.1016/0370-2693(90)92010-G} {\bibfield  {journal} {\bibinfo  {journal}
  {Phys. Lett. B}\ }\textbf {\bibinfo {volume} {234}},\ \bibinfo {pages} {103}
  (\bibinfo {year} {1990})}\BibitemShut {NoStop}%
\bibitem [{\citenamefont {Wilczek}(1990)}]{wilczek}%
  \BibitemOpen
  \bibfield  {author} {\bibinfo {author} {\bibfnamefont {F.}~\bibnamefont
  {Wilczek}},\ }\href {} {\emph {\bibinfo {title} {Fractional Statistics
  and Anyon Superconductivity}}}\ (\bibinfo  {publisher} {World Scientific},\
  \bibinfo {address} {Singapore},\ \bibinfo {year} {1990})\BibitemShut
  {NoStop}%
\bibitem [{\citenamefont {Ashcroft}\ and\ \citenamefont
  {Mermin}(1976)}]{ashcroftmermin}%
  \BibitemOpen
  \bibfield  {author} {\bibinfo {author} {\bibfnamefont {N.}~\bibnamefont
  {Ashcroft}}\ and\ \bibinfo {author} {\bibfnamefont {D.}~\bibnamefont
  {Mermin}},\ }\href {} {\emph {\bibinfo {title} {Solid State Physics}}}\
  (\bibinfo  {publisher} {Holt, Rinehart, Winston},\ \bibinfo {address} {New
  York},\ \bibinfo {year} {1976})\BibitemShut {NoStop}%
\bibitem [{\citenamefont {Klauder}\ and\ \citenamefont
  {Onorfi}(1989)}]{gwn100}%
  \BibitemOpen
  \bibfield  {author} {\bibinfo {author} {\bibfnamefont {J.~R.}\ \bibnamefont
  {Klauder}}\ and\ \bibinfo {author} {\bibfnamefont {E.}~\bibnamefont
  {Onorfi}},\ }\bibfield  {title} {\bibinfo {title} {Landau levels and
  geometric quantization},\ }\href {https://doi.org/10.1142/S0217751X89001606} {\bibfield  {journal} {\bibinfo
  {journal} {Int. Journal of Modern Physics A}\ }\textbf {\bibinfo {volume}
  {4}},\ \bibinfo {pages} {3939} (\bibinfo {year} {1989})}\BibitemShut
  {NoStop}%
\bibitem [{\citenamefont {Klauder}(1988)}]{gwn101}%
  \BibitemOpen
  \bibfield  {author} {\bibinfo {author} {\bibfnamefont {J.~R.}\ \bibnamefont
  {Klauder}},\ }\bibfield  {title} {\bibinfo {title} {Quantization is geometry,
  after all},\ }\href {https://doi.org/10.1016/0003-4916(88)90092-9} {\bibfield  {journal} {\bibinfo  {journal}
  {Annals of physics}\ }\textbf {\bibinfo {volume} {188}},\ \bibinfo {pages}
  {120} (\bibinfo {year} {1988})}\BibitemShut {NoStop}%
\bibitem [{\citenamefont {von Klitzing}\ \emph {et~al.}(1980)\citenamefont {von
  Klitzing}, \citenamefont {Dorda},\ and\ \citenamefont
  {Pepper}}]{klitzing1980}%
  \BibitemOpen
  \bibfield  {author} {\bibinfo {author} {\bibfnamefont {K.}~\bibnamefont {von
  Klitzing}}, \bibinfo {author} {\bibfnamefont {G.}~\bibnamefont {Dorda}},\
  and\ \bibinfo {author} {\bibfnamefont {M.}~\bibnamefont {Pepper}},\
  }\bibfield  {title} {\bibinfo {title} {New method for high-accuracy
  determination of the fine-structure constant based on quantized {H}all
  resistance},\ }\href {https://doi.org/10.1103/PhysRevLett.45.494} {\bibfield  {journal} {\bibinfo  {journal}
  {Phys. Rev. Lett.}\ }\textbf {\bibinfo {volume} {45}},\ \bibinfo {pages}
  {494} (\bibinfo {year} {1980})}\BibitemShut {NoStop}%
\bibitem [{\citenamefont {Tsui}\ \emph {et~al.}(1982)\citenamefont {Tsui},
  \citenamefont {St{\"o}rmer},\ and\ \citenamefont {Gossard}}]{tsui1982}%
  \BibitemOpen
  \bibfield  {author} {\bibinfo {author} {\bibfnamefont {D.~C.}\ \bibnamefont
  {Tsui}}, \bibinfo {author} {\bibfnamefont {H.~L.}\ \bibnamefont
  {St{\"o}rmer}},\ and\ \bibinfo {author} {\bibfnamefont {A.~C.}\ \bibnamefont
  {Gossard}},\ }\bibfield  {title} {\bibinfo {title} {Two-dimensional
  magnetotransport in the extreme quantum limit},\ }\href {https://doi.org/10.1103/PhysRevLett.48.1559} {\bibfield
  {journal} {\bibinfo  {journal} {Phys. Rev. Lett.}\ }\textbf {\bibinfo
  {volume} {48}},\ \bibinfo {pages} {1559} (\bibinfo {year}
  {1982})}\BibitemShut {NoStop}%
\bibitem [{\citenamefont {Jacak}(2021)}]{annals2021}%
  \BibitemOpen
  \bibfield  {author} {\bibinfo {author} {\bibfnamefont {J.~E.}\ \bibnamefont
  {Jacak}},\ }\bibfield  {title} {\bibinfo {title} {Topological approach to
  electron correlations at fractional quantum {H}all effect},\ }\href {https://doi.org/10.1016/j.aop.2021.168493}
  {\bibfield  {journal} {\bibinfo  {journal} {Annals of Physics}\ }\textbf
  {\bibinfo {volume} {430}},\ \bibinfo {pages} {168493} (\bibinfo {year}
  {2021})}\BibitemShut {NoStop}%
\bibitem [{\citenamefont {Prange}\ and\ \citenamefont {Girvin}(1990)}]{prange}%
  \BibitemOpen
  \bibfield  {author} {\bibinfo {author} {\bibfnamefont {R.~E.}\ \bibnamefont
  {Prange}}\ and\ \bibinfo {author} {\bibfnamefont {S.~M.}\ \bibnamefont
  {Girvin}},\ }\href {} {\emph {\bibinfo {title} {The Quantum Hall
  Effect}}}\ (\bibinfo  {publisher} {Springer Verlag},\ \bibinfo {address} {New
  York},\ \bibinfo {year} {1990})\BibitemShut {NoStop}%
\bibitem [{\citenamefont {Ciftja}\ and\ \citenamefont
  {Wexler}(2003)}]{montecarlo1}%
  \BibitemOpen
  \bibfield  {author} {\bibinfo {author} {\bibfnamefont {O.}~\bibnamefont
  {Ciftja}}\ and\ \bibinfo {author} {\bibfnamefont {C.}~\bibnamefont
  {Wexler}},\ }\bibfield  {title} {\bibinfo {title} {Monte {C}arlo simulation
  method for {L}aughlin-like states in a disk geometry},\ }\href {https://doi.org/10.1103/PhysRevB.67.075304}
  {\bibfield  {journal} {\bibinfo  {journal} {Phys. Rev. B}\ }\textbf {\bibinfo
  {volume} {67}},\ \bibinfo {pages} {075304} (\bibinfo {year}
  {2003})}\BibitemShut {NoStop}%
\bibitem [{\citenamefont {Morf}\ and\ \citenamefont
  {Halperin}(1987)}]{montecarlo2}%
  \BibitemOpen
  \bibfield  {author} {\bibinfo {author} {\bibfnamefont {R.}~\bibnamefont
  {Morf}}\ and\ \bibinfo {author} {\bibfnamefont {B.~I.}\ \bibnamefont
  {Halperin}},\ }\bibfield  {title} {\bibinfo {title} {Monte carlo evaluation
  of trial wavefunctions for the fractional quantized hall effect: Spherical
  geometry},\ }\href {https://doi.org/10.1007/BF01304256} {\bibfield  {journal} {\bibinfo  {journal} {Z.
  Phys. B Condensed Matter}\ }\textbf {\bibinfo {volume} {68}},\ \bibinfo
  {pages} {391} (\bibinfo {year} {1987})}\BibitemShut {NoStop}%
\bibitem [{\citenamefont {Pan}\ \emph {et~al.}(2003)\citenamefont {Pan},
  \citenamefont {St{\"o}rmer}, \citenamefont {Tsui}, \citenamefont {Pfeiffer},
  \citenamefont {Baldwin},\ and\ \citenamefont {West}}]{pan2003}%
  \BibitemOpen
  \bibfield  {author} {\bibinfo {author} {\bibfnamefont {W.}~\bibnamefont
  {Pan}}, \bibinfo {author} {\bibfnamefont {H.~L.}\ \bibnamefont
  {St{\"o}rmer}}, \bibinfo {author} {\bibfnamefont {D.~C.}\ \bibnamefont
  {Tsui}}, \bibinfo {author} {\bibfnamefont {L.~N.}\ \bibnamefont {Pfeiffer}},
  \bibinfo {author} {\bibfnamefont {K.~W.}\ \bibnamefont {Baldwin}},\ and\
  \bibinfo {author} {\bibfnamefont {K.~W.}\ \bibnamefont {West}},\ }\bibfield
  {title} {\bibinfo {title} {Fractional quantum {H}all effect of composite
  fermions},\ }\href {https://doi.org/10.1103/PhysRevLett.90.016801} {\bibfield  {journal} {\bibinfo  {journal} {Phys.
  Rev. Lett.}\ }\textbf {\bibinfo {volume} {90}},\ \bibinfo {pages} {016801}
  (\bibinfo {year} {2003})}\BibitemShut {NoStop}%
\bibitem [{\citenamefont {Laughlin}(1983)}]{laughlin2}%
  \BibitemOpen
  \bibfield  {author} {\bibinfo {author} {\bibfnamefont {R.~B.}\ \bibnamefont
  {Laughlin}},\ }\bibfield  {title} {\bibinfo {title} {Anomalous quantum {H}all
  effect: an incompressible quantum fluid with fractionally charged
  excitations},\ }\href {https://doi.org/10.1103/PhysRevLett.50.1395} {\bibfield  {journal} {\bibinfo  {journal}
  {Phys. Rev. Lett.}\ }\textbf {\bibinfo {volume} {50}},\ \bibinfo {pages}
  {1395} (\bibinfo {year} {1983})}\BibitemShut {NoStop}%
\bibitem [{\citenamefont {Haldane}(1983)}]{hh1}%
  \BibitemOpen
  \bibfield  {author} {\bibinfo {author} {\bibfnamefont {F.~D.~M.}\
  \bibnamefont {Haldane}},\ }\bibfield  {title} {\bibinfo {title} {Fractional
  quantization of the {H}all effect:{A} hierarchy of incompressible quantum
  fluid states},\ }\href {https://doi.org/10.1103/PhysRevLett.51.605} {\bibfield  {journal} {\bibinfo  {journal}
  {Phys. Rev. Lett.}\ }\textbf {\bibinfo {volume} {51}},\ \bibinfo {pages}
  {605} (\bibinfo {year} {1983})}\BibitemShut {NoStop}%
\bibitem [{\citenamefont {Halperin}(1984)}]{hh2}%
  \BibitemOpen
  \bibfield  {author} {\bibinfo {author} {\bibfnamefont {B.~I.}\ \bibnamefont
  {Halperin}},\ }\bibfield  {title} {\bibinfo {title} {Statistics of
  quasiparticles and the hierarchy of fractional quantized {H}all states},\
  }\href {https://doi.org/10.1103/PhysRevLett.52.1583} {\bibfield  {journal} {\bibinfo  {journal} {Phys. Rev. Lett.}\
  }\textbf {\bibinfo {volume} {52}},\ \bibinfo {pages} {1583} (\bibinfo {year}
  {1984})}\BibitemShut {NoStop}%
\bibitem [{\citenamefont {Jain}(1989)}]{jain}%
  \BibitemOpen
  \bibfield  {author} {\bibinfo {author} {\bibfnamefont {J.~K.}\ \bibnamefont
  {Jain}},\ }\bibfield  {title} {\bibinfo {title} {Composite-fermion approach
  for the fractional quantum {H}all effect},\ }\href {https://doi.org/10.1103/PhysRevLett.63.199} {\bibfield
  {journal} {\bibinfo  {journal} {Phys. Rev. Lett.}\ }\textbf {\bibinfo
  {volume} {63}},\ \bibinfo {pages} {199} (\bibinfo {year} {1989})}\BibitemShut
  {NoStop}%
\bibitem [{\citenamefont {Halperin}(1983)}]{halperin1983}%
  \BibitemOpen
  \bibfield  {author} {\bibinfo {author} {\bibfnamefont {B.~I.}\ \bibnamefont
  {Halperin}},\ }\bibfield  {title} {\bibinfo {title} {Theory of the quantized
  {H}all conductance},\ }\href {} {\bibfield  {journal} {\bibinfo
  {journal} {Helv. Phys. Acta}\ }\textbf {\bibinfo {volume} {56}},\ \bibinfo
  {pages} {75} (\bibinfo {year} {1983})}\BibitemShut {NoStop}%
\bibitem [{\citenamefont {Jain}(2007)}]{jain2007}%
  \BibitemOpen
  \bibfield  {author} {\bibinfo {author} {\bibfnamefont {J.~K.}\ \bibnamefont
  {Jain}},\ }\href {} {\emph {\bibinfo {title} {Composite Fermions}}}\
  (\bibinfo  {publisher} {Cambridge UP},\ \bibinfo {address} {Cambridge},\
  \bibinfo {year} {2007})\BibitemShut {NoStop}%
\bibitem [{\citenamefont {Jacak}(2018)}]{pra}%
  \BibitemOpen
  \bibfield  {author} {\bibinfo {author} {\bibfnamefont {J.~E.}\ \bibnamefont
  {Jacak}},\ }\bibfield  {title} {\bibinfo {title} {Application of the path
  integral quantization to indistinguishable particle systems topologically
  confined by a magnetic field},\ }\href {https://doi.org/10.1103/PhysRevA.97.012108} {\bibfield  {journal}
  {\bibinfo  {journal} {Phys. Rev. A}\ }\textbf {\bibinfo {volume} {97}},\
  \bibinfo {pages} {012108} (\bibinfo {year} {2018})}\BibitemShut {NoStop}%
\bibitem [{\citenamefont {Jacak}(2022)}]{sr2022}%
  \BibitemOpen
  \bibfield  {author} {\bibinfo {author} {\bibfnamefont {J.~E.}\ \bibnamefont
  {Jacak}},\ }\bibfield  {title} {\bibinfo {title} {Formal derivation of the
  {L}aughlin function and its generalization for other topological phases of
  {FQHE}},\ }\href {https://doi.org/10.1038/s41598-021-04672-z}
  {\bibfield  {journal} {\bibinfo  {journal} {Scientific Reports}\ }\textbf
  {\bibinfo {volume} {12}},\ \bibinfo {pages} {616} (\bibinfo {year}
  {2022})}\BibitemShut {NoStop}%
\bibitem [{\citenamefont {Jacak}(2017)}]{sr1ccc}%
  \BibitemOpen
  \bibfield  {author} {\bibinfo {author} {\bibfnamefont {J.}~\bibnamefont
  {Jacak}},\ }\bibfield  {title} {\bibinfo {title} {Unconventional fractional
  quantum {H}all effect in bilayer graphene},\ }\href {https://doi.org/10.1038/s41598-017-09166-5} {\bibfield
  {journal} {\bibinfo  {journal} {Scientific Reports}\ }\textbf {\bibinfo
  {volume} {7}},\ \bibinfo {pages} {8720} (\bibinfo {year} {2017})}\BibitemShut
  {NoStop}%
\bibitem [{\citenamefont {Dean}\ \emph {et~al.}(2011)\citenamefont {Dean},
  \citenamefont {Young}, \citenamefont {Cadden-Zimansky}, \citenamefont {Wang},
  \citenamefont {Ren}, \citenamefont {Watanabe}, \citenamefont {Taniguchi},
  \citenamefont {Kim}, \citenamefont {Hone},\ and\ \citenamefont
  {Shepard}}]{dean2011}%
  \BibitemOpen
  \bibfield  {author} {\bibinfo {author} {\bibfnamefont {C.~R.}\ \bibnamefont
  {Dean}}, \bibinfo {author} {\bibfnamefont {A.~F.}\ \bibnamefont {Young}},
  \bibinfo {author} {\bibfnamefont {P.}~\bibnamefont {Cadden-Zimansky}},
  \bibinfo {author} {\bibfnamefont {L.}~\bibnamefont {Wang}}, \bibinfo {author}
  {\bibfnamefont {H.}~\bibnamefont {Ren}}, \bibinfo {author} {\bibfnamefont
  {K.}~\bibnamefont {Watanabe}}, \bibinfo {author} {\bibfnamefont
  {T.}~\bibnamefont {Taniguchi}}, \bibinfo {author} {\bibfnamefont
  {P.}~\bibnamefont {Kim}}, \bibinfo {author} {\bibfnamefont {J.}~\bibnamefont
  {Hone}},\ and\ \bibinfo {author} {\bibfnamefont {K.~L.}\ \bibnamefont
  {Shepard}},\ }\bibfield  {title} {\bibinfo {title} {Multicomponent fractional
  quantum {H}all effect in graphene},\ }\href {https://doi.org/10.1038/nphys2007} {\bibfield  {journal}
  {\bibinfo  {journal} {Nature Physics}\ }\textbf {\bibinfo {volume} {7}},\
  \bibinfo {pages} {693} (\bibinfo {year} {2011})}\BibitemShut {NoStop}%
\bibitem [{\citenamefont {Amet}\ \emph {et~al.}(2015)\citenamefont {Amet},
  \citenamefont {Bestwick}, \citenamefont {Williams}, \citenamefont {Balicas},
  \citenamefont {Watanabe}, \citenamefont {Taniguchi},\ and\ \citenamefont
  {{Goldhaber-Gordon}}}]{amet}%
  \BibitemOpen
  \bibfield  {author} {\bibinfo {author} {\bibfnamefont {F.}~\bibnamefont
  {Amet}}, \bibinfo {author} {\bibfnamefont {A.~J.}\ \bibnamefont {Bestwick}},
  \bibinfo {author} {\bibfnamefont {J.~R.}\ \bibnamefont {Williams}}, \bibinfo
  {author} {\bibfnamefont {L.}~\bibnamefont {Balicas}}, \bibinfo {author}
  {\bibfnamefont {K.}~\bibnamefont {Watanabe}}, \bibinfo {author}
  {\bibfnamefont {T.}~\bibnamefont {Taniguchi}},\ and\ \bibinfo {author}
  {\bibfnamefont {D.}~\bibnamefont {{Goldhaber-Gordon}}},\ }\bibfield  {title}
  {\bibinfo {title} {Composite fermions and broken symmetries in graphene},\
  }\href {https://doi.org/10.1038/ncomms6838} {\bibfield  {journal} {\bibinfo  {journal} {Nat. Commun.}\
  }\textbf {\bibinfo {volume} {6}} (\bibinfo {year} {2015})}\BibitemShut
  {NoStop}%
\bibitem [{\citenamefont {Ki}\ \emph {et~al.}(2014)\citenamefont {Ki},
  \citenamefont {Falko}, \citenamefont {Abanin},\ and\ \citenamefont
  {Morpurgo}}]{bil}%
  \BibitemOpen
  \bibfield  {author} {\bibinfo {author} {\bibfnamefont {D.~K.}\ \bibnamefont
  {Ki}}, \bibinfo {author} {\bibfnamefont {V.~I.}\ \bibnamefont {Falko}},
  \bibinfo {author} {\bibfnamefont {D.~A.}\ \bibnamefont {Abanin}},\ and\
  \bibinfo {author} {\bibfnamefont {A.}~\bibnamefont {Morpurgo}},\ }\bibfield
  {title} {\bibinfo {title} {Observation of even denominator fractional quantum
  {H}all effect in suspended bilayer graphene},\ }\href {https://doi.org/10.1021/nl5003922} {\bibfield
  {journal} {\bibinfo  {journal} {Nano Lett.}\ }\textbf {\bibinfo {volume}
  {14}},\ \bibinfo {pages} {2135} (\bibinfo {year} {2014})}\BibitemShut
  {NoStop}%
\bibitem [{\citenamefont {Jacak}\ and\ \citenamefont
  {Jacak}(2015)}]{jetpllaaa}%
  \BibitemOpen
  \bibfield  {author} {\bibinfo {author} {\bibfnamefont {J.}~\bibnamefont
  {Jacak}}\ and\ \bibinfo {author} {\bibfnamefont {L.}~\bibnamefont {Jacak}},\
  }\bibfield  {title} {\bibinfo {title} {The commensurability condition and
  fractional quantum {H}all effect hierarchy in higher {L}andau levels},\
  }\href {https://doi.org/10.1134/S0021364015130044} {\bibfield  {journal} {\bibinfo  {journal} {JETP Letters}\
  }\textbf {\bibinfo {volume} {102}},\ \bibinfo {pages} {19} (\bibinfo {year}
  {2015})}\BibitemShut {NoStop}%
\bibitem [{\citenamefont {Diankov}\ \emph {et~al.}(2016)\citenamefont
  {Diankov}, \citenamefont {Liang}, \citenamefont {Amet}, \citenamefont
  {Gallagher}, \citenamefont {Lee}, \citenamefont {Bestwick}, \citenamefont
  {Tharratt}, \citenamefont {Coniglio}, \citenamefont {Jaroszynski},
  \citenamefont {Watanabe}, \citenamefont {Taniguchi},\ and\ \citenamefont
  {Goldhaber-Gordon}}]{bil1}%
  \BibitemOpen
  \bibfield  {author} {\bibinfo {author} {\bibfnamefont {G.}~\bibnamefont
  {Diankov}}, \bibinfo {author} {\bibfnamefont {C.-T.}\ \bibnamefont {Liang}},
  \bibinfo {author} {\bibfnamefont {F.}~\bibnamefont {Amet}}, \bibinfo {author}
  {\bibfnamefont {P.}~\bibnamefont {Gallagher}}, \bibinfo {author}
  {\bibfnamefont {M.}~\bibnamefont {Lee}}, \bibinfo {author} {\bibfnamefont
  {A.~J.}\ \bibnamefont {Bestwick}}, \bibinfo {author} {\bibfnamefont
  {K.}~\bibnamefont {Tharratt}}, \bibinfo {author} {\bibfnamefont
  {W.}~\bibnamefont {Coniglio}}, \bibinfo {author} {\bibfnamefont
  {J.}~\bibnamefont {Jaroszynski}}, \bibinfo {author} {\bibfnamefont
  {K.}~\bibnamefont {Watanabe}}, \bibinfo {author} {\bibfnamefont
  {T.}~\bibnamefont {Taniguchi}},\ and\ \bibinfo {author} {\bibfnamefont
  {D.}~\bibnamefont {Goldhaber-Gordon}},\ }\bibfield  {title} {\bibinfo {title}
  {Robust fractional quantum {H}all effect in the n=2 {L}andau level in bilayer
  graphene},\ }\href {https://doi.org/10.1038/ncomms13908} {\bibfield  {journal} {\bibinfo  {journal} {Nature
  Comm.}\ }\textbf {\bibinfo {volume} {7}},\ \bibinfo {pages} {13908} (\bibinfo
  {year} {2016})}\BibitemShut {NoStop}%
\bibitem [{\citenamefont {Maher}\ \emph {et~al.}(2014)\citenamefont {Maher},
  \citenamefont {Wang}, \citenamefont {Gao}, \citenamefont {Forsythe},
  \citenamefont {Taniguchi}, \citenamefont {Watanabe}, \citenamefont {Abanin},
  \citenamefont {Papi{\'c}}, \citenamefont {Cadden-Zimansk}, \citenamefont
  {Hone}, \citenamefont {Kim},\ and\ \citenamefont {Dean}}]{maher}%
  \BibitemOpen
  \bibfield  {author} {\bibinfo {author} {\bibfnamefont {P.}~\bibnamefont
  {Maher}}, \bibinfo {author} {\bibfnamefont {L.}~\bibnamefont {Wang}},
  \bibinfo {author} {\bibfnamefont {Y.}~\bibnamefont {Gao}}, \bibinfo {author}
  {\bibfnamefont {C.}~\bibnamefont {Forsythe}}, \bibinfo {author}
  {\bibfnamefont {T.}~\bibnamefont {Taniguchi}}, \bibinfo {author}
  {\bibfnamefont {L.}~\bibnamefont {Watanabe}}, \bibinfo {author}
  {\bibfnamefont {D.}~\bibnamefont {Abanin}}, \bibinfo {author} {\bibfnamefont
  {Z.}~\bibnamefont {Papi{\'c}}}, \bibinfo {author} {\bibfnamefont
  {P.}~\bibnamefont {Cadden-Zimansk}}, \bibinfo {author} {\bibfnamefont
  {J.}~\bibnamefont {Hone}}, \bibinfo {author} {\bibfnamefont {P.}~\bibnamefont
  {Kim}},\ and\ \bibinfo {author} {\bibfnamefont {C.~R.}\ \bibnamefont
  {Dean}},\ }\bibfield  {title} {\bibinfo {title} {Tunable fractional quantum
  {H}all phases in bilayer graphene},\ }\href {https://doi.org/10.1126/science.1252875} {\bibfield  {journal}
  {\bibinfo  {journal} {Science}\ }\textbf {\bibinfo {volume} {345}},\ \bibinfo
  {pages} {61} (\bibinfo {year} {2014})}\BibitemShut {NoStop}%
\bibitem [{\citenamefont {Schwarzschild}(1916)}]{schwarzschild}%
  \BibitemOpen
  \bibfield  {author} {\bibinfo {author} {\bibfnamefont {K.}~\bibnamefont
  {Schwarzschild}},\ }\bibfield  {title} {\bibinfo {title} {Über das
  {G}ravitationsfeld eines {M}assenpunktes nach der {E}insteinschen
  {T}heorie},\ }\href {} {\bibfield  {journal} {\bibinfo  {journal}
  {Sitzungsberichte der Königlich Preussischen Akademie der Wissenschaften}\
  }\textbf {\bibinfo {volume} {7}},\ \bibinfo {pages} {189} (\bibinfo {year}
  {1916})}\BibitemShut {NoStop}%
\bibitem [{\citenamefont {Novikov}\ and\ \citenamefont
  {Thorne}(1973)}]{novikov}%
  \BibitemOpen
  \bibfield  {author} {\bibinfo {author} {\bibfnamefont {I.~D.}\ \bibnamefont
  {Novikov}}\ and\ \bibinfo {author} {\bibfnamefont {K.~S.}\ \bibnamefont
  {Thorne}},\ }\href {} {\emph {\bibinfo {title} {Black Holes (Les Astres
  Occlus) by C. DeWitt and B. S. DeWitt}}}\ (\bibinfo  {publisher} {Gordon and
  Breach Science Publishers},\ \bibinfo {address} {London},\ \bibinfo {year}
  {1973})\BibitemShut {NoStop}%
\bibitem [{\citenamefont {Kruskal}(1960)}]{kruskal}%
  \BibitemOpen
  \bibfield  {author} {\bibinfo {author} {\bibfnamefont {M.~D.}\ \bibnamefont
  {Kruskal}},\ }\bibfield  {title} {\bibinfo {title} {Maximal extension of
  {S}chwarzschild metric},\ }\href {https://doi.org/10.1103/PhysRev.119.1743} {\bibfield  {journal} {\bibinfo
  {journal} {Physical Review}\ }\textbf {\bibinfo {volume} {119}},\ \bibinfo
  {pages} {1743} (\bibinfo {year} {1960})}\BibitemShut {NoStop}%
\bibitem [{\citenamefont {Szekeres}(1960)}]{szekeres}%
  \BibitemOpen
  \bibfield  {author} {\bibinfo {author} {\bibfnamefont {G.}~\bibnamefont
  {Szekeres}},\ }\bibfield  {title} {\bibinfo {title} {On the singularities of
  a {R}iemannian manifold},\ }\href {https://doi.org/10.5486/PMD.1960.7.1-4.26} {\bibfield  {journal} {\bibinfo
  {journal} {Publ. Math. Debrecen}\ }\textbf {\bibinfo {volume} {7}},\ \bibinfo
  {pages} {285} (\bibinfo {year} {1960})}\BibitemShut {NoStop}%
\bibitem [{\citenamefont {Pauli}(1940)}]{pauli}%
  \BibitemOpen
  \bibfield  {author} {\bibinfo {author} {\bibfnamefont {W.}~\bibnamefont
  {Pauli}},\ }\bibfield  {title} {\bibinfo {title} {The connection between spin
  and statistics},\ }\href {https://doi.org/10.1103/PhysRev.58.716} {\bibfield  {journal} {\bibinfo  {journal}
  {Physical Review}\ }\textbf {\bibinfo {volume} {58}},\ \bibinfo {pages} {716}
  (\bibinfo {year} {1940})}\BibitemShut {NoStop}%
\bibitem [{\citenamefont {Chandrasekhar}(1931)}]{chandrasekhar}%
  \BibitemOpen
  \bibfield  {author} {\bibinfo {author} {\bibfnamefont {S.}~\bibnamefont
  {Chandrasekhar}},\ }\bibfield  {title} {\bibinfo {title} {The maximum mass of
  ideal white dwarfs},\ }\href {https://doi.org/10.1086/143324} {\bibfield  {journal} {\bibinfo
  {journal} {The Astrophysical Journal}\ }\textbf {\bibinfo {volume} {74}},\
  \bibinfo {pages} {81} (\bibinfo {year} {1931})}\BibitemShut {NoStop}%
\bibitem [{\citenamefont {Tolman}(1939)}]{tolman}%
  \BibitemOpen
  \bibfield  {author} {\bibinfo {author} {\bibfnamefont {R.~C.}\ \bibnamefont
  {Tolman}},\ }\bibfield  {title} {\bibinfo {title} {Static solutions of
  {E}instein’s field equations for spheres of fluid},\ }\href {https://doi.org/10.1103/PhysRev.55.364}
  {\bibfield  {journal} {\bibinfo  {journal} {Physical Review}\ }\textbf
  {\bibinfo {volume} {55}},\ \bibinfo {pages} {364} (\bibinfo {year}
  {1939})}\BibitemShut {NoStop}%
\bibitem [{\citenamefont {Oppenheimer}\ and\ \citenamefont
  {Volkoff}(1939)}]{volkoff}%
  \BibitemOpen
  \bibfield  {author} {\bibinfo {author} {\bibfnamefont {J.~R.}\ \bibnamefont
  {Oppenheimer}}\ and\ \bibinfo {author} {\bibfnamefont {G.~M.}\ \bibnamefont
  {Volkoff}},\ }\bibfield  {title} {\bibinfo {title} {On massive neutron
  cores},\ }\href {https://doi.org/10.1103/PhysRev.55.374} {\bibfield  {journal} {\bibinfo  {journal} {Physical
  Review}\ }\textbf {\bibinfo {volume} {55}},\ \bibinfo {pages} {374} (\bibinfo
  {year} {1939})}\BibitemShut {NoStop}%
\bibitem [{\citenamefont {Yakovlev}\ \emph {et~al.}(2013)\citenamefont
  {Yakovlev}, \citenamefont {Haensel}, \citenamefont {Baym},\ and\
  \citenamefont {Pethick}}]{olandau}%
  \BibitemOpen
  \bibfield  {author} {\bibinfo {author} {\bibfnamefont {D.~G.}\ \bibnamefont
  {Yakovlev}}, \bibinfo {author} {\bibfnamefont {P.}~\bibnamefont {Haensel}},
  \bibinfo {author} {\bibfnamefont {G.}~\bibnamefont {Baym}},\ and\ \bibinfo
  {author} {\bibfnamefont {C.}~\bibnamefont {Pethick}},\ }\bibfield  {title}
  {\bibinfo {title} {Lev {L}andau and the concept of neutron stars},\
  }\href {https://doi.org/10.3367/UFNe.0183.201303f.0307} {\bibfield  {journal} {\bibinfo  {journal} {Phys. Usp}\
  }\textbf {\bibinfo {volume} {56}},\ \bibinfo {pages} {289} (\bibinfo {year}
  {2013})}\BibitemShut {NoStop}%
\bibitem [{\citenamefont {Abrikosov}\ \emph {et~al.}(1975)\citenamefont
  {Abrikosov}, \citenamefont {Gorkov},\ and\ \citenamefont
  {Dzialoshinskii}}]{abrikosov1975}%
  \BibitemOpen
  \bibfield  {author} {\bibinfo {author} {\bibfnamefont {A.~A.}\ \bibnamefont
  {Abrikosov}}, \bibinfo {author} {\bibfnamefont {L.~P.}\ \bibnamefont
  {Gorkov}},\ and\ \bibinfo {author} {\bibfnamefont {I.~E.}\ \bibnamefont
  {Dzialoshinskii}},\ }\href {} {\emph {\bibinfo {title} {Methods of
  Quantum Field Theory in Statistical Physics}}}\ (\bibinfo  {publisher} {Dover
  Publ. Inc.},\ \bibinfo {address} {Dover},\ \bibinfo {year}
  {1975})\BibitemShut {NoStop}%
\bibitem [{\citenamefont {Luttinger}(1960)}]{luttinger}%
  \BibitemOpen
  \bibfield  {author} {\bibinfo {author} {\bibfnamefont {J.~M.}\ \bibnamefont
  {Luttinger}},\ }\bibfield  {title} {\bibinfo {title} {Fermi surface and some
  simple equilibrium properties of a system of interacting fermions},\
  }\href {https://doi.org/10.1103/PhysRev.119.1153} {\bibfield  {journal} {\bibinfo  {journal} {Physical Review}\
  }\textbf {\bibinfo {volume} {119}},\ \bibinfo {pages} {1153} (\bibinfo {year}
  {1960})}\BibitemShut {NoStop}%
\bibitem [{\citenamefont {Landau}\ and\ \citenamefont
  {Lifshitz}(1972)}]{landau1972}%
  \BibitemOpen
  \bibfield  {author} {\bibinfo {author} {\bibfnamefont {L.~D.}\ \bibnamefont
  {Landau}}\ and\ \bibinfo {author} {\bibfnamefont {E.~M.}\ \bibnamefont
  {Lifshitz}},\ }\href {} {\emph {\bibinfo {title} {Quantum Mechanics:
  nonrelativistic Theory}}}\ (\bibinfo  {publisher} {Nauka},\ \bibinfo
  {address} {Moscow},\ \bibinfo {year} {1972})\BibitemShut {NoStop}%
\bibitem [{\citenamefont {Hawking}(1974)}]{hawking}%
  \BibitemOpen
  \bibfield  {author} {\bibinfo {author} {\bibfnamefont {S.~W.}\ \bibnamefont
  {Hawking}},\ }\bibfield  {title} {\bibinfo {title} {Black hole explosions?},\
  }\href {https://doi.org/10.1038/248030a0} {\bibfield  {journal} {\bibinfo
   {journal} {Nature}\ }\textbf {\bibinfo {volume} {248}},\ \bibinfo {pages}
  {30} (\bibinfo {year} {1974})}\BibitemShut {NoStop}%
\bibitem [{\citenamefont {Bekenstein}(1972)}]{bekenstein}%
  \BibitemOpen
  \bibfield  {author} {\bibinfo {author} {\bibfnamefont {A.}~\bibnamefont
  {Bekenstein}},\ }\bibfield  {title} {\bibinfo {title} {Black holes and the
  second law},\ }\href {https://doi.org/10.1007/BF02757029} {\bibfield
  {journal} {\bibinfo  {journal} {Lettere al Nuovo Cimento}\ ,\ \bibinfo
  {pages} {737}} (\bibinfo {year} {1972})}\BibitemShut {NoStop}%
\bibitem [{\citenamefont {Unruh}(1976)}]{unruh}%
  \BibitemOpen
  \bibfield  {author} {\bibinfo {author} {\bibfnamefont {W.~G.}\ \bibnamefont
  {Unruh}},\ }\bibfield  {title} {\bibinfo {title} {Notes on black-hole
  evaporation},\ }\href {https://doi.org/10.1103/PhysRevD.14.870}
  {\bibfield  {journal} {\bibinfo  {journal} {Phys. Rev. D}\ }\textbf {\bibinfo
  {volume} {14}},\ \bibinfo {pages} {870} (\bibinfo {year} {1976})}\BibitemShut
  {NoStop}%
\bibitem [{\citenamefont {Hawking}(1976)}]{hawking1}%
  \BibitemOpen
  \bibfield  {author} {\bibinfo {author} {\bibfnamefont {S.~W.}\ \bibnamefont
  {Hawking}},\ }\bibfield  {title} {\bibinfo {title} {Breakdown of
  predictability in gravitational collapse},\ }\href 
  {https://doi.org/10.1103/PhysRevD.14.2460} {\bibfield  {journal}
  {\bibinfo  {journal} {Phys. Rev. D}\ }\textbf {\bibinfo {volume} {14}},\
  \bibinfo {pages} {2460} (\bibinfo {year} {1976})}\BibitemShut {NoStop}%
\bibitem [{\citenamefont {Almheiri}\ \emph {et~al.}(2013)\citenamefont
  {Almheiri}, \citenamefont {Marolf}, \citenamefont {Polchinski},\ and\
  \citenamefont {J.~Sully}}]{polczinski}%
  \BibitemOpen
  \bibfield  {author} {\bibinfo {author} {\bibfnamefont {A.}~\bibnamefont
  {Almheiri}}, \bibinfo {author} {\bibfnamefont {D.}~\bibnamefont {Marolf}},
  \bibinfo {author} {\bibfnamefont {J.}~\bibnamefont {Polchinski}},\ and\
  \bibinfo {author} {\bibfnamefont {J.}~\bibnamefont {J.~Sully}},\ }\bibfield
  {title} {\bibinfo {title} {Black holes: complementarity or firewalls},\
  }\href {https://doi.org/10.1007/JHEP02(2013)062} {\bibfield  {journal}
  {\bibinfo  {journal} {Journal of High Energy Physics}\ }\textbf {\bibinfo
  {volume} {62}} (\bibinfo {year} {2013})}\BibitemShut {NoStop}%
\bibitem [{\citenamefont {Almheiri}\ \emph {et~al.}(2021)\citenamefont
  {Almheiri}, \citenamefont {Hartman}, \citenamefont {Maldacena}, \citenamefont
  {Shaghoulian},\ and\ \citenamefont {Tajdini}}]{maldac}%
  \BibitemOpen
  \bibfield  {author} {\bibinfo {author} {\bibfnamefont {A.}~\bibnamefont
  {Almheiri}}, \bibinfo {author} {\bibfnamefont {T.}~\bibnamefont {Hartman}},
  \bibinfo {author} {\bibfnamefont {J.}~\bibnamefont {Maldacena}}, \bibinfo
  {author} {\bibfnamefont {E.}~\bibnamefont {Shaghoulian}},\ and\ \bibinfo
  {author} {\bibfnamefont {A.}~\bibnamefont {Tajdini}},\ }\bibfield  {title}
  {\bibinfo {title} {The entropy of hawking radiation},\ }\href
  {https://doi.org/10.1103/RevModPhys.93.035002} {\bibfield  {journal} {\bibinfo  {journal}
  {Rev. of Mod. Phys.}\ }\textbf {\bibinfo {volume} {93}},\ \bibinfo {pages}
  {035002} (\bibinfo {year} {2021})}\BibitemShut {NoStop}%
\bibitem [{\citenamefont {Shakura}\ and\ \citenamefont
  {Sunyaev}(1973)}]{sunaev}%
  \BibitemOpen
  \bibfield  {author} {\bibinfo {author} {\bibfnamefont {N.~I.}\ \bibnamefont
  {Shakura}}\ and\ \bibinfo {author} {\bibfnamefont {N.~R.~A.}\ \bibnamefont
  {Sunyaev}},\ }\bibfield  {title} {\bibinfo {title} {Black holes in binary
  systems. observational appearance},\ }\href {} {\bibfield  {journal}
  {\bibinfo  {journal} {Astronomy and Astrophysics}\ }\textbf {\bibinfo
  {volume} {24}},\ \bibinfo {pages} {337} (\bibinfo {year} {1973})}\BibitemShut
  {NoStop}%
\bibitem [{\citenamefont {Salpeter}(1964)}]{salpet}%
  \BibitemOpen
  \bibfield  {author} {\bibinfo {author} {\bibfnamefont {E.~E.}\ \bibnamefont
  {Salpeter}},\ }\bibfield  {title} {\bibinfo {title} {Accretion of
  interstellar matter by massive objects},\ }\href {https://doi.org/10.1086/147973} {\bibfield
  {journal} {\bibinfo  {journal} {The Astrophysical Journal}\ }\textbf
  {\bibinfo {volume} {140}},\ \bibinfo {pages} {796} (\bibinfo {year}
  {1964})}\BibitemShut {NoStop}%
\bibitem [{\citenamefont {Zeldovich}(1964)}]{zeld}%
  \BibitemOpen
  \bibfield  {author} {\bibinfo {author} {\bibfnamefont {Y.~B.}\ \bibnamefont
  {Zeldovich}},\ }\bibfield  {title} {\bibinfo {title} {The fate of a star, and
  the liberation of gravitation energy in accretion},\ }\href {}
  {\bibfield  {journal} {\bibinfo  {journal} {Dokl. Akad. Nauk SSSR}\ }\textbf
  {\bibinfo {volume} {155}},\ \bibinfo {pages} {67} (\bibinfo {year}
  {1964})}\BibitemShut {NoStop}%
\bibitem [{\citenamefont {Frank}\ \emph {et~al.}(2002)\citenamefont {Frank},
  \citenamefont {King},\ and\ \citenamefont {Derek}}]{ksiazka}%
  \BibitemOpen
  \bibfield  {author} {\bibinfo {author} {\bibfnamefont {J.}~\bibnamefont
  {Frank}}, \bibinfo {author} {\bibfnamefont {A.}~\bibnamefont {King}},\ and\
  \bibinfo {author} {\bibfnamefont {D.}~\bibnamefont {Derek}},\ }\href {}
  {\emph {\bibinfo {title} {Accretion Power in Astrophysics: Third Edition}}}\
  (\bibinfo  {publisher} {Cambridge UP},\ \bibinfo {address} {Cambridge},\
  \bibinfo {year} {2002})\BibitemShut {NoStop}%
\bibitem [{\citenamefont {Merloni}\ \emph {et~al.}(2000)\citenamefont
  {Merloni}, \citenamefont {Fabian},\ and\ \citenamefont {Ross}}]{merloni}%
  \BibitemOpen
  \bibfield  {author} {\bibinfo {author} {\bibfnamefont {A.}~\bibnamefont
  {Merloni}}, \bibinfo {author} {\bibfnamefont {A.~C.}\ \bibnamefont
  {Fabian}},\ and\ \bibinfo {author} {\bibfnamefont {R.~R.}\ \bibnamefont
  {Ross}},\ }\bibfield  {title} {\bibinfo {title} {On the interpretation of the
  multicolour disc model for black hole candidates},\ }\href {https://doi.org/10.1046/j.1365-8711.2000.03226.x} {\bibfield
   {journal} {\bibinfo  {journal} {Monthly Notes of the Royal Astronomical
  Society}\ }\textbf {\bibinfo {volume} {313}},\ \bibinfo {pages} {193}
  (\bibinfo {year} {2000})}\BibitemShut {NoStop}%
\bibitem [{\citenamefont {Shapiro}\ \emph {et~al.}(1976)\citenamefont
  {Shapiro}, \citenamefont {Lightman},\ and\ \citenamefont
  {Eardley}}]{shapiro}%
  \BibitemOpen
  \bibfield  {author} {\bibinfo {author} {\bibfnamefont {S.~L.}\ \bibnamefont
  {Shapiro}}, \bibinfo {author} {\bibfnamefont {A.~P.}\ \bibnamefont
  {Lightman}},\ and\ \bibinfo {author} {\bibfnamefont {D.~M.}\ \bibnamefont
  {Eardley}},\ }\bibfield  {title} {\bibinfo {title} {A two-temperature
  accretion disk model for {C}ygnus {X}-1: structure and spectrum},\ }\href
  {https://doi.org/10.1086/154162} {\bibfield  {journal} {\bibinfo  {journal}
  {The Astrophysical Journal}\ }\textbf {\bibinfo {volume} {204}},\ \bibinfo
  {pages} {187} (\bibinfo {year} {1976})}\BibitemShut {NoStop}%
\bibitem [{\citenamefont {Brightman}\ \emph {et~al.}(2019)\citenamefont
  {Brightman}, \citenamefont {Bachetti}, \citenamefont {Earnshaw},
  \citenamefont {F{\"u}rst}, \citenamefont {García}, \citenamefont
  {Grefenstette}, \citenamefont {Heida}, \citenamefont {Kara}, \citenamefont
  {Madsen}, \citenamefont {Middleton}, \citenamefont {Stern}, \citenamefont
  {Tombesi},\ and\ \citenamefont {Walton}}]{supereddington}%
  \BibitemOpen
  \bibfield  {author} {\bibinfo {author} {\bibfnamefont {M.}~\bibnamefont
  {Brightman}}, \bibinfo {author} {\bibfnamefont {M.}~\bibnamefont {Bachetti}},
  \bibinfo {author} {\bibfnamefont {H.~P.}\ \bibnamefont {Earnshaw}}, \bibinfo
  {author} {\bibfnamefont {F.}~\bibnamefont {F{\"u}rst}}, \bibinfo {author}
  {\bibfnamefont {J.}~\bibnamefont {García}}, \bibinfo {author} {\bibfnamefont
  {B.}~\bibnamefont {Grefenstette}}, \bibinfo {author} {\bibfnamefont
  {M.}~\bibnamefont {Heida}}, \bibinfo {author} {\bibfnamefont
  {E.}~\bibnamefont {Kara}}, \bibinfo {author} {\bibfnamefont {K.~K.}\
  \bibnamefont {Madsen}}, \bibinfo {author} {\bibfnamefont {M.~J.}\
  \bibnamefont {Middleton}}, \bibinfo {author} {\bibfnamefont {D.}~\bibnamefont
  {Stern}}, \bibinfo {author} {\bibfnamefont {F.}~\bibnamefont {Tombesi}},\
  and\ \bibinfo {author} {\bibfnamefont {D.~J.}\ \bibnamefont {Walton}},\
  }\bibfield  {title} {\bibinfo {title} {Breaking the limit: {Super-Eddington}
  accretion onto black holes and neutron stars},\ }\href {} {\bibfield
  {journal} {\bibinfo  {journal} {Bulletin of the American Astronomical
  Society}\ }\textbf {\bibinfo {volume} {51}},\ \bibinfo {pages} {352}
  (\bibinfo {year} {2019})}\BibitemShut {NoStop}%
\bibitem [{\citenamefont {Fragile}\ \emph {et~al.}(2018)\citenamefont
  {Fragile}, \citenamefont {Etheridge}, \citenamefont {Anninos}, \citenamefont
  {Mishra},\ and\ \citenamefont {Klu{\'z}niak}}]{6p}%
  \BibitemOpen
  \bibfield  {author} {\bibinfo {author} {\bibfnamefont {P.~C.}\ \bibnamefont
  {Fragile}}, \bibinfo {author} {\bibfnamefont {S.~M.}\ \bibnamefont
  {Etheridge}}, \bibinfo {author} {\bibfnamefont {P.}~\bibnamefont {Anninos}},
  \bibinfo {author} {\bibfnamefont {B.}~\bibnamefont {Mishra}},\ and\ \bibinfo
  {author} {\bibfnamefont {W.}~\bibnamefont {Klu{\'z}niak}},\ }\bibfield
  {title} {\bibinfo {title} {Relativistic, viscous, radiation hydrodynamic
  simulations of geometrically thin disks. {I. T}hermal and other
  instabilities},\ }\href {https://doi.org/10.3847/1538-4357/aab788} {\bibfield
   {journal} {\bibinfo  {journal} {The Astrophysical Journal}\ }\textbf
  {\bibinfo {volume} {857}},\ \bibinfo {pages} {1} (\bibinfo {year}
  {2018})}\BibitemShut {NoStop}%
\bibitem [{\citenamefont {Farrah}\ \emph {et~al.}(2022)\citenamefont {Farrah},
  \citenamefont {Efstathiou}, \citenamefont {Afonso}, \citenamefont
  {Bernard-Salas}, \citenamefont {Cairns}, \citenamefont {Clements},
  \citenamefont {Croker}, \citenamefont {Hatziminaoglou}, \citenamefont
  {Joyce}, \citenamefont {Lacy}, \citenamefont {Lebouteiller}, \citenamefont
  {Lieblich}, \citenamefont {Lonsdale}, \citenamefont {Oliver}, \citenamefont
  {Pearson}, \citenamefont {Petty}, \citenamefont {Pitchford}, \citenamefont
  {Rigopoulou}, \citenamefont {Rowan-Robinson}, \citenamefont {Runburg},
  \citenamefont {Spoon}, \citenamefont {Verma},\ and\ \citenamefont
  {Wang}}]{farahzest}%
  \BibitemOpen
  \bibfield  {author} {\bibinfo {author} {\bibfnamefont {D.}~\bibnamefont
  {Farrah}}, \bibinfo {author} {\bibfnamefont {A.}~\bibnamefont {Efstathiou}},
  \bibinfo {author} {\bibfnamefont {J.}~\bibnamefont {Afonso}}, \bibinfo
  {author} {\bibfnamefont {J.}~\bibnamefont {Bernard-Salas}}, \bibinfo {author}
  {\bibfnamefont {J.}~\bibnamefont {Cairns}}, \bibinfo {author} {\bibfnamefont
  {D.}~\bibnamefont {Clements}}, \bibinfo {author} {\bibfnamefont
  {K.}~\bibnamefont {Croker}}, \bibinfo {author} {\bibfnamefont
  {E.}~\bibnamefont {Hatziminaoglou}}, \bibinfo {author} {\bibfnamefont
  {M.}~\bibnamefont {Joyce}}, \bibinfo {author} {\bibfnamefont
  {M.}~\bibnamefont {Lacy}}, \bibinfo {author} {\bibfnamefont {V.}~\bibnamefont
  {Lebouteiller}}, \bibinfo {author} {\bibfnamefont {A.}~\bibnamefont
  {Lieblich}}, \bibinfo {author} {\bibfnamefont {C.}~\bibnamefont {Lonsdale}},
  \bibinfo {author} {\bibfnamefont {S.}~\bibnamefont {Oliver}}, \bibinfo
  {author} {\bibfnamefont {C.}~\bibnamefont {Pearson}}, \bibinfo {author}
  {\bibfnamefont {S.}~\bibnamefont {Petty}}, \bibinfo {author} {\bibfnamefont
  {L.}~\bibnamefont {Pitchford}}, \bibinfo {author} {\bibfnamefont
  {D.}~\bibnamefont {Rigopoulou}}, \bibinfo {author} {\bibfnamefont
  {M.}~\bibnamefont {Rowan-Robinson}}, \bibinfo {author} {\bibfnamefont
  {J.}~\bibnamefont {Runburg}}, \bibinfo {author} {\bibfnamefont
  {H.}~\bibnamefont {Spoon}}, \bibinfo {author} {\bibfnamefont
  {A.}~\bibnamefont {Verma}},\ and\ \bibinfo {author} {\bibfnamefont
  {L.}~\bibnamefont {Wang}},\ }\bibfield  {title} {\bibinfo {title} {Stellar
  and black hole assembly in {$ z < 0.3$} infrared-luminous mergers:
  intermittent starbursts vs. super-{E}ddington accretion},\ }\href
  {https://doi.org/10.1093/mnras/stac980} {\bibfield  {journal} {\bibinfo
  {journal} {Monthly Notices of the Royal Astronomical Society}\ }\textbf
  {\bibinfo {volume} {513}},\ \bibinfo {pages} {4770} (\bibinfo {year}
  {2022})}\BibitemShut {NoStop}%
\bibitem [{\citenamefont {Dexter}\ \emph {et~al.}(2021)\citenamefont {Dexter},
  \citenamefont {Scepi},\ and\ \citenamefont {Begelman}}]{24p}%
  \BibitemOpen
  \bibfield  {author} {\bibinfo {author} {\bibfnamefont {J.}~\bibnamefont
  {Dexter}}, \bibinfo {author} {\bibfnamefont {N.}~\bibnamefont {Scepi}},\ and\
  \bibinfo {author} {\bibfnamefont {M.~C.}\ \bibnamefont {Begelman}},\
  }\bibfield  {title} {\bibinfo {title} {Radiation {GRMHD} simulations of the
  hard state of black hole {X}-ray binaries and the collapse of a hot accretion
  flow},\ }\href {https://doi.org/10.3847/2041-8213/ac2608} {\bibfield
  {journal} {\bibinfo  {journal} {The Astrophysical Journal Letters}\ }\textbf
  {\bibinfo {volume} {919}},\ \bibinfo {pages} {L20} (\bibinfo {year}
  {2021})}\BibitemShut {NoStop}%
\bibitem [{\citenamefont {Laha}\ \emph {et~al.}(2022)\citenamefont {Laha},
  \citenamefont {Meyer}, \citenamefont {Roychowdhury}, \citenamefont
  {{BecerraGonzalez}}, \citenamefont {Acosta–Pulido}, \citenamefont {Thapa},
  \citenamefont {Ghosh}, \citenamefont {Behar}, \citenamefont {Gallo},
  \citenamefont {Kriss}, \citenamefont {Panessa}, \citenamefont {Bianchi},
  \citenamefont {{La Franca}}, \citenamefont {Scepi}, \citenamefont {Begelman},
  \citenamefont {Longinotti}, \citenamefont {Lusso}, \citenamefont {Oates},
  \citenamefont {Nicholl},\ and\ \citenamefont {Cenko}}]{ga1flare}%
  \BibitemOpen
  \bibfield  {author} {\bibinfo {author} {\bibfnamefont {S.}~\bibnamefont
  {Laha}}, \bibinfo {author} {\bibfnamefont {E.}~\bibnamefont {Meyer}},
  \bibinfo {author} {\bibfnamefont {A.}~\bibnamefont {Roychowdhury}}, \bibinfo
  {author} {\bibfnamefont {J.}~\bibnamefont {{BecerraGonzalez}}}, \bibinfo
  {author} {\bibfnamefont {J.~A.}\ \bibnamefont {Acosta–Pulido}}, \bibinfo
  {author} {\bibfnamefont {A.}~\bibnamefont {Thapa}}, \bibinfo {author}
  {\bibfnamefont {R.}~\bibnamefont {Ghosh}}, \bibinfo {author} {\bibfnamefont
  {E.}~\bibnamefont {Behar}}, \bibinfo {author} {\bibfnamefont {L.~C.}\
  \bibnamefont {Gallo}}, \bibinfo {author} {\bibfnamefont {G.~A.}\ \bibnamefont
  {Kriss}}, \bibinfo {author} {\bibfnamefont {F.}~\bibnamefont {Panessa}},
  \bibinfo {author} {\bibfnamefont {S.}~\bibnamefont {Bianchi}}, \bibinfo
  {author} {\bibfnamefont {F.}~\bibnamefont {{La Franca}}}, \bibinfo {author}
  {\bibfnamefont {N.}~\bibnamefont {Scepi}}, \bibinfo {author} {\bibfnamefont
  {M.~C.}\ \bibnamefont {Begelman}}, \bibinfo {author} {\bibfnamefont {A.~L.}\
  \bibnamefont {Longinotti}}, \bibinfo {author} {\bibfnamefont
  {E.}~\bibnamefont {Lusso}}, \bibinfo {author} {\bibfnamefont
  {S.}~\bibnamefont {Oates}}, \bibinfo {author} {\bibfnamefont
  {M.}~\bibnamefont {Nicholl}},\ and\ \bibinfo {author} {\bibfnamefont {S.~B.}\
  \bibnamefont {Cenko}},\ }\bibfield  {title} {\bibinfo {title} {A radio,
  optical, {UV}, and {X}-ray view of the enigmatic changing-look active
  galactic nucleus {1ES} 1927+654 from its pre- to postflare states},\
  }\href {https://doi.org/10.3847/1538-4357/ac63aa} {\bibfield  {journal} {\bibinfo  {journal} {The Astrophysical
  Journal}\ }\textbf {\bibinfo {volume} {931}},\ \bibinfo {pages} {5} (\bibinfo
  {year} {2022})}\BibitemShut {NoStop}%
\bibitem [{\citenamefont {Blandford}\ and\ \citenamefont
  {Znajek}(1977)}]{znajek}%
  \BibitemOpen
  \bibfield  {author} {\bibinfo {author} {\bibfnamefont {R.~D.}\ \bibnamefont
  {Blandford}}\ and\ \bibinfo {author} {\bibfnamefont {R.~L.}\ \bibnamefont
  {Znajek}},\ }\bibfield  {title} {\bibinfo {title} {Electromagnetic extraction
  of energy from {K}err black holes},\ }\href {https://doi.org/10.1093/mnras/179.3.433} {\bibfield  {journal}
  {\bibinfo  {journal} {Monthly Notices of the Royal Astronomical Society}\
  }\textbf {\bibinfo {volume} {179}},\ \bibinfo {pages} {433} (\bibinfo {year}
  {1977})}\BibitemShut {NoStop}%
\bibitem [{\citenamefont {Goerbig}(2011)}]{geb}%
  \BibitemOpen
  \bibfield  {author} {\bibinfo {author} {\bibfnamefont {M.~O.}\ \bibnamefont
  {Goerbig}},\ }\bibfield  {title} {\bibinfo {title} {Electronic properties of
  graphene in a strong magnetic field},\ }\href {https://doi.org/10.1103/RevModPhys.83.1193} {\bibfield  {journal}
  {\bibinfo  {journal} {Rev. Mod. Phys.}\ }\textbf {\bibinfo {volume} {83}},\
  \bibinfo {pages} {1193} (\bibinfo {year} {2011})}\BibitemShut {NoStop}%
\bibitem [{\citenamefont {Landau}\ and\ \citenamefont
  {Lifshitz}(2001)}]{lanfield}%
  \BibitemOpen
  \bibfield  {author} {\bibinfo {author} {\bibfnamefont {L.~D.}\ \bibnamefont
  {Landau}}\ and\ \bibinfo {author} {\bibfnamefont {E.~M.}\ \bibnamefont
  {Lifshitz}},\ }\href {} {\emph {\bibinfo {title} {Field Theory}}}\
  (\bibinfo  {publisher} {Fizmatlit},\ \bibinfo {address} {Moscow},\ \bibinfo
  {year} {2001})\BibitemShut {NoStop}%
\bibitem [{\citenamefont {Landau}\ and\ \citenamefont
  {Lifshitz}(1978)}]{landaushort}%
  \BibitemOpen
  \bibfield  {author} {\bibinfo {author} {\bibfnamefont {L.~D.}\ \bibnamefont
  {Landau}}\ and\ \bibinfo {author} {\bibfnamefont {E.~M.}\ \bibnamefont
  {Lifshitz}},\ }\href {} {\emph {\bibinfo {title} {Short course of
  theoretical physics: {Q}uantum mechanics}}}\ (\bibinfo  {publisher} {Nauka},\
  \bibinfo {address} {Moscow},\ \bibinfo {year} {1978})\BibitemShut {NoStop}%
\bibitem [{\citenamefont {Duck}\ and\ \citenamefont {Sudarshan}(1998)}]{duck1}%
  \BibitemOpen
  \bibfield  {author} {\bibinfo {author} {\bibfnamefont {I.}~\bibnamefont
  {Duck}}\ and\ \bibinfo {author} {\bibfnamefont {E.~C.~G.}\ \bibnamefont
  {Sudarshan}},\ }\bibfield  {title} {\bibinfo {title} {Toward an understanding
  of the spin-statistics theorem},\ }\href {https://doi.org/10.1119/1.18860} {\bibfield  {journal}
  {\bibinfo  {journal} {Am. J. Phys.}\ }\textbf {\bibinfo {volume} {66}},\
  \bibinfo {pages} {284} (\bibinfo {year} {1998})}\BibitemShut {NoStop}%
\bibitem [{\citenamefont {Duck}\ and\ \citenamefont {Sudarshan}(1997)}]{duck2}%
  \BibitemOpen
  \bibfield  {author} {\bibinfo {author} {\bibfnamefont {I.}~\bibnamefont
  {Duck}}\ and\ \bibinfo {author} {\bibfnamefont {E.~C.~G.}\ \bibnamefont
  {Sudarshan}},\ }\href {} {\emph {\bibinfo {title} {Pauli and the
  Spin-Statistics Theorem}}}\ (\bibinfo  {publisher} {World Sc.},\ \bibinfo
  {address} {Singapore},\ \bibinfo {year} {1997})\BibitemShut {NoStop}%
\bibitem [{\citenamefont {Leinaas}\ and\ \citenamefont
  {Myrheim}(1977)}]{leinaas1977}%
  \BibitemOpen
  \bibfield  {author} {\bibinfo {author} {\bibfnamefont {J.~M.}\ \bibnamefont
  {Leinaas}}\ and\ \bibinfo {author} {\bibfnamefont {J.}~\bibnamefont
  {Myrheim}},\ }\bibfield  {title} {\bibinfo {title} {On the theory of
  identical particles},\ }\href {https://doi.org/10.1007/BF02727953} {\bibfield  {journal} {\bibinfo
  {journal} {Nuovo Cimmento}\ }\textbf {\bibinfo {volume} {37B}},\ \bibinfo
  {pages} {1} (\bibinfo {year} {1977})}\BibitemShut {NoStop}%
\bibitem [{\citenamefont {Rumer}\ and\ \citenamefont {Fet}(1970)}]{rumerfet1}%
  \BibitemOpen
  \bibfield  {author} {\bibinfo {author} {\bibfnamefont {Y.~R.}\ \bibnamefont
  {Rumer}}\ and\ \bibinfo {author} {\bibfnamefont {A.~I.}\ \bibnamefont
  {Fet}},\ }\href {} {\emph {\bibinfo {title} {Theory of Unitary
  Symmetry}}}\ (\bibinfo  {publisher} {Nauka},\ \bibinfo {address} {Moscow},\
  \bibinfo {year} {1970})\BibitemShut {NoStop}%
\end{thebibliography}


\begin{appendix}

\section{Size of cyclotron orbits for 2D electrons}
\label{A}
Let us consider a 2D electron in a perpendicular magnetic field $B$. The $x$ and $y$ components of its kinematic momentum are (for vector potential in Landau gauge $\mathbf{A}=(0,-Bx,0)$, $\mathbf{B}=rot \mathbf{A}=(0,0,B)$),
\begin{equation}
	\begin{array}{l}
	P_x=-i\hbar\frac{\partial}{\partial x},\\
	P_y=-i\hbar\frac{\partial}{\partial y}-eBx,\\
	\end{array}
	\end{equation}
The commutator $[P_x,P_y]_-=i\hbar eB$ (it is invariant with respect to magnetic field gauge). Two variables $Y=\frac{P_x}{eB}$ and $P_y$ are conjugated, as $[Y,P_y]_-=i\hbar$, and they can be considered as generalised 1D position and momentum. By application of Bohr-Sommerfeld rule \cite{landau1972} to this generalised 1D phase space, one obtains,
\begin{equation}
	\label{gwncond}
	\Delta S_{Y,P_y}=\Delta \oint P_y dY= h,
	\end{equation}
where $\Delta S_{Y,P_y}$ is the smallest portion of the 1D phase space $(Y,P_y)$, $h=2\pi \hbar$. 
The phase space $(Y,P_y)$ is in fact the renormalized space $(P_x,P_y)$. The closed trajectory in $(P_x,P_y)$ space is repeated in the space $(x,y)$, with the renormalization factor $\frac{1}{(eB)^2}$ and turned by the angle $\pi/2$, because of the Lorentz force, $\mathbf{F}=\frac{d\mathbf{P}}{dt}=e\frac{d\mathbf{r}}{dt}\times \mathbf{B}$, where $\mathbf{r}=(x,y)$ and $\mathbf{P}=(P_x,P_y)$. 

Thus, from (\ref{gwncond}) one obtains,
\begin{equation}
	\label{gwnkwant}
	\Delta S_{x,y}=\Delta \oint y dx=\frac{h}{eB},
	\end{equation}
which is the smallest 2D cyclotron orbit in $(x,y)$ of an electron ($\Delta S_{x,y}B =\Phi_0=\frac{h}{e}$ is the magnetic field flux quantum).

The cyclotron orbit (\ref{gwnkwant}) is single-loop. If, however, the elementary trajectory between turning points in the Bohr-Sommefeld rule is the path with additional $k$ loops, then the smallest piece of 1D phase space $(Y,P_y)$ is $(2k+1)h$ \cite{pra}. Such a situation happens when loop-less $\sigma_i$ braids cannot match the nearest neighbors in Wigner lattice. The elementary trajectory in this case can be $\sigma_i^{2k+1}$, i.e.,  the cyclotron braid with additional $k$ loops, and such a simplest trajectory must be taken in the Bohr-Sommerfeld rule instead of $\sigma_i$, which results in $(2k+1)$ times larger size of the phase space orbit.
Hence, instead of (\ref{gwnkwant}) one obtains,
\begin{equation}
	\label{gwnkwant1}
	\Delta S_{x,y}=\frac{(2k+1)h}{eB},
\end{equation}
which defines the size of $(2k+1)$-loop cyclotron orbit $\left(\sigma_i^{2k+1}\right)^2$ (the effective magnetic field flux quantum in the correlated state of electrons defined by braid generators $\sigma_i^{2k+1}$, equals to $\Phi_k =\frac{(2k+1)h}{e}$).	

The sizes of cyclotron orbits, single-loop (\ref{gwnkwant}) and $(2k+1)$-loop (\ref{gwnkwant1}), refer to the LLL. In higher LLs cyclotron orbits are proportional to the factor $(2n+1)$, with $n$ -- Landau index, because the kinetic energy in consecutive LLs scales linearly with Landau index, just $\sim(2n+1)$.
Hence the cyclotron orbit size in arbitrary LLs equals to,
\begin{equation}
	\label{gwnkwant2}
	\Delta S_{x,y}=(2n+1)(2k+1)\frac{h}{eB},
	\end{equation}
where $n$ -- Landau index, $(2k+1)$ -- number of loops in the orbit.

Note, that the size of cyclotron orbit in particular LLs is immune to interparticle interaction in a collective state of electrons, because the Bohr-Sommerfeld rule holds independently of interaction  (as the quasiclassical approximation is not perturbative with respect to interaction). In 2D systems of interacting electrons, cyclotron orbits can be deformed but they surface  (\ref{gwnkwant2}) is conserved and universal.

\section{General form of cyclotron braid generators for admissible trajectories for 2D interacting electrons in magnetic field and corresponding wave functions for FQHE hierarchy in the LLL}
\label{B}

	The general homotopy invariant for cyclotron electron correlations of 2D electrons has the form as given by Eq. (\ref{gwn11}),
	\begin{equation}
		\label{invariantogolny}
		\frac{BS}{N}=\frac{h}{x_1e}\pm\frac{h}{x_2e}\pm\dots\pm\frac{h}{x_qe},
	\end{equation}
	where $q=2k+1$ is the number of loops of cyclotron orbit and $x_i$ indicates the fraction of next-nearest neighbors in Wigner lattice commensurate with the $i$-th loop. The form of the invariant (\ref{invariantogolny}) results from the commensurability condition for a single-loop cyclotron orbit with next-nearest neighbors of some rank. If the total number of these neighbors is $\frac{N}{x}$, then the single-loop commensurability condition is $\frac{BS}{N/x}=\frac{h}{e}$, which gives the form of components (each per single loop) in (\ref{invariantogolny}). $\pm$ in (\ref{invariantogolny}) indicates a possible inverted ($-$) or congruent ($+$) circulation of a loop with respect to the preceding one. 
	
	To the invariant (\ref{invariantogolny}) it corresponds to the filling rate given by Eq. (\ref{gwn12}), i.e., 
	\begin{equation}
		\label{fillingrateogolne}
		\nu=\left(\frac{1}{x_1}\pm\frac{1}{x_2}\pm\dots \pm \frac{1}{x_q}\right)^{-1},
	\end{equation}
	because $\nu=\frac{N}{N_0}$ and the degeneracy of LLs, $N_0=\frac{BSe}{h}$. 

	To the invariant (\ref{invariantogolny})
	there correspond  generators for related cyclotron subgroup, which have  the following form,
	\begin{equation}
		\label{generatorogolne}
		\begin{array}{l}
			b_j= (\sigma_j\sigma_{j+1}\dots\sigma_{j+x_{1}-2}\sigma_{j+x_1-1}\sigma_{j+x_1-2}^{-1}\dots\sigma_j^{-1})\\
			(	\sigma_j\sigma_{j+1}\dots\sigma_{j+x_{1}-2}\sigma_{j+x_1-1}\sigma_{j+x_1-2}^{-1}\dots\sigma_j^{-1})^{\pm 1}\\
			\dots\\
			(\sigma_j\sigma_{j+1}\dots\sigma_{j+x_{q}-2}\sigma_{j+x_q-1}\sigma_{j+x_q-2}^{-1}\dots\sigma_j^{-1})^{\pm 1},\\
			j=1,\dots, N'\;\;N'=N-max(x_i),\\
		\end{array}
	\end{equation}
 where the segment 
	\begin{equation}
		\label{segment}
		(\sigma_j\sigma_{j+1}\dots\sigma_{j+x_{i}-2}\sigma_{j+x_i-1}\sigma_{j+x_i-2}^{-1}\dots\sigma_j^{-1})
	\end{equation} 
	defines the exchange of the electron $j$-th with $(j+x_i)$-th one as prescribed in (\ref{invariantogolny}) for $i$-th loop of $q$-loop cyclotron orbit. For $x_i=1$ (the nearest neighbors) this segment (\ref{segment}) is simply $\sigma_j$.  
	
	The generators (\ref{generatorogolne}) define elementary exchanges of electrons. Not all transpositions are possible but only those defined by the generators. Scalar unitary representation of generators (\ref{generatorogolne})
	is $e^{i (1\pm 1 \pm \dots \pm 1 )\pi}$, as for original electrons $\sigma_j\rightarrow e^{i\pi}$ and $\sigma_j^{-1}\rightarrow e^{-i\pi}$. Therefore, the segment (\ref{segment}) must generate  the polynomial factor to the multiparticle wave function, 
	\begin{equation}
		\prod_{j=1,k=1;j<mod(j,x_i,1)+(k-1)x_i}^{N', N/x_i} (z_j-z_{mod(j,x_i,1)+(k-1)x_i}),
\end{equation}
($N'$ is the collection of admissible values of $j$ at which the generator (\ref{generatorogolne}) can be defined, it is equal to $N-max(x_i)$ for $x_i$ entering (\ref{generatorogolne}))
as the projective scalar unitary  representation of this segment is $e^{i\pi}$ (or $e^{-i \pi}$ if it enters as an inverted operator). In the above formula $mod(j,x_i,1)$ is the rest of the division of $j$ by $x_i$ with offset $1$. 
Thus the total multiparticle wave function corresponding to generators (\ref{generatorogolne}) acquires the form,
\begin{equation}
\label{functionogolne}
\begin{array}{lll}
	\Psi(z_1, \dots, z_N)
	&=& {\cal{A}}	\prod_{j=1,k=1;j<mod(j,x_1,1)+(k-1)x_1}^{N', N/x_1}\\
	&& (z_j-z_{mod(j,x_1,1)+(k-1)x_1})	 \\
	&\times& 	\prod_{j=1,k=1;j<mod(j,x_2,1)+(k-1)x_2}^{N', N/x_2}\\
	&& (z_j-z_{mod(j,x_2,1)+(k-1)x_2})\\
	&\times&\dots \\
	&\times	&\prod_{j=1,k=1;j<mod(j,x_q,1)+(k-1)x_q}^{N', N/x_q} \\
	&&(z_j-z_{mod(j,x_q,1)+(k-1)x_q})	\\
	& \times& e^{-i\sum_{i=1}^N|z_{i}|^2/4l_B^2}, \\
\end{array}
\end{equation} 
for both two possibilities of scalar unitary representations related to $\pm$ in (\ref{invariantogolny}) causing only unimportant change of sign.

One can notice from Eq. (\ref{functionogolne}) that in the case of $x_1=x_2=\dots=x_q=1$ and only $+$ instead of $\pm$ in (\ref{invariantogolny}), the Laughlin function is reproduced for $\nu=\frac{1}{q}$. The envelope part of function (\ref{functionogolne}), $e^{-i\sum_{i=1}^N|z_{i}|^2/4l_B^2}$, is correct only in GaAs  and this envelope changes in graphene according to the different explicit form of single electron LL functions in graphene \cite{geb}. 

The examples of the wave function (\ref{functionogolne})
for various $\nu$ from the general FQHE hierarchy given by Eq. (\ref{fillingrateogolne}) are presented in Ref.  
\cite{sr2022} including the related energy calculation.

	\section{Homotopy of particle trajectories  in the vicinity of the event horizon in Schwarzschild metric: Derivation }
	\label{D}
	Let us consider trajectories of particles (with mass $m$ vanishingly small in comparison to central mass $M$) in the upper neighborhood of the Schwarzschild event horizon. These trajectories coincide with geodesics in the metric (\ref{metryka1}).  Because of the spherical symmetry of the gravitational field described by (\ref{metryka1}) these geodesics must lie in planes and without any loss of generality one can consider the geodesic plane $\theta=\frac{\pi}{2}$.
	The geodesics for a particle  with the mass $m$ can be determined in various equivalent classical dynamics formulations, e.g.,  by  solution of the Hamilton-Jacobi equation,
	\begin{equation}
		\label{geodesics}
		g^{ik}\frac{\partial S}{\partial x^i}\frac{\partial S}{\partial x^k}-m^2c^2=0,
	\end{equation}
	with $g^{ik}$ metric tensor components corresponding to metric (\ref{metryka1})  \cite{lanfield}.
	Eq. (\ref{geodesics}) attains  for the Schwarzschild metric  (\ref{metryka1})  the following form,
	\begin{equation}
		\label{jacobi1}
		\begin{array}{l}
		\left(1-\frac{r_s}{r}\right)^{-1}\left(  \frac{\partial S}{c\partial t}\right)^2\\
		-\left(1-\frac{r_s}{r}\right)\left(\frac{\partial S}{\partial r}\right)^2
		-\frac{1}{r^2}\left(\frac{\partial S}{\partial \phi}\right)^2-m^2c^2=0,\\
		\end{array}
	\end{equation}
	with the	function $S$ in the form,
	\begin{equation}
		\label{jacobi}
		S=-{\cal{E}}_0t+{\cal{L}}\phi+S_r(r).
	\end{equation}
	In the above formula  the quantities  ${\cal{E}}_0$ and  ${\cal{L}}$ are the particle energy and its angular momentum, respectively. ${\cal{E}}_0$ and  ${\cal{L}}$ are constants of motion.
	Eq. (\ref{geodesics}) can be also applied to define trajectories of photons assuming in (\ref{geodesics}) $m=0$.

	If one  substitutes Eq. (\ref{jacobi}) into Eq. (\ref{jacobi1}), then one can find $\frac{\partial S_r}{\partial r}$. By the integration of this formula one obtains, 
	\begin{equation}
		\begin{array}{l}	
		S_r=\int dr\left[ \frac{{\cal{E}}_0^2}{c^2}\left(1-\frac{r_s}{r}\right)^{-2} \right.\\
	\left.	-\left(m^2c^2
		+\frac{{\cal{L}}^2}{r^2}\right)\left(1-\frac{r_s}{r}\right)^{-1}\right]^{1/2}.\\
		\end{array}
	\end{equation}

	Geodesics are thus defined  by the condition $\frac{\partial S}{\partial{\cal{E}}_0}=const.$, which gives radial dependence of the trajectory  $r=r(t)$, and by the condition    $\frac{\partial S}{\partial{\cal{L}}}=const.$, determining the angular dependence $\phi=\phi(t)$                                of the particle trajectory. 
	The condition $\frac{\partial S}{\partial{\cal{E}}_0}=const.$ gives,
	\begin{equation}
		\label{promien}
		ct=\frac{{\cal{E}}_0}{mc^2}\int \frac{dr}{(1-\frac{r_s}{r})\sqrt{\left(\frac{{\cal{E}}_0}{mc^2}\right)^2-\left(1+\frac{{\cal{L}}^2}{m^2c^2r^2}\right)\left(1-\frac{r_s}{r}\right)}}.
	\end{equation}
	The  condition $\frac{\partial S}{\partial{\cal{L}}}=const.$ results in the relation,
	\begin{equation}
		\label{phase}
		\phi=\int dr\frac{{\cal{L}}}{r^2}\left[\frac{{\cal{E}}_0^2}{c^2}-\left(m^2c^2+\frac{{\cal{L}}^2}{r^2}\right)
		\left(1-\frac{r_s}{r}\right)\right]^{-1/2}.
	\end{equation}

	Eq. (\ref{promien}) can be rewritten in a differential form,
	\begin{equation}
		\label{differential}
		\frac{1}{1-r_s/r}\frac{dr}{cdt}=\frac{1}{{\cal{E}}_0}\left[{\cal{E}}_0^2-U^2(r)\right]^{1/2},
	\end{equation}
	with the  effective potential,
	\begin{equation}
		\label{potencjal}
		U(r)=mc^2\left[\left(1-\frac{r_s}{r}\right)\left(1+\frac{{\cal{L}}^2}{m^2c^2r^2}\right)\right]^{1/2},
	\end{equation}
	where ${\cal{E}}_0$ and ${\cal{L}}$
	are energy and angular momentum of the particle, respectively.

	The equation (\ref{differential}) allows for the definition of an accessible region for the motion via the following condition, ${\cal{E}}_0\geq  U(r)	$.
	Moreover, the condition 
	${\cal{E}}_0=  U(r)	$ defines circular orbits.
	Limiting circular orbits can be thus found  by the determination of extrema of $U(r)$.  Maxima of $U(r)$ define unstable orbits, whereas minima stable ones (depending on parameters ${\cal{E}}_0$ and ${\cal{L}}$, which are integrals of the motion). The conditions $U(r)={\cal{E}}_0$ and $\frac{\partial U(r)}{\partial r}=0$ (for extreme) attain the explicit  form,
	\begin{equation}
		\label{branches}
		\begin{array}{l}
			{\cal{E}}_0={\cal{L}}c\sqrt\frac{2}{rr_s}\left(1-\frac{r_s}{r}\right),\\
			\frac{r}{r_s}=\frac{{\cal{L}}^2}{m^2c^2r_s^2}\left[1\pm \sqrt{1-\frac{3m^2c^2r_s^2}{{\cal{L}}^2}}\right],
		\end{array}
	\end{equation}
	where the sign $+$ in the second equation  corresponds to stable orbits (minima of $U(r)$) and the sign $-$ to unstable ones (maxima of $U(r)$). Positions of stable and unstable circular orbits depend on energy ${\cal{E}}_0$ and angular momentum ${\cal{L}}$. 
	
	This is illustrated in Fig. \ref{photosphere} -- the upper curve (red one in this figure) gives positions of stable circular orbits (with respect to angular momentum ${\cal{L}}$) and the lower curve (blue one) gives positions of unstable circular orbits (also with respect to ${\cal{L}}$). The related value of ${\cal{E}}_0$ is given by the first equation of the system (\ref{branches}). 	
	
	\section{Pauli theorem on connection between statistics and spin}
\label{C}
Pauli theorem on spin-statistics connection \cite{pauli} states that quantum statistics of particles with half spin must be of fermionic type, while of particles with integer spin -- of bosonic type. This theorem is  supported by the quantum relativistic reasoning that within  the Dirac  electrodynamics for by spinor described particles, the Hamiltonian formulation is admitted for simultaneously particles and antiparticles and to assure positively defined kinetic energy of free particles,  the field operators defining particles (and antiparticles) must  anticommute, thus are of fermionic type \cite{landaushort}. Such a proof is confined, however, to free particles and to 3D position space -- the manifold on which particles are located. A wide discussion of  Pauli theorem on spin-statistics connection including various trials of its proof are presented in \cite{duck1,duck2}, where also a rigorous  proof of this theorem in the case of  noninteracting particles is provided. Pauli theorem holds, however, also for arbitrarily strongly interacting particles. This is visible  in terms of topology as the actual proof of Pauli theorem must invoke homotopy type reasoning in view of braid group based quantum statistics definition \cite{leinaas1977}.   Pauli theorem  follows   from the coincidence of unitary irreducible representations of the rotation group, which define quantization of spin or angular momentum  \cite{rumerfet1} with unitary representations of braid groups nominating quantum statistics \cite{mermin1979}. The agreement between unitary representations of both groups arises  due to the overlap of some elements of the braid group and the rotation group. The representations which are uniform on group generators must thus agree for whole groups. The half spin representation of the rotation group  must agree  with the odd representation  $e^{i\pi}=-1$ of the braid group for 3D manifold (and fermions),  whereas the integer angular momentum representation of the rotation group must agree with even  $e^{i0}=1$ representation of the permutation group (and bosons). Such an approach allows simultaneously for  the extension of Pauli theorem onto 2D manifolds with  anyons (which, in general are neither fermions nor bosons). For 2D position space the rotation group is Abelian, which causes that spin in 2D is not quantized and perfectly agrees with the  continuous  scalar unitary representations of the Artin braid group defining anyon fractional statistics. Such a topological proof of  Pauli theorem is immune to  particle interaction.

To be more specific, let us note that for 3D manifolds, the rotation group  $O(3)$  has  the  covering group  $SU(2)$  and  the  irreducible unitary representations of $SU(2)$ fall into two classes assigning integer and half-integer angular momenta. These two classes agree with only two  possible scalar unitary representations of the permutation group $S_N$, which is the  braid group for 3D manifolds. The representations of both groups coincide as they have some common elements.  However, for 2D manifolds the rotation group $O(2)$  is Abelian and isomorphic with $U(1)$ group possessing just the same continuous  unitary representations $e^{i\alpha}$, $\alpha\in[0,2\pi)$,  as the Artin group, which is the braid group for $M=R^2$. Thus, in two dimensional space   Pauli theorem also holds  for not quantized spin assigned by $s=\frac{\alpha}{2\pi}$ and similarly continuously changing anyon statistics defined by $e^{i\alpha}$ numbered by  $\alpha\in[0,2\pi)$.

Quantum  statistics and spin, though coincide via the agreement between unitary representations of rotation and braid groups, are in fact independent to some extent, and one can imagine a situation when the spin is still defined but the statistics not, as  in the case of the absence of a nontrivial braid group. Such a situation occurs in an extremely strong gravitational field inside the black hole and close to its event horizon, beneath the photon sphere rim,   as is demonstrated  in  the present paper.

	\section{Assessment of efficiency of the collapse of Fermi spheres of electrons and protons in the accretion disk close to event horizon of a quasar including general-relativistic corrections}
	\label{E}
	The matter falling onto the black holes of super-luminous quasars  must convert up to ca. 30 \% of their mass into radiation to explain their observable luminosity and simultaneously the rate of the increase of central black hole mass over a long time period  of their activity to be consistent with observed masses of supermassive   black holes in galaxies. As the  giant black holes in closer galaxies are of size at most of the order of billions masses of the Sun, the estimation of the mass consumption rate for   quasars with luminosity of order of $10^{40}$ W is typically ca. $10$ Sun mass per year (i.e., of order of $0.1$  Earth mass per second), in extreme case of $1000$ Sun mass per year ($10$ Earth mass per second).  
	
	Central black holes in quasars vary between $ 10^5 - 10^9$ of solar masses, as  have been measured using a reverberation mapping. Several dozen nearby large galaxies, including our own Milky Way, that do not have an active center and do not show any activity similar to a quasar, are confirmed to contain similar supermassive black holes in their  centers. Thus it is now thought that all large galaxies have giant black holes of this kind, but only a small fraction have sufficient matter in the right kind of orbit at their center to become active and power radiation in such a way as to be seen as quasars.

	For concreteness of the estimation let us assume that the central black hole in quasar consumes 5.6 $M_{\odot}$ per year, i.e., ca.  $0.06$ Earth mass per second. Let us assume the stable uniform in time process of matter accretion. The transport of matter across the disk is steady,  thus we can perform calculation e.g., per a single second. Using Eqs (\ref{fm}) and (\ref{en}) one can assess the energy stored in the Fermi spheres for electrons and protons, if all the electrons and protons from the gas mass equalled to $0.06 $ Earth mass, are compressed to the spatial volume $V$  per second. The local Fermi momentum
	\begin{equation}
		\label{fmm}
		p_F(r)=\hbar(3\pi^2 \rho(r))^{1/3}=\hbar\left(3\pi^2 \frac{n}{V(r)}\right)^{1/3},
	\end{equation}
	where $r$ is the distance from the center. $p_F(r)$  is constant in time and grows across the disk with increasing local concentration $\rho(r)=\frac{dn}{dV}=\frac{n}{V(r)}$, the same for electrons and protons. The latter equality holds for steady accretion and $n$ is the total number of electrons (or protons) per second, compressed in total to the volume $V(r)$ at the distance $r$ from the origin with central gravitational singularity. This means that  portions $dn$ of electrons and protons in infinitely small consecutive periods $dt$ incoming in radial direction towards the central singularity compressed to $dV$ at radius $r$ add up in due of a single second time period to the total constant flow of mass (in the example, of $0.06 \times M_Z$ kg/s, the Earth mass $M_Z=5.97 \times 10^{24}$ kg) and as the whole is compressed locally at $r$ to the volume $V(r)$. The locally accumulated energy in the Fermi spheres of electrons and protons  grows with lowering $r$ due to the increase of the compression caused by the gravitational field. This energy is proportional to $V(r)$ and, moreover, depends on $V(r)$ via the local  Fermi momentum  (\ref{fmm}) and in accordance  with  Eq. (\ref{en}) can be written as,
	\begin{equation}
		\label{enn}
		\begin{array}{l}
			E(r)=E_e(r)+E_p(r),\\
			E_e(r)=\frac{V(r)}{2 \pi^2 \hbar^3}\int_o^{p_F(r)}dp p^2 \left( \sqrt{p^2 c^2 +m_e^2 c^4}-m_e c^2\right),\\ 
			E_p(r)=\frac{V(r)}{2 \pi^2 \hbar^3}\int_o^{p_F(r)}dp p^2 \left(\sqrt{p^2 c^2 +m_p^2 c^4}-m_p c^2\right),\\
		\end{array}
	\end{equation}
	where the energy $E_{e(p)}$ refers to electrons (protons).

	At  the critical radius $r^*$ close to Schwarzschild zone (we argue that $r^*=1.5r_s$)  the decay of  quantum statistics takes place due to  the topological reason  and both Fermi spheres of electrons and protons collapse. The amount of energy given by Eq. (\ref{enn}) per one second can be thus released in vicinity of the Schwarzschild zone due to collapse of Fermi spheres. This released energy per second can contribute to the observed luminosity of quasar ($10^{40}$ W). This undergoes by portions $dn$ of particle flow incoming to $r^*$ region in infinite small time periods $dt$, adding up in total to  $0.06 \times  M_Z$ kg per second. The value  of the energy released depends on local  Fermi momentum and  attains  $10^{40}$ J at sufficiently high level of compression, i.e., at sufficiently small $V(r^*)$ determined from the self-consistent system of Eqs (\ref{fmm}) and (\ref{enn}) if one assumes $E(r^*)=10^{40}$ J.

	To the initial mass of a gas (assuming to be composed of hydrogen H)
	contribute mostly protons (ca. 2000 times more massive than electrons), thus the total number of electrons, the same as the number of protons, equals to, $n\simeq 0.06 M_Z /m_p\simeq 2.14 \times 10^{50}$ per second. Simultaneously solving Eqs (\ref{fmm}) and (\ref{enn}), assuming $n=2.14 \times  10^{50}$  in volume $V(r^*)$ and released energy $E(r^*)=10^{40}$ J,  we find the volume of plasma compression $V(r^*)=0.5 \times 10^5$ m$^3$ and electron or proton Fermi sphere radius $p_F(r^*)=5.4 \times 10^{-19}$ kg m/s. Electrons and protons (their amount per second) are compressed to the same volume $V(r^*)$ (due to neutrality of plasma), hence,  their  concentration at $r^*$, $\rho(r^*)=4.3 \times 10^{45}$ 1/m$^3$. The mass density at $r^*$ (including mass equivalent to the energy stored up in Fermi spheres of electrons and protons) is thus  $\xi(r^*)=\frac{0.06 M_Z}{V(r^*)}+\frac{E(r^*)}{c^2 V(r^*)}\simeq 9 \times 10^{18}$ kg/m$^3$, similar to mass density in neutron stars at Tolman-Oppenheimer-Volkoff limit (being of order of hadron density in atom nuclei). This is the uppermost mass density at the critical $r^*$, which evidences the self-consistency of the model. This limit regulates the matter consumption by a black hole, when the supply of the matter to an accretion disk is unlimited in the black hole surroundings.   The  released energy of $E(r^*)=10^{40}$ J is equivalent to 30 \% of the infalling mass of $0.06$ Earth mass (per second). It means that the compressed plasma with degenerate Fermi liquid of electrons (and also of protons) is at $r=r^*$ by 30 \% more massive than initial remote diluted gas. This increase of mass is caused by the gravitational field of the central black hole, which compresses both  systems of fermions and accumulates the energy in their Fermi spheres. 
	
	The energy of the gravitational field is accumulated in Fermi spheres of electrons and protons.  The ratio of total Fermi sphere energies of electrons and protons is $\frac{E_n(r^*)}{E_p(r^*)}\simeq 1.4$. The Fermi energy of electrons with Fermi momentum $p_F(r^*)=5.48 \times  10^{-19}$ kg m/s equals to $\varepsilon_F= 1 $ GeV (it is the upper possible energy of emitted photons), which in thermal scale (in units of $k_B=1$) is of order of $10^{13}$ K -- this makes the electron liquid quantumly degenerated at lower temperatures (quasars are not source of thermal gamma radiation, thus their actual temperatures are much lower). The Fermi energy of protons with Fermi momentum $p_F(r^*)=5.48 \times  10^{-19}$ kg m/s equals to $\varepsilon_F= 0.4 $ GeV (it is the upper possible energy of emitted photons by jumping of protons), which in thermal scale (in units of $k_B=1$) is of order of $4 \times 10^{12}$ K  -- thus for the temperature of plasma of order of $10^{6-9}$ K (the most realistic is $10^6$) protons also form the degenerated Fermi liquid. 
	
	The release  of energy  due to the  collapse of the  Fermi sphere of charged particles
	undergoes according to the  Fermi golden rule scheme for quantum transitions \cite{landau1972}, when such transitions are admitted by the local revoking of Pauli exclusion principle.  Charged carries (electrons and protons) couple to electromagnetic field and the matrix element of this coupling between an individual particle state in the Fermi sphere and its ground state  is the kernel of the Fermi golden rule formula for transition  probability per time unit for this particle.
	This interaction depends also on electromagnetic field strength, thus the increasing number of excited photons strengthens the coupling in the similar manner as at stimulated emission (known from e.g.,  laser action) and accelerates quantum transition of the Fermi sphere collapse.

	Note that the above estimation of the energy accumulated in Fermi spheres at critical distance from the gravitational singularity  has been done in conventional rigid coordinates, time and space-spherical coordinates  $(t,r,\theta,\phi)$ of the remote observer. Schwarzschild metric (\ref{metryka1}), though  written in the rigid and stationary coordinates, describes the folded spacetime.
	Even if in the Eq. (\ref{fmm}) one  replaces  $V(r)$  by the proper volume at the distance $r$ from the central singularity, then according to the Schwarzschild metric (\ref{metryka1})
	one obtains for the elementary proper volume the formula,
	\begin{equation}
		\label{correction}
		d{\cal{V}}=\left(1-\frac{r_s}{r}\right)^{-1/2}drr^2 sin\theta d\theta d\phi,
	\end{equation}
	which is only by the factor $\left(1-\frac{r}{r_s}\right)^{-1/2}$ greater than $dV=drr^2sin\theta d\theta d\phi$ in the remote observer coordinates neglecting curvature. At $r^*=1.5r_s$ this factor is ca. 1.7, which gives the reduction  of $p_F$ caused by gravitation curvature by factor ca. $1.7^{-1/3}\simeq0.84$, which does not change orders in the above estimations. The change of $p_F$ by one order of the magnitude would need the closer approaching the Schwarzschild horizon, at $r\simeq 1.000001 r_s$, i.e., rather distant from the critical $r^*=1.5r_s$. Hence, for the rough estimation of the effect of Fermi sphere collapse  the correction (\ref{correction})
	is unimportant and can be included as the factor $0.84$ to the right-hand side of Eq. (\ref{fmm}), which does not change the orders in the energy estimation.
	
	\end{appendix}

\end{document}